\documentclass[12pt,preprint]{aastex}

\def\aj{{AJ}}

\def\apj{{ApJ}}
\def\apjl{{ApJL}}
\def\apjs{{ApJS}}

\def\mnras{{MNRAS}}

\received{}
\accepted{25 July 2005}
\slugcomment{}
\shorttitle{A {\it Chandra} Survey of ULIRGs}
\shortauthors{Teng et al.}
\journalid{}{}
\articleid{}{}

\begin{document}

\title{A {\it Chandra} X-Ray Survey of Ultraluminous Infrared Galaxies}

\author{Stacy H. Teng{\footnote{Contacting author: stacyt@astro.umd.edu}}, A. S. Wilson, S. Veilleux}
\affil{Department of Astronomy, University of Maryland, College Park, MD 20742}
\author{A. J. Young}
\affil{Center for Space Research, Massachusetts Institute of Technology,  Cambridge, MA 02139}
\author{D. B. Sanders}
\affil{Institute for Astronomy, University of Hawaii, 2680 Woodlawn Drive, Honolulu, HI 96822}
\and
\author{N. M. Nagar}
\affil{Kapteyn Institute, Landleven 12, 9747 AD Groningen, The Netherlands \\
\& Astronomy Group, Departamento de F\'isica, Universidad de Concepci\'on, Casilla 160-C, Concepci\'on, Chile}

\begin{abstract}
We present results from {\it Chandra} observations of 14 ultraluminous infrared galaxies (ULIRGs; log($\rm{L_{IR}/L_\odot) \geq 12}$) with redshifts between 0.04 and 0.16.  The goals of the observations were to investigate any correlation between infrared color or luminosity and the properties of the X-ray emission and to attempt to determine whether these objects are powered by starbursts or active galactic nuclei (AGNs).   The sample contains approximately the same number of high and low luminosity objects and ``warm'' and ``cool'' ULIRGs.  All 14 galaxies were detected by {\it Chandra}.  Our analysis shows that the X-ray emission of the two Seyfert~1 galaxies in our sample are dominated by AGN.  The remaining 12 sources are too faint for conventional spectral fitting to be applicable.  Hardness ratios were used to estimate the spectral properties of these faint sources.  The photon indices, $\Gamma$'s, for our sample plus the {\it Chandra}--observed sample from \citet{ptak} peak in the range of 1.0--1.5, consistent with expectations for X-ray binaries in a starburst, an absorbed AGN, or hot bremsstrahlung from a starburst or AGN.  The values of $\Gamma$ for the objects in our sample classified as Seyferts (type 1 or 2) are larger than 2, while those classified as HII regions or LINERs tend to be less than 2.  The hard X-ray to far-infrared ratios for the 12 weak sources are similar to those of starbursts, but we cannot rule out the possibility of absorbed, possibly Compton-thick, AGNs in some of these objects.  Two of these faint sources were found to have X-ray counterparts to their double optical and infrared nuclei. 
\end{abstract}


\section{Introduction}
\label{sec:intro}
Ultraluminous infrared galaxies (ULIRGs) are defined as galaxies with $\rm{L_{IR}=L_{8-1000 \mu m} > 10^{12}~L_\odot}$ (H$\rm{_0 =}$75~km~s$^{-1}$~Mpc$^{-1}$, q$\rm{_0 = }$0).  Ground-based observations have shown that almost all of these galaxies are undergoing mergers (e.g. Sanders et al. 1988); these galactic mergers are thought to be the progenitors of some elliptical galaxies (e.g. Genzel et al. 2001; Veilleux et al. 2002) and may be a phase through which galaxies pass before a quasar is formed.  ULIRGs in the local universe may be compared with submillimeter sources at z = 1--4 observed with the SCUBA instrument (e.g. Smail et al. 1997; Hughes et al. 1998).  The mean properties (L$\rm{_{IR}}$, M(H$_2$), and near-infrared colors) of the two classes are remarkably similar.  Integration of the light from the ULIRG/SCUBA population shows that it may account for most or all of the submillimeter/far-infrared background, as a result of the strong cosmological evolution of these sources.

It is thus important to study the nature of ULIRGs at modest redshifts in order to understand their evolution and star formation at high redshifts.  One fundamental question that needs to be addressed is whether the high luminosity of these galaxies results from starbursts or accretion onto supermassive black holes (SMBHs).   Optical and infrared emission-line spectra suggest the energy source of ULIRGs is mostly from starbursts (e.g. Veilleux et al. 1997; Genzel et al. 1998) while the ``warm'' infrared colors of some objects, especially the more luminous galaxies, suggest black hole driven activity (e.g. Surace \& Sanders 1999).  There may exist an evolutionary sequence of merger-induced starburst galaxies (``cool'' ULIRGs), then ``warm'' ULIRGs, and then eventually Quasi-Stellar Objects (QSOs).  If this sequence is valid, one would expect dominance by active galactic nuclei (AGNs) in ``warm'' ULIRGs and, indeed, they tend to have Seyfert-like optical and near-infrared spectra (see e.g. Veilleux et al. 1995, 1999a,b).  Such an evolutionary sequence can also be tested with X-ray observations.

The nuclei of ULIRGs may be very heavily obscured.  Therefore, observations in UV, optical, near-infrared and even the far-infrared may not penetrate through the dust to the nucleus.  High resolution, high frequency radio observations can penetrate the high columns and are excellent probes of whether an AGN is present \citep{nagar}.  However, the bolometric luminosity in the radio band is insignificant, and thus radio observations cannot prove that accretion onto a SMBH is the dominant energy source.  The remaining option is to observe ULIRGs in hard X-rays.  While the ratio of hard X-ray luminosity to infrared luminosity is small in nearby ULIRGs, it is not very much smaller than that of radio-quiet QSOs.  Pure starburst galaxies, at low redshifts, do not exhibit unresolved hard X-ray (2.0-8.0~keV) nuclei.  Starbursts, such as M82, have extended hard X-ray emission from both diffuse gas and X-ray binaries \citep{griffiths}.  At the typical distances of the 1-Jy sample (z $\sim$ 0.1), the angular extent of the diffuse, hard X-ray emission in M82 would be $\simeq$0\farcs05 ($\simeq$ 100~pc) and that of the X-ray binaries would be $\simeq$0\farcs5 ($\simeq$ 1~kpc).  Starbursts in ULIRGs also have typical extents of $\lesssim$1~kpc \citep{soifer}, so their hard X-ray emissions in our sample may be difficult to resolve with {\it Chandra}.  The situation may be further complicated by the possible presence of a large column density of gas (N$_{\rm H}$ $\gtrsim$ 10$^{23-25}$ cm$^{-2}$), which can strongly attenuate directly viewed X-rays from an AGN.  If such is the case, X-rays emitted along the polar axis of a disk-like gaseous structure can be electron scattered into the line of sight, with the signature of an Fe~K$\alpha$ line of large (a few keV) equivalent width.  Discovery of such Fe~K$\alpha$ lines may be the best determinant of an energetically dominant AGN.  

Previous studies of X-ray emission from ULIRGs have been made by \citet{ptak} with {\it Chandra} and \citet{frances} with {\it XMM-Newton}.  The \citet{ptak} sample is a volume limited sample in which the redshifts of the galaxies do not exceed 0.045.  \citet{frances} selected their sample from the 15 ULIRGs observed by \citet{genz98} which included only the brightest nearby ULIRGs and only one ULIRG with redshift greater than 0.082.  Our sample encompasses ULIRGs with greater redshifts (0.043 $\le$ z $\le$ 0.163), and is selected to cover uniformly the {\it IRAS} color-luminosity plane.

The organization of this paper is as follows: \S~\ref{sec:sample} discusses the sample selection, \S~\ref{sec:obs} the observations and data reduction, \S~\ref{sec:analysis} the analysis and results concerning X-ray structure and spectra, \S~\ref{sec:dis} a discussion of some astrophysical consequences, and \S~\ref{sec:summary} a summary of our conclusions.  We will assume H$\rm{_0 = }$75~km~s$^{-1}$~Mpc$^{-1}$, q$\rm {_0 = }$0 throughout this paper.

\section{Sample Selection} 
\label{sec:sample}

The ``1-Jy sample'' of ULIRGs comprises a sample of {\it IRAS} galaxies with fluxes at 60~$\mu$m exceeding 1~Jy, L$\rm{_{IR} > 10^{12}}$~L$\rm{_\odot}$, galactic latitude $\mid$b$\mid$ $> 30^\circ$, f(60~$\mu$m) $>$ f(12~$\mu$m) (to avoid stars), {\it IRAS} color log($\rm{f(60~\mu m)/f(100~\mu m)}$) $>$ --0.3, and redshift 0 $< \rm{z} <$ 0.28 (e.g. Kim \& Sanders, 1998; Veilleux et al. 1999a,b; Kim et al. 2002; and Veilleux et al. 2002).  As part of our sample, we selected 13 galaxies from the 1-Jy sample.  Also observed were F17208-0014 and F23365+3604 which satisfy all the criteria except the galactic latitude one (they have $\mid$b$\mid$ $<$ 30$^\circ$), and F15250+3609 which meets all the criteria except, marginally, the luminosity one (it has L$\rm{_{IR}}$ = 10$^{11.99}$~L$_\odot$).  

The galaxies were selected to cover the full range in the key parameters L$\rm{_{IR}}$ and $\rm{f(25~\mu m)/f(60~\mu m)}$.  Specifically, we have chosen galaxies that are approximately equally distributed over L$\rm{_{IR}}$ and $\rm{f(25~\mu m)/f(60~\mu m)}$ in the following 4 bins: log($\rm{L_{IR}/L_\odot}$) $<$ 12.3, log($\rm{L_{IR}/L_\odot}$) $>$ 12.3, $\rm{f(25~\mu m)/f(60~\mu m)}$ $<$ 0.2 (``cool'' ULIRGs), and $\rm{f(25~\mu m)/f(60~\mu m)}$ $>$ 0.2 (``warm'' ULIRGs).  The sample size of 16 galaxies is large enough to adequately sample the range of infrared luminosities and infrared colors that characterize the class of ULIRGs.  

Of these 16 galaxies, only 14 were scheduled to be observed with {\it Chandra}.  Figure~\ref{fig:fluxvl} depicts the distribution of the entire 1-Jy sample in the log$(\rm{f(25~\mu m)/f(60~\mu m)})$ versus log($\rm{L_{IR}/L_\odot}$) plane.  Also indicated are those galaxies in the 1-Jy sample which have been previously observed with {\it Chandra} \citep{ptak}, the 14 galaxies observed by us with {\it Chandra}, and those observed by \citet{frances} with {\it XMM-Newton}.  It is notable that previous {\it Chandra} observations have focused preferentially on ULIRGs with low infrared luminosities (log($\rm{L_{IR}/L_{\odot}}$) $<$ 12.3), whereas our sample contains equal numbers of objects below and above 10$^{12.3}$ L$\rm{_{\odot}}$.  Furthermore, our sample contains approximately equal numbers in each of the four quadrants of Figure~\ref{fig:fluxvl}.  This distribution allows us to test whether objects with certain infrared colors and luminosities are powered preferentially by stars or by AGNs.  It is notable that essentially all objects in our sample are classified as ongoing or old mergers based on a comparison between the optical/near-infrared images and published numerical simulations of galaxy interactions \citep{vei02}.  

\section{Observations and Data Reduction}
\label{sec:obs}

The 14 galaxies were observed between December 2002 and September 2003.  Each galaxy was observed in a single exposure using the ACIS S3 CCD chip with the standard frame time of 3.2 seconds.  Total exposure times, actual dates of observations, and some properties of the sources are summarized in Table~\ref{tab:obssum}.  

Most of the data reduction and analysis was done using CIAO v2.3 with CALDB 2.23 and XSPEC v11.2.  Only a comparison of the radial profiles of two sources with models of the point spread function (PSF) was done using CIAO v3.0.2 and CALDB 2.25.  The effects of the CIAO and calibration updates since v2.3 are negligible for CCD resolution observations of our sources due to the low signal-to-noise ratios.  The data reduction followed the procedures outlined in the Science Analysis Threads for ACIS data on the CIAO webpages {\footnote{http://cxc.harvard.edu/ciao/.}}. 

The position of each X-ray source was determined using the IDL routine CNTRD.  The routine returns the X and Y positions of the centroid of a point source starting from user-provided initial guess positions.  The R.A. and Dec were then determined using the ds9 software from SAO based on the X and Y output of the CNTRD routine.

Nuclear spectra were extracted for the two bright X-Ray sources F01572+0009 and Z11598-0112 using the CIAO tool PSEXTRACT, which creates a source spectrum, a background spectrum, and associated response matrices.  PSEXTRACT also bins output spectra to a specified minimum number of counts per bin.  ACISABS was then applied to correct for the degradation in the low energy response of the ACIS chips as a result of deposition of contaminants on the pre-CCD filter or the CCDs.  We have ignored channels below 0.5~keV (where the instrumental calibration is uncertain) and above 8.0~keV (where there are few counts) in modeling the spectra.  The data were binned to both at least 15~counts per bin and at least 3~counts per bin.  The spectra were then modeled using the XSPEC package (\S~\ref{sec:bright}).  The 15~counts per bin spectra were modeled using $\chi^2$ statistics, while the 3~counts per bin spectra were modeled in c-stat mode using Poissonian statistics.  Although the c-stat fitting approach was devised for unbinned spectra, c-stat in XSPEC performs better if the data are binned to at least 1 count per bin.  This ensures that there are no bins with zero counts or any mis-match between the source and background spectra.  Therefore, the data were binned to 3~counts per bin for the c-stat mode.  For the other 12 sources, hardness ratios were calculated using the counts in a soft energy band (0.5--2.0~kev) and a hard energy band (2.0--8.0~kev), using equation~\ref{eq:hreq} in \S~\ref{sec:weak}.  The hardness ratios were then compared with power law and MEKAL models, photoelectrically absorbed by an intervening column.  This method provides estimates of the spectral parameters (\S~\ref{sec:weak}).  Due to the low number of counts from most of our galaxies, we cannot place meaningful constraints on more complex models.

\section{Analysis and Results}
\label{sec:analysis}

Here, we compare the positions of the X-ray sources with the optical and near-infrared positions in \S~\ref{sec:astro}.  Then we describe the X-ray structures in \S~\ref{sec:struc}.  All 14 sources observed were detected with {\it Chandra}, but only two were bright enough for detailed spectral modeling to be performed.  An analysis of the spectra of these two bright sources is presented in \S~\ref{sec:bright}, while the spectra of the others are discussed in \S~\ref{sec:weak}. 

\subsection{Astrometry}
\label{sec:astro}

The positions of the X-ray peaks are offset from the infrared and optical peaks by typically $\la$ 1\arcsec, which is consistent with the errors of measurement in the three wavebands.  Table~\ref{tab:coord} details the positions and offsets of each source.

\subsection{X-Ray Structure}
\label{sec:struc}

Two (F10190+1322 and F12112+0305) of the three sources in our sample that have double near-infrared and optical nuclei were found to have double X-ray nuclei.  The X-ray separations of these nuclei agree to within $\simeq$ 1$\arcsec$ of that of their infrared counterparts.  Figure~\ref{fig:galpair} shows X-ray grey scales of F10190+1322 and F12112+0305 with infrared and optical contours.  The infrared and optical positions of the midpoints between the two nuclei were shifted to match those of the X-ray midpoints.  The magnitude of the R.A. and Dec shift applied for F10190+1322 was 1\farcs2 and 0\farcs6, respectively, in the infrared, and 1\farcs42 and 0\farcs55, respectively, in the optical.  The R.A. and Dec shift applied for F12112+0305 was 0\farcs08 and 0\farcs9, respectively, in the infrared, and 0\farcs0 and 0\farcs9, respectively, in the optical.  There is thus weak evidence that the X-ray peaks of F10190+1322 are offset from the infrared and optical peaks by a fraction of an arcsecond.  The X-ray peaks of F12112+0305 are probably consistent with the locations of the infrared and optical peaks, after the small spatial shift has been applied.  The linear separations of the two X-ray peaks for F10190+1322 and F12112+0305 are approximately 5.6~kpc and 3.7~kpc, respectively.

The X-ray emissions of the two bright sources F01572+0009 and Z11598-0112 are concentrated in the central regions of these galaxies.  Figure~\ref{fig:bright} shows X-ray grey scale representations of these Seyfert~1 galaxies with infrared and optical contours superposed.  The infrared and optical images have been shifted so that the infrared and optical peaks match the positions of the X-ray peaks.  There is a suggestion of an E--W extension in F01572+0009 in the infrared and optical, as well as in the X-ray.  The morphologies of the X-ray emission have been investigated by comparing the radial profiles of the X-ray sources with the PSF models in the standard calibration library.  The X-ray spectra show that most of the observed flux from both of these bright sources is concentrated in the range of 0.5--2.0~keV.  Therefore, we compared the azimuthally averaged radial profiles of these sources in this energy range with PSFs evaluated at 1.0~keV.  This comparison (Figure~\ref{fig:psf}) shows that the soft X-ray emission is unresolved or, at best, marginally resolved.  

Most of the remaining 10 sources appear to be unresolved, with the exceptions of F16090-0139 and F17208-0014.  F16090-0139 appears to be extended in the NW--SE direction.  Its linear extent is approximately 8.2~kpc (3\farcs0).  F17208-0014 seems to be resolved with a linear diameter of approximately 5.2~kpc (6\farcs2).  The upper limits to the linear sizes of the rest of the sources fall in the range of 0.5--6.3~kpc.

\subsection{X-Ray Spectra}
\label{sec:spectra}

\subsubsection{The Bright Sources}
\label{sec:bright}

It is not surprising that F01572+0009 and Z11598-0112 are bright X-ray sources: they are the only type~1 Seyferts in our sample.  Using the 15~counts per bin data, one can use $\chi^2$ statistics to evaluate models of the continuum emission of these sources.  The spectra were first modeled with single power laws.  Due to the high flux of the soft component and consequently the high signal-to-noise ratio in the soft energy bins, such single power law models underestimate the flux in the hard energy band.  Therefore, a two component model was needed to describe the continuum spectra: a hard power law and a soft component represented by another power law or a MEKAL model were used.  If most of the flux in the soft band is produced by starbursts, then the soft band flux could plausibly be represented by a MEKAL model (for a hot diffuse gas).  The results of our modeling are listed in Table~\ref{tab:fits}, and the spectrum and the double power law model of F01572+0009 are shown in Figure~\ref{fig:f01572}.

The 3~counts per bin data were used to determine if there are weak emission lines in the spectra.  The continua of the data were first modeled using two power laws.  The spectral indices from these fits are consistent with those obtained from binning the data to at least 15 counts per bin and using $\chi^2$ statistics (see Table~\ref{tab:fits}).  This agreement indicates that binning the data to at least 3~counts per bin did not introduce any biases.  The F01572+0009 spectrum shows an excess above the power law continuum at around 6.0~keV (Figure~\ref{fig:f01572}), but this suggestion of an emission line(s) is not significant. The spectrum of Z11598-0112 has a possible emission line at an energy consistent with redshifted Fe~K$\alpha$ ({Figure~\ref{fig:z11598}).  We modeled the spectrum with a double power law as we had previously done with the 15~counts per bin data.  Then a narrow Gaussian feature was added to the continuum to represent the emission line.  Using the c-stat statistics option in XSPEC, the best fit model suggests that the line is located at a rest energy of 7.0~keV with an equivalent width of 1.0$^{+1.2}_{-0.7}$~keV (Table~\ref{tab:fits}).  Emission at 7~keV in the source frame would require the iron to be highly ionized.  The photon indices for the soft and hard power laws are 3.47 and 0.99, respectively. Figure~\ref{fig:z11598} shows the observed spectrum of Z11598-0112 together with the model components. 

The significance of the line cannot be tested using the F-test because the test is only valid for Gaussian statistics.  Therefore, simulated ``fake'' spectra, constructed in XSPEC, were used to determine the likelihood that the emission line seen in Z11598-0112 is real.  A set of 500 fake spectra were created using the FAKEIT command in XSPEC.  The task uses the response matrices associated with the real spectra and the best-fit continuum model to create artificial source and background spectra.  The simulated source and background spectra were also binned to at least 3~counts per bin.  As a sanity check, we modeled the simulated spectra.  The distribution of the photon indices in the 500 simulated spectra were consistent with the distribution of the photon indices for the observation of Z11598-0112, modeled with 3 counts per bin.  Therefore, we are confident that binning the data to an arbitrary small number of counts per bin did not introduce any biases.  We found that only 3 of the 500 spectra showed a flux at the energy of the line exceeding the measured line flux minus its error bar (a conservative measure of the line flux).  Thus, the line is significant at above the 99\% level.

It is also important to note that Z11598-0112 is considered to be a Narrow-Line Seyfert~1 (NLS1) galaxy whose H${\beta}$ line width (FWHM) is 770 km s$^{-1}$, based on the data presented in \citet{vei99a}.  NLS1s tend to have steeper soft X-ray spectra than normal Seyfert~1 galaxies, as shown independently by {\it ASCA} observations analyzed by \citet{leighly} and \citet{vaughan}.  Of the 24 NLS1's studied by \citet{leighly} and \citet{vaughan}, 79\% have soft flux in excess of the power law model that fits the individual spectra at high energies; the excess flux dominates the spectra at energies $\lesssim$ 1.5~keV.  The nominal power law photon indices of NLS1s over the 0.6--10~keV energy band span the range of 1.6--2.5, larger than normal Seyfert~1 galaxies \citep{vaughan}.  The X-ray spectral properties of Z11598-0112, in particular the steep soft X-ray spectrum and the flat hard X-ray spectrum, are consistent with a NLS1 classification.

\subsubsection{The Faint Sources}
\label{sec:weak}

Twelve of the fourteen galaxies that we have observed with {\it Chandra} do not have enough counts for the usual spectral modeling procedure.  These sources have total counts in the 0.5--8.0~keV band ranging from 3 to 92.  In order to determine the properties of these sources, we used hardness ratios to estimate model parameters from XSPEC.  The hardness ratio (HR) is defined as
\begin{equation}
\rm{
HR=\frac{H-S}{H+S}},
\label{eq:hreq}
\end{equation}
where H is the number of counts in the hard band (0.5 - 2.0~keV) and S is the number of counts in the soft band (2.0-8.0~keV).  The hardness ratios calculated from the data may then be compared with hardness ratios generated from models (such as a power law or a MEKAL) to determine the model parameters which describe the observations.

Two models were assumed -- a single power law and a single temperature MEKAL.  For a single power law, the photon index ($\Gamma$) was varied, while the temperature (kT) was varied in the MEKAL model.  In both models, photoelectric absorption by cold gas was included.  For each column density (N$\rm{_H}$) and model parameter ($\Gamma$ or kT) pair, XSPEC generated a model spectrum which was then multiplied by the effective area at each energy (obtained from the response matrices for the actual data) and sampled appropriately.  The output was thus a model of the number of photons detected per second as a function of energy, which could be compared with the observation.  These simulated data were then used to calculate the hardness ratio as a function of N$_{\rm H}$ and $\Gamma$ or N$\rm{_H}$ and kT.  One can then plot contours of constant hardness ratio on a diagram of N$\rm{_H}$ versus $\Gamma$ (Figure~\ref{fig:gamma}) or N$\rm{_H}$ versus kT (Figure~\ref{fig:kt}).  In each panel of Figures~\ref{fig:gamma} and \ref{fig:kt}, the middle curve represents the observed hardness ratio and the two other curves represent the observed hardness ratio plus and minus the error.  We have made use of 1-$\sigma$ values in Tables~1 and 2 of \citet{stat} to estimate the errors in our measurements.  Numerical values of model parameters are listed in Table~\ref{tab:hrs}, and the footnotes describe how the errors in the hardness ratios, $\Gamma$, and kT were obtained.

The reliability of this hardness ratio method can be tested by comparing its results with those given by the more traditional method of fitting models to the observed spectra.  For this comparison, we used the two bright sources F01572+0009 and Z11598-0112.  The hardness ratio method systematically underestimated the values of the photon indices compared with spectral fitting of single power law models when the full energy band was considered (compare Tables~\ref{tab:fits} and~\ref{tab:hrs}).  This is because, as previously mentioned in \S~\ref{sec:bright}, a single power law does not adequately describe the data for these two sources; their spectra are steeper at lower energies (Figures~\ref{fig:f01572} \& \ref{fig:z11598}).  Tests were made in which the two methods were compared over narrower spectral bands (within which a power law is a good representation of the spectra) and the results were found to be consistent to within the errors.

The reliability of our hardness ratio method can be further tested by comparing our results for the three galaxies that were also in the samples of \citet{ptak} and \citet{frances} with theirs.  Our results for F12112+0305, F15250+3609, and F17208-0014 agree with these previous {\it XMM-Newton} and {\it Chandra} observations to within the errors.

\section{Discussion}
\label{sec:dis}

\subsection{The Two Bright Sources}
\label{sec:brightdis}

Based on the radial profiles of F01572+0009 and Z11598-0112 (Figure~\ref{fig:psf}), the nuclear soft X-ray emissions are probably unresolved (F01572+0009 may be slightly extended -- see \S~\ref{sec:struc}).  This is consistent with the X-ray emission being dominated by the Seyfert~1 nuclei. 

The spectra of these two sources cannot be described by a single power law.  At least two components are needed: a hard power law and a soft component represented by a power law or a MEKAL model.  These two models describe the data equally well.  We find that the best-fit MEKAL model for the soft component in both Seyfert~1 galaxies has kT $\sim$250~eV.  \citet{ptak99} found that similar models applied to starbursts usually have a temperature greater than 600~eV.  Therefore, the low temperatures of the Seyfert~1 galaxies suggest that starburst activity may not be the dominant energy source of the soft component.  The same conclusion was drawn by \citet{boller02} for F01572+0009.

Following the analysis done on the {\it XMM-Newton} observations of F01572+0009 by \citet{boller02}, we can further support our claim that F01572+0009 and Z11598-0112 are AGN dominated through a quantitative comparison.  According to \citet{boller96}, the ratio of soft X-ray (0.1--2.4~keV) to far-infrared (40--120~$\mu$m) fluxes is $\rm{F_{SX1}/F_{FIR}}$ $\simeq$ 10$^{-2.5}$ for an unabsorbed starburst in equilibrium and 10$^{-1}$ for an unabsorbed Seyfert~1 galaxy.  Here we use the notation SX1 for the 0.1--2.4~keV ({\it ROSAT}) band, SX for the 0.5--2.0~keV ({\it Chandra}) band, HX1 for the 2--10~keV band, and HX for the 2--8~keV ({\it Chandra}) band.  The far-infrared fluxes in the 40--120~$\mu$m band can be estimated using Equation~1 in \citet{helou} transcribed here:
\begin{equation}
\rm{F_{FIR}=1.26\times10^{-14}\times[2.58f_\nu(60 \mu m) + f_\nu(100 \mu m)]},
\label{eq:firflux}
\end{equation}
where ${\rm f_\nu}$ are flux densities in Jy, and $\rm{F_{FIR}}$ is in W m$^{-2}$.  Using the flux densities in the {\it IRAS} 60 and 100~$\mu$m bands, we estimate the far-infrared fluxes for F01572+0009 and Z11598-0112 to be $9.93\times10^{-11}$~ergs~cm$^{-2}$~s$^{-1}$ and $1.13\times10^{-10}$~ergs~cm$^{-2}$~s$^{-1}$, respectively.  The {\it Chandra} soft X-ray fluxes are ${\rm{F_{SX}}}$ = $1.54\times10^{-12}$~ergs~cm$^{-2}$~s$^{-1}$ for F01572+0009 and $5.79\times10^{-13}$~ergs~cm$^{-2}$~s$^{-1}$ for Z11598-0112.  The $\rm{F_{SX}}$ values can be scaled to $\rm{F_{SX1}}$ values based on the photon indices of the soft power law continuum models:
\begin{equation}
\rm{F_{SX1}=F_{SX}\times\biggl(\frac{2.4^{-(\alpha-1)}-0.1^{-(\alpha-1)}}{2.0^{-(\alpha-1)}-0.5^{-(\alpha-1)}}\biggr)},
\label{eq:xconvert}
\end{equation}
where the spectral index $\alpha$ = $\Gamma-1$ and $\rm{f_{\nu}\propto \nu^{-\alpha}}$.  From these estimates, we find that both F01572+0009 and Z11598-0112 have $\rm{F_{SX1}/F_{FIR}}$ $\simeq$ 10$^{-1.2}$.  These flux ratios are approximately consistent with the value (10$^{-1.4}$) found by \citet{boller02}.  The soft X-ray to far infrared flux ratios thus indicate that the two Seyfert~1 galaxies in our sample are energetically dominated by AGNs. 

Given the large obscuration to the nuclei of ULIRGs, the X-ray flux in the soft band may be heavily attenuated by photoelectric absorption.  Therefore, a better method of determining whether a source is starburst or AGN dominated is to compare its hard X-ray flux to its bolometric flux.  \citet{review} suggested that, on average, the bolometric flux (F$\rm{_{bol}}$) of ULIRGs is 1.15 times the infrared flux (F$\rm{_{IR}}$) over the 8--1000 $\mu$m band.  The bolometric flux of our sources can be estimated using Equation~3 of \citet{kim98} and the {\it IRAS} flux densities taken directly from the {\it IRAS} Faint Source Catalog:
\begin{equation}
\rm{F_{IR} = 1.8\times 10^{-14} \times [13.48 \times f_\nu(12 \mu m) + 5.16 \times f_\nu(25 \mu m) + 2.58 \times f_\nu(60 \mu m) + f_\nu(100 \mu m)]}, 
\label{eq:fir}
\end{equation}
where ${\rm f_\nu}$ are flux densities in Jy, and $\rm{F_{FIR}}$ is in W m$^{-2}$.
  
A study of 109 quasars from the Palomar-Green survey by \citet{san89} suggests that the hard X-ray (HX1, 2--10~keV) to bolometric luminosity ratio $\gtrsim$ 10$^{-4}$ for these quasars.  Equations ``0'' and 1 of \citet{frances} suggest $\rm{F_{HX1}/F_{bol}}$ $<$ 10$^{-4}$ for starbursts.  Therefore, a $\rm{F_{HX1}/F_{bol}}$ value $\gtrsim$ 10$^{-4}$ implies AGN dominance.  $\rm{F_{HX1}/F_{bol}}$ for F01572+0009 and Z11598-0112 are 10$^{-2.2}$ and 10$^{-3.1}$, respectively.  This result further emphasizes that the two Seyfert~1 galaxies in our sample are AGN dominated.  This is consistent with what Veilleux et al. (1999a,b) concluded based on the broad line region (BLR) luminosity to bolometric luminosity ratio.  It appears that detection of an optical/near-infrared BLR in a ULIRG is a sufficient condition to predict AGN dominance in a ULIRG.  

\subsection{The Twelve Faint Sources}
\label{sec:weakdis}
Applying the nominal photon index derived from the hardness ratio, we estimated the 0.5--2.0~keV soft X-ray fluxes of the weak sources.  Using Equations~\ref{eq:firflux} and \ref{eq:xconvert}, we calculated the soft X-ray to far-infrared flux ratio ($\rm{F_{SX1}/F_{FIR}}$) for each galaxy in our sample, and found $\rm{F_{SX1}/F_{FIR}}$ $\lesssim$ 10$^{-3.5}$ for all sources, well below the values quoted above for both an unabsorbed starburst and an unabsorbed Seyfert~1 galaxy.  Furthermore, all of the weak sources have $\rm{F_{HX1}/F_{bol}}$ $<$ 10$^{-4.0}$.  The results are provided in Table~\ref{tab:xrfir}.  The large error bars in the flux ratios result from uncertainties in our estimation of the photon index.  

\subsection{Correlations with Infrared Color and Luminosity}
\label{sec:coraltn}
In Figure~\ref{fig:lumin} we plot the values of $\rm{F_{SX1}/F_{FIR}}$ and $\rm{F_{HX1}/F_{FIR}}$ for our sample of galaxies as a function of the infrared luminosities and the log of the {\it IRAS} 25-to-60~$\mu$m flux ratio.  It appears that both Seyfert~1's have ``warm'' colors, high infrared luminosity, and high X-ray to far-infrared flux ratios, as expected.  If having ``warm'' colors and high infrared luminosity is a pre-requisite for AGN dominance, then the other source (F12072-0444) in our sample in this quadrant of Figure~\ref{fig:fluxvl}, which has a low X-ray to far-infrared flux ratio, could be a Compton-thick AGN.  

Figure~\ref{fig:gamma_distr} shows the distribution of photon index {\it assuming the Galactic column} as a function of infrared color and luminosity.  With the exception of one LINER, all the non-Seyfert galaxies have photon indices less than 2.  There is no obvious correlation between $\Gamma$ and infrared color or luminosity.  It is evident that the Seyferts have larger $\Gamma$s than the rest of the sample.

If an intrinsic spectral shape is assumed, we can calculate N$_{\rm H}$ from the observed spectra.  By fixing $\Gamma$ at 1.7, we used the hardness ratio curves in Figure~\ref{fig:gamma} to estimate N$\rm_{H}$.  The results are tabulated in column~(8) of Table~\ref{tab:hrs}.  None of the sources appears to be Compton-thick.  This implies that the power source for the non-Seyfert~1 galaxies could simply be intrinsically weak AGNs.

\subsection{Comparison with Previous Work}
\label{sec:pastwork}
Following Figure~5 of \citet{ptak}, we have plotted the ratio of hard X-ray to far-infrared flux as a function of the {\it IRAS} 25-to-60~$\mu$m flux ratio.  We reproduced the \citet{ptak} plot and added our results to their figure (Figure~\ref{fig:ptakplot}).  The two type 1 Seyferts in our sample lie within the region occupied by other Seyferts and composites.  The dotted line in Figure~\ref{fig:ptakplot} represents the average $\mathrm{F_{2 - 10~keV}/F_{FIR}}$ of the pure starbursts.  Our data agree with \citet{ptak} in that the ratios of the hard X-ray to FIR fluxes of ULIRGs are usually similar to those of pure starbursts, suggesting that most ULIRGs are powered by starbursts.

Figure~\ref{fig:hist} compares the photon indices of a single-power-law fit to our galaxy spectra with those from the \citet{ptak} sample.  This histogram shows that our spectra tend to have higher photon indices.  The \citet{ptak} single power law fits are very poor models of the data based on the statistical values they reported.  Therefore, we have also compared our estimated photon indices with the photon indices from their two component (plasma + power law) fits (Figure~\ref{fig:hist2}).  The histograms in Figure~\ref{fig:hist2} peak at $\Gamma \sim$1.0--1.5 for both samples.  

The far-infrared luminosity is a good measure of the star formation rate (SFR) in dusty systems like ULIRGs.  Comparison of the SFR from X-ray measurements with the SFR from the FIR measurements will indicate if there is any energy contribution from sources other than the starburst (e.g. an AGN).  If the galaxy is powered purely by a starburst, then one would expect its ``SFR in X-rays'' to equal its ``SFR in FIR''.  There are many references in the literature that relate the 2--10~keV hard X-ray luminosity to SFR$_{\mathrm {2 - 10~keV}}$.  We have chosen to adopt the \citet{ranalli} and \citet{persic2} relations because they appear to be the best-fits to pure starbursts \citep[Figure~6b]{horn}.  The two SFR$_{\mathrm {2 - 10~keV}}$ values will give approximate upper and lower limits as a function of SFR$_{\rm FIR}$ {\footnote{It should be noted here that both the \citet{ranalli} and \citet{persic2} relations are calibrated based on the \citet{ken} relation for FIR.  \citet{ken} defined the FIR band to be 8--1000$\mu$m, which is our definition of the IR band.  However, since subsequent SFR relations are calibrated assuming the \citet{ken} relation is for the wavelength range of 40--120 $\mu$m, we will use our FIR values to determine SFR$_{\rm FIR}$ to be consistent with the literature.}}.   Figure~\ref{fig:sfr} relates the SFR from the \citet{ranalli} and \citet{persic2} relations to the SFR from the FIR luminosity.  We have included the \citet{ptak} sample in the plot.  From the figure, it is evident that our two Seyfert~1 galaxies have X-ray luminosities in excess of that expected from a starburst (as discussed above).  The only other source from our sample located above the line of equality is the LINER F04103-2838.  The four galaxies \citet{ptak} determined to have AGN contributions in their spectra (Mrk~231, Mrk~273, IRAS~05189-2524, and NGC~6240) are also located far above the line of equality.  The rest of the ULIRGs are most likely powered by starbursts.

\subsection{Emission Processes}
\label{sec:emiss}
In this section, we explore the possible emission processes that may be responsible for the detected X-ray emission.  

\subsubsection{X-ray Binaries}
\label{sec:xrb}
\citet{persic} suggested that X-ray binaries dominate the 2--15~keV luminosity in the absence of an AGN.  There are two types of X-ray binaries.  The high mass type (HMXB) produces X-ray emission from the accretion of wind material of an OB star onto its neutron star or black hole companion whereas the low mass type (LMXB) produces X-ray emission from accretion onto a neutron star or black hole via Roche lobe overflow from a low mass companion \citep{persic}.  Assuming the lifetime of a single burst of star formation is approximately 10$^8$ years, the X-ray emission from HMXBs are expected to dominate the hard spectra because the low mass stars have not had time to evolve away from the main sequence and to come into Roche lobe contact \citep{persic2}.  As a result, the spectra of the galaxies should reflect the properties of HMXBs with average $\Gamma \simeq$ 1.0--1.4 \citep[and references therein]{persic2}.  

However, in mergers, multiple events of starbursts may occur due to recurrent tidal interactions \citep{persic2}.  In this scenario, the low mass companions in binaries formed in the earlier starburst events have had time to evolve and, therefore, the X-ray emission from LMXBs may also contribute to the hard X-ray spectra.  To determine the significance of this contribution, mass estimates from \citet{vei02} were used in conjunction with Equation~7 of \citet{colbert} to approximate the SFR based on the global hard X-ray luminosity.  The LMXB-subtracted SFRs are within $\sim$3\% of the SFRs found without subtraction of the contribution of LMXBs.  Therefore, the LMXBs do not contributed significantly to the X-ray luminosity.

Based on the peaks of the histograms in Figure~\ref{fig:hist2} and the above discussion, the spectra of our sources may well be dominated by contributions from HMXBs.  This implies that the weak sources are starburst dominated, as suggested by their values of log($\mathrm{F_{2 - 10~keV}/F_{FIR}}$) (Figures~\ref{fig:ptakplot} and \ref{fig:sfr}).  If we assume that the luminosity in the hard X-ray band is solely due to X-ray binaries with luminosities $\gtrsim$ 10$^{37}$ ergs s$^{-1}$, then the galaxies in our sample contain 10$^3$--10$^5$ binaries that contribute to the X-ray emission.  Assuming a universal stellar initial mass function and star formation rate, these values are 4--170 times the number of X-ray binaries in the nearby starburst galaxies M82 and NGC~253 which in turn have 4--16 times the number of binaries in the Milky Way \citep{persic}.  If ultraluminous X-ray sources (ULXs) contribute to the emission, then a smaller number of X-ray binaries would be required.

\subsubsection{Thermal Bremsstrahlung}
\label{sec:brem}
On the other hand, the hard X-ray emission may result from thermal bremsstrahlung from a hot wind driven by either a starburst or an AGN.  For low abundances, the dominant emission process of a thermal gas is, of course, bremsstrahlung.  \citet{rupke02} and Rupke et al.(2005a,b,c) have shown evidence for galactic winds in the 1-Jy sample.  The spectrum of thermal bremsstrahlung is almost ``flat'' with $\Gamma \sim$1.2 for E~$\lesssim$~kT.  This value of $\Gamma$ coincides with the peak of the histograms of observed photon indices in Figures~\ref{fig:hist} and \ref{fig:hist2}.  For a given luminosity, a relationship between the size of the emitting region and electron density can be determined.  Assuming the emitting region is spherical, the ion density equals that of the electrons, and a gas temperature of $10^7 - 10^8$~K, the radius of the emitting region based on Equation~5.14b and Figure~5.3 of \citet{randl} is
\begin{equation}
\rm{R \approx 1.5 \times 10^7 L_{ff}^{\frac{1}{3}} {\it f}^{-\frac{1}{3}} n_e^{-\frac{2}{3}}}~\rm{cm},
\label{eq:brem}
\end{equation}
where {\it f} is the filling factor for the hot gas (L$\propto {\rm R}^3 f \epsilon_{\rm ff}$).  If the luminosity of these sources in the 0.5--8.0~keV band is completely due to thermal bremsstrahlung, assuming an electron density of $\rm{n_e = 1}$ cm$^{-3}$ and $f = 1$, the size of the emitting regions ranges from 0.21 to 0.50~kpc.  These sizes of the X-ray emitting regions, deduced assuming dominance of bremsstrahlung emission, are consistent with our finding that most of our faint sources are unresolved by {\it Chandra}.

\subsubsection{Absorbed Active Galactic Nuclei}
\label{sec:ctn}

As discussed in Section~\ref{sec:coraltn}, if we assume $\Gamma \simeq 1.7$ (the photon index of a typical unabsorbed AGN), the column density can be estimated from the hardness ratios.  For many of the galaxies in our sample, the total column density estimated in this way exceeds the Galactic value (see Table~\ref{tab:hrs}).  This suggests that many of the galaxies in our sample may be absorbed AGNs with total N$_{\rm H}$ about 2--16 times the Galactic value -- i.e. up to N$_{\rm H} \simeq 5 \times 10^{22}$~cm$^{-2}$.  These column density estimates could be inaccurate if absorption is patchy or if scattering from an ionized medium is significant, as in the cases of NGC~1068 \citep{matt} and NGC~6240 \citep{ptak}.  Nevertheless, Compton-thick AGNs have been detected in several ULIRGs -- e.g. IRAS~19254-7245 \citep{braito03}, Mrk~231 \citep{braito04, ptak}.  If the intervening column density is $\sim$1000  or more times the Galactic value, then Compton scattering will become important.  The quality of our data does not allow us to rule out this possibility.

\section{Summary}
\label{sec:summary}
We have obtained and analyzed X-ray observations of 14 ULIRGs with the {\it Chandra X-Ray Observatory}.  Although all 14 galaxies were detected in the 0.5--8.0~keV energy range, only two were bright enough for traditional spectral fitting to be applicable.  Spectral analysis of these two galaxies (F01572+0009 and Z11598-0112) with Seyfert~1 type optical spectra shows that their soft X-ray emissions are unresolved.  There is a suggestion of an emission line at 7.0~keV in the rest frame of Z11598-0112.  Monte Carlo simulations of the spectrum showed that this line is significant at more than the 99\% confidence level. The soft X-ray to far-infrared flux ratios and the hard X-ray to bolometric flux ratios of F01572+0009 and Z11598-0112 indicate that these two galaxies are energetically dominated by AGNs. This result is consistent with the statement by Veilleux et al. (1999a,b) that detection of an optical/near-infrared BLR in a ULIRG is a sufficient condition to predict AGN dominance in a ULIRG. 

Unfortunately, the rest (and majority) of our sample is too faint for conventional spectral fitting.  Instead, we used hardness ratios to estimate their spectral parameters.  The soft X-ray to far-infrared flux ratios and the hard X-ray to bolometric flux ratios suggest that these galaxies are not energetically dominated by AGNs.  A comparison between star formation rates derived from hard X-ray and far-infrared luminosities seems to support the idea that these objects are powered by starbursts, although the large uncertainties on the empirical relation between hard X-ray luminosities and star formation rates prevent us from making a definitive statement.  A histogram of the photon indices of the X-ray spectra peaks at $\Gamma$ = 1.0--1.5, consistent with the spectrum expected from high mass X-ray binaries or bremsstrahlung from a hot, starburst-- (or AGN--) driven wind.  However, we cannot rule out the possibility that these very hard X-ray spectra and the very low hard X-ray to far-infrared flux ratios are produced by absorbed AGNs.

\acknowledgments

We thank the anonymous referee for useful comments and suggestions.  This paper makes use of data products from the Two Micron All Sky Survey, which is a joint project of the University of Massachusetts and the Infrared Processing and Analysis Center/California Institute of Technology, funded by the National Aeronautics and Space Administration and the National Science Foundation.  This research has also made use of the NASA/IPAC Extragalactic Database (NED), which is operated by the Jet Propulsion Laboratory, California Institute of Technology, under contract with the National Aeronautics and Space Administration.  This research was supported by NASA through Chandra General Observer grants GO 34146 to the Universities of Hawaii and Maryland and GO 23147X to the University of Maryland.  ASW and SV were partially supported by NASA LTSA grants NAG 513065 and 56547, respectively.  Thanks are also due to Chris Reynolds who provided suggestions on an early version of this paper.

\clearpage
\begin{center}

\begin{deluxetable}{cccccccc}
\tabletypesize{\scriptsize}
\tablecolumns{8}
\tablewidth{0pc}
\tablecaption{Some Properties of the Sample and the {\it Chandra} Observations\tablenotemark{a}}
\tablehead{
\colhead{Source}&\colhead{z}&\colhead{log$_{10}$($\rm{\frac{L_{IR}}{L_{\odot}}}$)}&\colhead{log$_{10}$($\rm{\frac{f_{25 \mu m}}{f_{60 \mu m}}}$)}&\colhead{Spectral}&\colhead{In 1-Jy}&\colhead{Observation}&\colhead{Exposure\tablenotemark{c}}\\
\colhead{Name \tablenotemark{b}}&\colhead{}&\colhead{}&\colhead{}&\colhead{Type}&\colhead{Sample?}&\colhead{Date}&\colhead{(ks)}\\}
\startdata
F00188-0856&0.128&12.33&--0.85&LINER&Yes&2003 Sep 4&9.80\\
F01004-2237&0.118&12.24&--0.54&H {\tiny II} galaxy&Yes&2003 Aug 3&9.40\\
F01572+0009&0.163&12.53&--0.61&Seyfert 1&Yes&2003 Aug 26&10.60\\
Z03521+0028&0.152&12.45&--1.06&LINER&Yes&2002 Dec 25&7.20\\
F04103-2838&0.118&12.15&--0.53&LINER&Yes&2003 Apr 28&10.00\\
F10190+1322&0.077&12.00&--0.94&H {\tiny II} galaxy&Yes&2003 Jan 31&9.40\\
Z11598-0112&0.151&12.43&--0.64&Seyfert 1&Yes&2003 Apr 13&10.20\\
F12072-0444&0.129&12.35&--0.66&Seyfert 2&Yes&2003 Feb 1&9.20\\
F12112+0305&0.073&12.28&--1.22&LINER &Yes&2003 Apr 15&10.00\\
F15130-1958&0.109&12.09&--0.69&Seyfert 2&Yes&2003 Jun 2&9.80\\
F15250+3609\tablenotemark{d}&0.055&11.99&--0.74&LINER&No&2003 Aug 27&9.20\\
F16090-0139&0.134&12.49&--1.14&LINER&Yes&2003 Feb 10&9.80\\
F17208-0014\tablenotemark{e}&0.043&12.39&--1.27&H {\tiny II} galaxy&No&2003 May 7&8.60\\
F23365+3604\tablenotemark{d}&0.064&12.10&--0.94&LINER&No&2003 Feb 3&10.20\\
\enddata
\tablenotetext{a}{Redshift and IR luminosity are taken from \citet{kim02}, unless otherwise noted.  Spectral types are taken from \citet{vei99a} and references therein, unless otherwise noted.}
\tablenotetext{b}{All source names should be preceded by IRAS.}
\tablenotetext{c}{Total good time interval after dead-time corrections.} 
\tablenotetext{d}{Redshift, IR luminosity, and spectral type are taken from \citet{surace} and references therein.}
\tablenotetext{e}{Redshift, IR luminosity, and spectral type are taken from  \citet{vei99b}.}
\label{tab:obssum}
\end{deluxetable}

\clearpage

\begin{deluxetable}{ccccccccccc}
\rotate
\tabletypesize{\scriptsize}
\tablecolumns{4}
\tablewidth{0pc}
\tablecaption{Astrometry of Our Sample\tablenotemark{a}}
\tablehead{
\colhead{Source}&\colhead{RA}&\colhead{Dec}&\colhead{RA}&\colhead{Dec}&\colhead{RA}&\colhead{Dec}&\colhead{$\Delta$RA\tablenotemark{b}}&\colhead{$\Delta$Dec\tablenotemark{b}}&\colhead{$\Delta$RA\tablenotemark{b}}&\colhead{$\Delta$Dec\tablenotemark{b}}\\
\colhead{Name}&\colhead{X-Ray}&\colhead{X-Ray}&\colhead{Opt.}&\colhead{Opt.}&\colhead{IR}&\colhead{IR}&\colhead{XR - Opt.}&\colhead{XR - Opt.}&\colhead{XR - IR}&\colhead{XR - IR}\\
\colhead{}&\colhead{(h m s)}&\colhead{($^{\circ}$ $\arcmin$ $\arcsec$)}&\colhead{(h m s)}&\colhead{($^{\circ}$ $\arcmin$ $\arcsec$)}&\colhead{(h m s)}&\colhead{($^{\circ}$ $\arcmin$ $\arcsec$)}&\colhead{($\arcsec$)}&\colhead{($\arcsec$)}&\colhead{($\arcsec$)}&\colhead{($\arcsec$)}\\}
\startdata
F00188-0856&00 21 26.54&--08 39 25.9&00 21 26.48&--08 39 27.1&00 21 26.48&--08 39 27.1&0.90&1.2&0.90&1.2\\
F01004-2237&01 02 49.99&--22 21 57.3&01 02 49.92&--22 21 57.0&01 02 49.94&--22 21 57.3&1.05&--0.3&0.75&0.0\\
F01572+0009&01 59 50.26&+00 23 40.9&01 59 50.22&+00 23 40.6&01 59 50.23&+00 23 40.5&0.60&0.3&0.45&0.4\\
Z03521+0028-E&03 54 42.22&+00 37 02.9&03 54 42.23&+00 37 02.4&03 54 42.25&+00 37 02.0&--0.15&0.5&--0.45&0.9\\
Z03521+0028-W&---&---&03 54 42.16&+00 37 02.4&03 54 42.15&+00 37 02.0&---&---&---&---\\
F04103-2838&04 12 19.43&--28 30 25.0&04 12 19.47&--28 30 24.4&04 12 19.53&--28 30 24.4&--0.60&--0.6&--1.50&--0.6\\
F10190+1322-E&10 21 42.73&+13 06 55.3&10 21 42.85&+13 06 55.3&10 21 42.81&+13 06 55.0&--1.80&0.0&--1.20&0.3\\
F10190+1322-W&10 21 42.48&+13 06 54.2&10 21 42.55&+13 06 53.1&10 21 42.56&+13 06 53.3&--1.05&1.1&--1.20&0.9\\
Z11598-0112&12 02 26.77&--01 29 15.4&12 02 26.76&--01 29 15.7&12 02 26.70&--01 29 15.8&0.15&0.3&1.05&0.4\\
F12072-0444&12 09 45.15&--05 01 13.8&12 09 45.12&--05 01 13.9&12 09 45.12&--05 01 13.9&0.45&0.1&0.45&0.1\\
F12112+0305-NE&12 13 46.06&+02 48 41.0&12 13 46.11&+02 48 42.4&12 13 46.07&+02 48 42.0&--0.75&--1.4&--0.15&--1.0\\
F12112+0305-SW&12 13 45.97&+02 48 39.0&12 13 45.92&+02 48 39.4&12 13 45.97&+02 48 39.8&0.75&--0.4&0.00&--0.8\\
F15130-1958&15 15 55.20&--20 09 16.9&15 15 55.16&--20 09 17.0&15 15 55.16&--20 09 17.2&0.60&0.1&0.60&0.3\\
F15250+3609\tablenotemark{c}&15 26 59.45&+35 58 37.1&15 26 59.48&+35 58 37.7&15 26 59.41&+35 58 37.3&--0.45&--0.6&0.60&--0.2\\
F16090-0139&16 11 40.43&--01 47 06.4&16 11 40.42&--01 47 06.5&16 11 40.42&--01 47 05.8&0.15&0.1&0.15&--0.6\\
F17208-0014\tablenotemark{c}&17 23 22.00&--00 16 59.8&17 23 21.99&--00 17 00.6&17 23 21.96&--00 17 00.8&0.15&0.8&0.6&1.0\\
F23365+3604\tablenotemark{c}&23 39 01.30&+36 21 08.4&23 39 01.25&+36 21 09.1&23 39 01.27&+36 21 08.6&0.75&--0.7&0.45&--0.2\\
\enddata
\tablenotetext{a}{Positions are given in J2000 coordinates.  X-ray positions are from this work.  Unless otherwise noted, optical and IR positions are taken from \citet{kim02}, who state that their typical positional error ($\simeq$ 2$\sigma$) is estimated to be less than 0\farcs5.}
\tablenotetext{b}{$\Delta $RA$(\lambda_1-\lambda_2)=[$RA$(\lambda_1)-$RA$(\lambda_2)] \times $Cos$[$Dec$(\lambda_1)]\times \rm[{15\arcsec/1 s}]$.  $\Delta $Dec$(\lambda_1-\lambda_2)=$ Dec$(\lambda_1)-$Dec$(\lambda_2)$.}
\tablenotetext{c}{These sources are not part of the IRAS 1-Jy sample studied by \citet{kim02}.  Optical positions are taken from the USNO A2.0 Catalog.  IR positions are taken from the 2MASS Point Source Catalog (PSC).  Both catalogs are accessible through the VizieR Service at http://vizier.u-strasbg.fr/viz-bin/VizieR.  The expected astrometric accuracy for stars in the USNO A2.0 Catalog is 0\farcs25.  The astrometric accuracy of the 2MASS PSC is $<$ 0\farcs2.}
\label{tab:coord}
\end{deluxetable}

\clearpage

\begin{deluxetable}{ccccccccc}
\tabletypesize{\tiny}
\tablecolumns{8}
\tablewidth{0pc}
\setlength{\tabcolsep}{0.02in}
\tablecaption{Spectral Models for F01572+0009 \& Z11598-0112\tablenotemark{a}}
\tablehead{\colhead{Source}&\colhead{Model}&\colhead{$\rm{N_H}$(Galactic)\tablenotemark{b}}&\colhead{kT}&\colhead{Z\tablenotemark{c}}&\colhead{$\Gamma$\tablenotemark{d}}&\colhead{$K$\tablenotemark{e}}&\colhead{$\chi^2 / dof$}\\
\colhead{Name}&\colhead{}&\colhead{(cm$^{-2}$)}&\colhead{(keV)}&\colhead{(Z$_\odot$)}&\colhead{}&\colhead{}&\colhead{}\\}
\startdata
\cutinhead{Spectra Binned to 15 Counts per Bin}
F01572+0009&PL&$2.6\times10^{20}$&---&---&$\Gamma_1$=2.4$^{+0.1}_{-0.1}$&$K_{PL_1}$=$7.3^{+0.2}_{-0.2}\times10^{-4}$&175/132\\
&PL+PL&$2.6\times10^{20}$&---&---&$\Gamma_1$=2.6$^{+0.1}_{-0.1}$&$K_{PL_1}$=$6.8^{+0.3}_{-0.4}\times10^{-4}$&142/131\\
&&&&&$\Gamma_2$=0.7$^{+0.4}_{-0.5}$&$K_{PL_2}$=$3.4^{+3.4}_{-1.9}\times10^{-5}$&\\
&PL+MEKAL&$2.6\times10^{20}$&0.3$^{+0.1}_{-0.1}$&1.0&$\Gamma_1$=2.1$^{+0.1}_{-0.1}$&$K_{PL_1}$=$6.2^{+0.3}_{-0.4}\times10^{-4}$&133/130\\
&&&&&&$K_M$=$3.1^{+0.8}_{-0.8}\times10^{-4}$&\\
\hline
Z11598-0112&PL&$2.3\times10^{20}$&---&---&$\Gamma_1$=3.4$^{+0.1}_{-0.1}$&$K_{PL_1}$=$2.4^{+0.1}_{-0.1}\times10^{-4}$&88/58\\
&PL+PL&$2.3\times10^{20}$&---&---&$\Gamma_1$=3.6$^{+0.1}_{-0.1}$ &$K_{PL_1}$=$2.2^{+0.1}_{-0.1}\times10^{-4}$&58/57\\
&&&&&$\Gamma_2$=0.2$^{+0.3}_{-0.2}$&$K_{PL_2}$=$7.6^{+10.4}_{-3.3}\times10^{-6}$&\\
&PL+MEKAL&$2.3\times10^{20}$&0.3$^{+0.1}_{-0.1}$&1.0&$\Gamma_1$=2.5$^{+0.3}_{-0.2}$&$K_{PL_1}$=$1.7^{+0.2}_{-0.2}\times10^{-4}$&60/56\\
&&&&&&$K_M$=$2.7^{+0.6}_{-0.7}\times10^{-4}$&\\
\hline
\hline
\colhead{Source}&\colhead{Model}&\colhead{$\rm{N_H}$(Galactic)\tablenotemark{b}}&\colhead{Line Energy\tablenotemark{f}}&\colhead{EW}&\colhead{$\Gamma$\tablenotemark{d}}&\colhead{$K$\tablenotemark{e}}&\colhead{c-stat$^{\rm g} / $PHA bins}\\
\colhead{Name}&\colhead{}&\colhead{(cm$^{-2}$)}&\colhead{(keV)}&\colhead{(keV)}&\colhead{}&\colhead{}&\colhead{}\\
\cutinhead{Spectra Binned to 3 Counts per Bin}
F01572+0009&PL+PL&$2.6\times10^{20}$&---&---&$\Gamma_1$=2.6$^{+0.8}_{-0.6}$&$\rm{K_{PL_1}}$=$7.1^{+2.6}_{-1.6}\times10^{-4}$&260/252\\
&&&&&$\Gamma_2$=0.5$^{+0.3}_{-0.3}$&$\rm{K_{PL_2}}$=$3.1^{+1.8}_{-2.4}\times10^{-5}$&\\
\hline
&&&&&&&\\
Z11598-0112&PL+PL+ZGAUSS&$2.3\times10^{20}$&7.0$^{+0.1}_{-0.1}$&1.0$^{+1.2}_{-0.7}$&$\Gamma_1$=3.5$^{+0.7}_{-0.5}$ &$\rm{K_{PL_1}}$=$2.3^{+0.1}_{-0.2}\times10^{-4}$&171/129\\
&&&&&$\Gamma_2$=1.0$^{+0.5}_{-0.7}$&$\rm{K_{PL_2}}$=$1.5^{+1.5}_{-0.8}\times10^{-5}$&\\
&&&&&& $\rm{K_{line}}$=3.3$^{+3.9}_{-2.4}\times10^{-6}$&\\
\enddata
\tablenotetext{a} {All errors are 90\% confidence for each parameter.  The models applied to the data are a single power law, double power law, and MEKAL plus a power law.  All models assume absorption by the Galactic column.  Unless otherwise mentioned, all models were applied to data in the energy range of 0.5--8.0 keV.}
\tablenotetext{b}{Parameter fixed to the Galactic column, obtained from the CIAO observing toolkit accessible through http://asc.harvard.edu/toolkit/colden.jsp.}
\tablenotetext{c}{Metallicity fixed at solar.}
\tablenotetext{d}{The photon index for the first power law ($\Gamma_{1}$) in the double power law model was determined using data in the soft band (0.5--2.0 keV).  This value was kept fixed when fitting the double power law model to the data over the whole band (0.5--8.0 keV).}
\tablenotetext{e}{The normalization of the power law models is $\rm{K_{PL}}$ = photons cm$^{-2}$ s$^{-1}$ keV$^{-1}$ at 1 keV.   The normalization of the MEKAL thermal plasma model is $\rm{K_M}$ = $\rm{\frac{10^{-14}}{4\pi[D_A(1+z)]^2} \int n_en_HdV}$ where $\rm{D_A}$ is the angular size distance (cm), and $\rm{n_e}$ and $\rm{n_H}$ are the electron and hydrogen densities (cm$^{-3}$), respectively.  The normalization of the ZGAUSS model is $K_{line}$ = total photons cm$^{-2}$ s$^{-1}$ in the line.}
\tablenotetext{f}{This is the energy of the line in the rest frame of the galaxy.}
\tablenotetext{g}{Cash-statistics option in the XSPEC fitting package.}
\label{tab:fits}
\end{deluxetable}

\clearpage

\begin{deluxetable}{ccccccccc}
\tabletypesize{\scriptsize}
\tablecolumns{9}
\tablewidth{0pc}
\tablecaption{Spectral Models Derived from Hardness Ratios}
\tablehead{
\colhead{Source}&\colhead{N$\rm{_H}$(Galactic)\tablenotemark{a}}&\colhead{Total}&\colhead{Hard}&\colhead{Soft}&\colhead{Hardness}&\colhead{$\Gamma$\tablenotemark{d}}&\colhead{N$\rm_{H}$($\Gamma$=1.7)\tablenotemark{e}}&\colhead{kT\tablenotemark{f}}\\
\colhead{Name}&\colhead{($10^{20}$ cm$^{-2}$)}&\colhead{Counts}&\colhead{Counts (H)\tablenotemark{b}}&\colhead{Counts (S)\tablenotemark{b}}&\colhead{Ratio (HR)\tablenotemark{c}}&\colhead{}&\colhead{($10^{20}$ cm$^{-2}$)}&\colhead{(keV)}\\
\colhead{(1)}&\colhead{(2)}&\colhead{(3)}&\colhead{(4)}&\colhead{(5)}&\colhead{(6)}&\colhead{(7)}&\colhead{(8)}&\colhead{(9)}\\}
\startdata
F00188-0856&3.21&16.0&6.0$^{+3.6}_{-2.4}$&10.0$^{+4.3}_{-3.1}$&--0.25$^{+0.36}_{-0.25}$&1.1$^{+0.5}_{-0.6}$&36$^{+64}_{-31}$&79.9$^{+ \textrm{\textemdash}}_{-73}$\\
F01004-2237&1.58&20.0&6.0$^{+3.6}_{-2.4}$&14.0$^{+4.8}_{-3.7}$&--0.40$^{+0.32}_{-0.24}$&1.4$^{+0.6}_{-0.6}$&16$^{+47}_{- \textrm{\textemdash}}$&15.8$^{+62}_{-12}$\\
F01572+0009&2.56&4386.0&674.0$^{+26}_{-26}$&3712.0$^{+61}_{-61}$&--0.69$^{+0.02}_{-0.02}$&2.17$^{+0.10}_{-0.07}$&---&2.8$^{+0.2}_{-0.2}$\\
Z03521+0028&12.5&3.0&1.0$^{+2.3}_{-0.8}$&2.0$^{+2.6}_{-1.3}$&--0.33$^{+1.23}_{-0.54}$&1.5$^{+1.9}_{-3.4}$&25$^{+564}_{- \textrm{\textemdash}}$&12.6$^{+65}_{-11}$\\
F04103-2838&2.45&30.0&12.0$^{+4.6}_{-3.4}$&18.0$^{+5.3}_{-4.2}$&--0.20$^{+0.24}_{-0.18}$&1.05$^{+0.35}_{-0.45}$&43$^{+40}_{-24}$&79.9$^{+ \textrm{\textemdash}}_{-61}$\\
F10190+1322&3.78&16.0&6.0$^{+3.6}_{-2.4}$&10.0$^{+4.3}_{-3.1}$&--0.25$^{+0.36}_{-0.25}$&1.18$^{+0.50}_{-0.68}$&35$^{+65}_{-30}$&79.9$^{+ \textrm{\textemdash}}_{-73}$\\
Z11598-0112&2.25&1481.0&130.0$^{+11}_{-11}$&1351.0$^{+37}_{-37}$&--0.82$^{+0.03}_{-0.03}$&2.7$^{+0.2}_{-0.1}$&---&1.8$^{+0.2}_{-0.2}$\\
F12072-0444&3.32&16.0&2.0$^{+2.6}_{-1.3}$&14.0$^{+4.8}_{-3.7}$&--0.75$^{+0.43}_{-0.25}$&2.5$^{+4.6}_{-1.1}$&---&2.0$^{+48}_{-1.7}$\\
F12112+0305&1.75&51.0&14.0$^{+4.8}_{-3.7}$&37.0$^{+7.1}_{-6.1}$&--0.45$^{+0.19}_{-0.15}$&1.5$^{+0.4}_{-0.4}$&11$^{+24}_{- \textrm{\textemdash}}$&7.9$^{+72}_{-3.9}$\\
F15130-1958&8.60&38.0&7.0$^{+3.8}_{-2.6}$&31.0$^{+6.6}_{-5.5}$&--0.63$^{+0.24}_{-0.19}$&2.15$^{+0.75}_{-0.65}$&$<$17&3.2$^{+5.7}_{-1.5}$\\
F15250+3609&1.56&37.0&5.0$^{+3.4}_{-2.2}$&32.0$^{+6.7}_{-5.6}$&--0.73$^{+0.25}_{-0.20}$&2.27$^{+1.24}_{-0.77}$&$<$7&2.5$^{+5.4}_{-1.4}$\\
F16090-0139&9.25&27.0&10.0$^{+4.3}_{-3.1}$&17.0$^{+4.2}_{-4.1}$&--0.41$^{+0.23}_{-0.20}$&1.57$^{+0.53}_{-0.45}$&15$^{+33}_{- \textrm{\textemdash}}$&7.9$^{+72}_{-4.7}$\\
F17208-0014&9.96&92.0&30.0$^{+6.5}_{-5.5}$&62.0$^{+8.9}_{-7.9}$&--0.35$^{+0.13}_{-0.11}$&1.43$^{+0.27}_{-0.23}$&23$^{+18}_{- \textrm{\textemdash}}$&11.2$^{+69}_{-5.6}$\\
F23365+3604&9.36&34.0&14.0$^{+4.8}_{-3.7}$&20.0$^{+5.6}_{-4.4}$&--0.18$^{+0.22}_{-0.17}$&1.10$^{+0.35}_{-0.25}$&50$^{+39}_{-27}$&79.9$^{+ \textrm{\textemdash}}_{-68}$\\
\enddata
\tablenotetext{a}{Column densities were obtained from the CIAO observing toolkit accessible through http://asc.harvard.edu/toolkit/colden.jsp.}
\tablenotetext{b}{The counting errors for the faint sources ($<$ 1000 total counts) are determined assuming Poisson statistics, using Tables 1 \& 2 of \citet{stat} and the total number of counts in each band.  The counting errors for the bright sources are simply $\sqrt{\rm{N}}$.}
\tablenotetext{c}{The hardness ratio is defined in Equation~\ref{eq:hreq}. The errors given were determined from error propagation based on the hard and soft band counts (columns 4 and 5).}
\tablenotetext{d}{The photon index for a power law model where n(E)$\propto$E$^{-\Gamma}$.  $\Gamma$ was calculated with N$\rm_{H}$ fixed at the Galactic column density.  The errors in the photon indices were determined from the hardness ratio limits.}
\tablenotetext{e}{Estimated total column density for $\Gamma$=1.7, the photon index typical of an unobscured AGN.  Listed are the column densities required to produce the observed spectra.  The column densities of several sources (F01572+0009, Z11598-0112, and F12072-0444) cannot be modified to reduce $\Gamma$ to 1.7, since N$\rm_{H}$ must always be greater than or equal to the Galactic value (column (2)).}
\tablenotetext{f}{The temperature in a MEKAL model, assuming solar abundances.  The model has an upper limit of kT=79.9 keV, beyond which the MEKAL model is indistinguishable from a thermal bremsstrahlung model.  The errors were determined from the hardness ratio limits.}
\label{tab:hrs}
\end{deluxetable}

\clearpage

\begin{deluxetable}{ccccccccc}
\tabletypesize{\scriptsize}
\rotate
\tablecolumns{6}
\tablewidth{0pc}
\tablecaption{X-Ray and Infrared Fluxes and Luminosities}
\tablehead{
\colhead{Source}&\colhead{$\rm{F_{FIR}}$}&\colhead{$\rm{F_{SX1}}$\tablenotemark{a}}&\colhead{$\rm{F_{HX1}}$\tablenotemark{b}}&\colhead{log$_{10}$($\rm{\frac{F_{SX1}}{F_{FIR}}}$)}&\colhead{log$_{10}$($\rm{\frac{F_{HX1}}{F_{bol}}}$)}&\colhead{L$\rm{_{SX1}}$\tablenotemark{c}}&\colhead{L$\rm{_{HX1}}$\tablenotemark{c}}&\colhead{L$\rm{_{bol}}$\tablenotemark{c}}\\
\colhead{Name}&\colhead{(ergs cm$^{-2}$ s$^{-1}$)}&\colhead{(ergs cm$^{-2}$ s$^{-1}$)}&\colhead{(ergs cm$^{-2}$ s$^{-1}$)}&\colhead{}&\colhead{}&\colhead{(ergs s$^{-1}$)}&\colhead{(ergs s$^{-1}$)}&\colhead{(ergs s$^{-1}$)}\\}
\startdata
F00188-0856&1.3$\pm{0.1}\times 10^{-10}$&7.2$^{+6.5}_{-6.0}\times 10^{-15}$&2.3$^{+2.4}_{-1.8}\times 10^{-14}$&--4.2$^{+0.4}_{-0.4}$&--4.1$^{+0.5}_{-0.5}$&2.5$^{+2.3}_{-2.1}\times 10^{41}$&8.0$^{+8.4}_{-6.3}\times 10^{41}$&1.0$\pm{0.7}\times 10^{46}$\\
F01004-2237&9.7$\pm{0.6}\times 10^{-11}$&1.1$^{+8.8}_{-1.1}\times 10^{-14}$&2.0$^{+1.7}_{-1.4}\times 10^{-14}$&--3.9$^{+3.4}_{-0.4}$&--4.2$^{+0.4}_{-0.3}$&3.4$^{+26}_{-3.4}\times 10^{41}$&6.0$^{+4.9}_{-4.2}\times 10^{41}$&8.8$\pm{0.7}\times 10^{45}$\\
F01572+0009&9.9$\pm{0.7}\times 10^{-11}$&5.8$^{+0.1}_{-0.1}\times 10^{-12}$&1.5$^{+0.1}_{-0.1}\times 10^{-12}$&--1.2$^{+0.001}_{-0.001}$&--2.2$^{+0.05}_{-0.05}$&3.5$^{+0.1}_{-0.1}\times 10^{44}$&9.0$^{+0.7}_{-0.7}\times 10^{43}$&1.5$\pm{0.1}\times 10^{46}$\\
Z03521+0028&6.6$\pm{0.5}\times 10^{-11}$&2.0$^{+5.2}_{-1.7}\times 10^{-15}$&3.7$^{+9.6}_{-3.2}\times 10^{-15}$&--4.5$^{+1.1}_{-0.4}$&--4.8$^{+1.1}_{-0.4}$&1.0$^{+2.7}_{-0.9}\times 10^{41}$&1.9$^{+4.9}_{-1.6}\times 10^{41}$&1.2$\pm{0.1}\times 10^{46}$\\
F04103-2838&8.1$\pm{0.4}\times 10^{-11}$&1.1$^{+7.1}_{-1.1}\times 10^{-14}$&3.8$^{+2.5}_{-2.2}\times 10^{-14}$&--3.9$^{+2.8}_{-0.4}$&--4.1$^{+0.3}_{-0.3}$&3.3$^{+21}_{-3.3}\times 10^{41}$&1.1$^{+0.7}_{-0.7}\times 10^{42}$&1.2$\pm{0.2}\times 10^{46}$\\
F10190+1322&1.8$\pm{0.1}\times 10^{-10}$&6.4$^{+5.7}_{-4.9}\times 10^{-15}$&1.8$^{+1.7}_{-1.4}\times 10^{-14}$&--4.4$^{+0.4}_{-0.3}$&--4.3$^{+0.4}_{-0.3}$&7.9$^{+6.9}_{-6.0}\times 10^{40}$&2.2$^{+2.1}_{-1.7}\times 10^{41}$&4.4$\pm{0.2}\times 10^{45}$\\
Z11598-0112&1.1$\pm{0.1}\times 10^{-10}$&7.0$^{+0.6}_{-0.6}\times 10^{-12}$&2.6$^{+0.1}_{-0.1}\times 10^{-13}$&--1.2$^{+0.05}_{-0.05}$&--3.1$^{+0.03}_{-0.03}$&3.5$^{+0.3}_{-0.3}\times 10^{44}$&1.3$^{+0.1}_{-0.1}\times 10^{43}$&1.7$\pm{0.1}\times 10^{46}$\\
F12072-0444&1.1$\pm{0.1}\times 10^{-10}$&2.3$^{+17}_{-2.3}\times 10^{-14}$&4.0$^{+3.6}_{-3.1}\times 10^{-15}$&--3.7$^{+3.3}_{-0.4}$&--4.8$^{+0.4}_{-0.3}$&8.2$^{+53}_{-8.2}\times 10^{41}$&1.4$^{+1.3}_{-1.1}\times 10^{41}$&9.9$\pm{0.6}\times 10^{45}$\\
F12112+0305&4.0$\pm{0.2}\times 10^{-10}$&2.3$^{+1.1}_{-1.1}\times 10^{-14}$&3.5$^{+1.7}_{-1.7}\times 10^{-14}$&--4.2$^{+0.2}_{-0.2}$&--4.3$^{+0.2}_{-0.2}$&2.6$^{+2.6}_{-1.2}\times 10^{41}$&3.8$^{+1.9}_{-1.9}\times 10^{41}$&8.2$\pm{0.4}\times 10^{45}$\\
F15130-1958&9.1$\pm{0.7}\times 10^{-11}$&3.0$^{+1.3}_{-1.7}\times 10^{-14}$&1.4$^{+0.8}_{-0.7}\times 10^{-14}$&--3.5$^{+0.2}_{-0.3}$&--4.2$^{+0.2}_{-0.2}$&7.7$^{+44}_{-40}\times 10^{41}$&3.6$^{+1.9}_{-1.8}\times 10^{41}$&5.8$\pm{0.4}\times 10^{45}$\\
F15250+3609&3.1$\pm{0.1}\times 10^{-10}$&2.2$^{+18}_{-2.2}\times 10^{-14}$&6.2$^{+5.8}_{-4.9}\times 10^{-15}$&--4.2$^{+3.5}_{-0.4}$&--5.1$^{+0.4}_{-0.4}$&1.3$^{+6.5}_{-1.3}\times 10^{41}$&3.8$^{+3.5}_{-3.0}\times 10^{40}$&4.3$\pm{0.2}\times 10^{45}$\\
F16090-0139&1.8$\pm{0.1}\times 10^{-10}$&1.2$^{+11}_{-1.2}\times 10^{-14}$&1.9$^{+1.6}_{-1.5}\times 10^{-14}$&--4.2$^{+3.8}_{-0.4}$&--4.3$^{+0.4}_{-0.4}$&4.8$^{+6.0}_{-4.8}\times 10^{41}$&7.4$^{+6.1}_{-5.8}\times 10^{41}$&1.4$\pm{0.1}\times 10^{46}$\\
F17208-0014&1.5$\pm{0.1}\times 10^{-9}$&4.4$^{+1.6}_{-1.5}\times 10^{-14}$&9.0$^{+3.7}_{-3.4}\times 10^{-14}$&--4.5$^{+0.2}_{-0.1}$&--5.5$^{+0.2}_{-0.2}$&1.6$^{+2.6}_{-2.5}\times 10^{41}$&3.3$^{+1.4}_{-1.3}\times 10^{41}$&9.7$\pm{0.4}\times 10^{46}$\\
F23365+3604&3.4$\pm{0.2}\times 10^{-10}$&1.1$^{+7.0}_{-1.1}\times 10^{-14}$&3.9$^{+2.6}_{-2.4}\times 10^{-14}$&--4.5$^{+2.8}_{-0.4}$&--4.2$^{+0.3}_{-0.3}$&9.2$^{+58}_{-9.2}\times 10^{40}$&3.3$^{+2.2}_{-2.0}\times 10^{41}$&5.6$\pm{0.3}\times 10^{45}$\\
\enddata
\tablenotetext{a}{Soft X-ray flux in the range of 0.1--2.4~keV scaled from the 0.5--2.0~keV flux using Equation~\ref{eq:xconvert}.  The 0.5--2.0~keV flux of the 12 weak sources is based on the power law model derived from the hardness ratio.}
\tablenotetext{b}{Hard X-ray flux in the range of 2.0--10.0~keV scaled from the 2.0--8.0~keV flux calculated using an equation similar to Equation~\ref{eq:xconvert}.  The 2.0--8.0~keV flux of the 12 weak sources is based on the power law model derived from the hardness ratio.}
\tablenotetext{c}{Luminosity distance calculated from Equation~2 of \citet{kim98}.}
\label{tab:xrfir}
\end{deluxetable}

\clearpage

\begin{figure}
\figurenum{1}
\epsscale{0.5}
\center
\rotatebox{-90}{\plotone{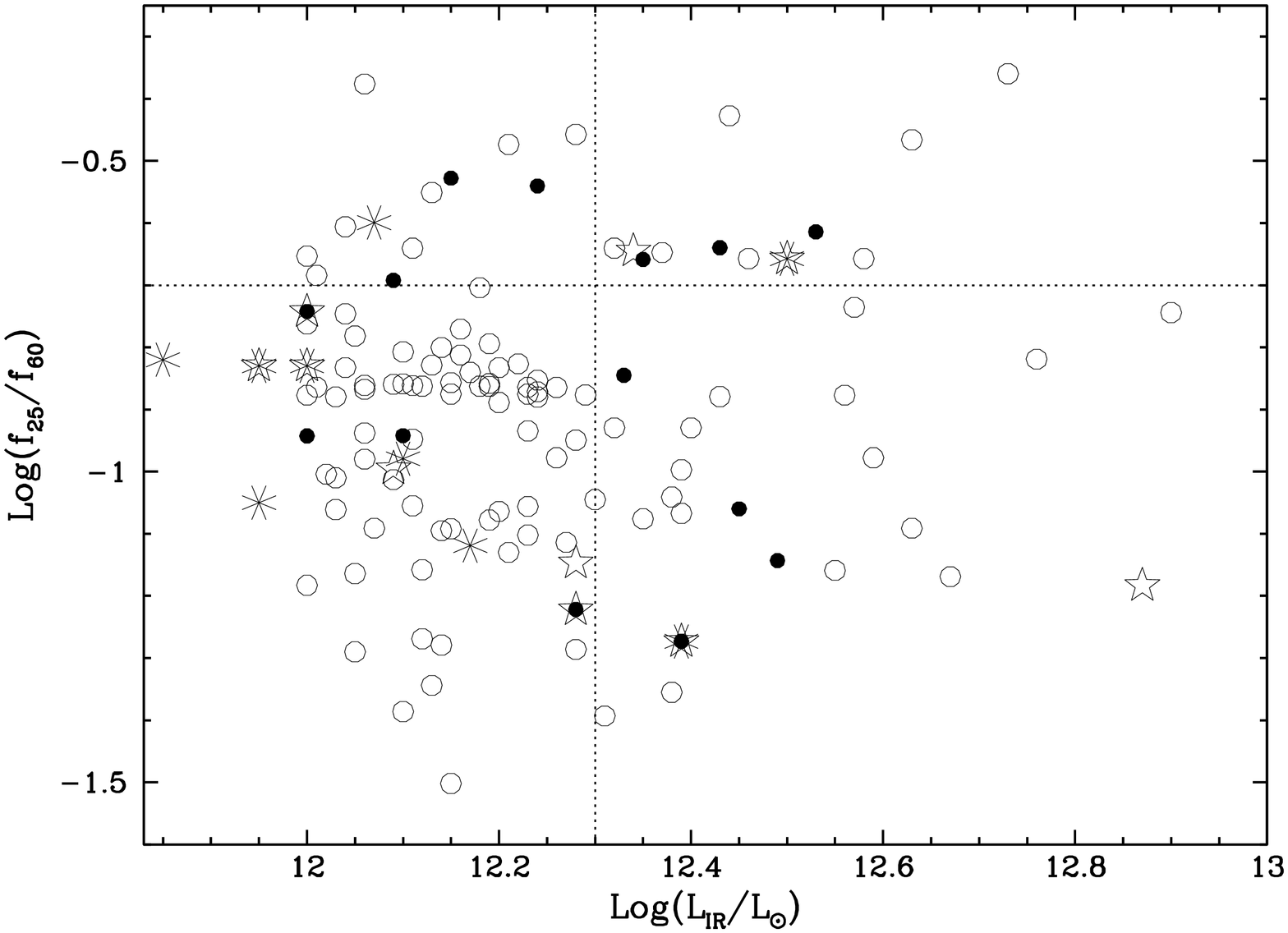}}
\linebreak
\rotatebox{-90}{\plotone{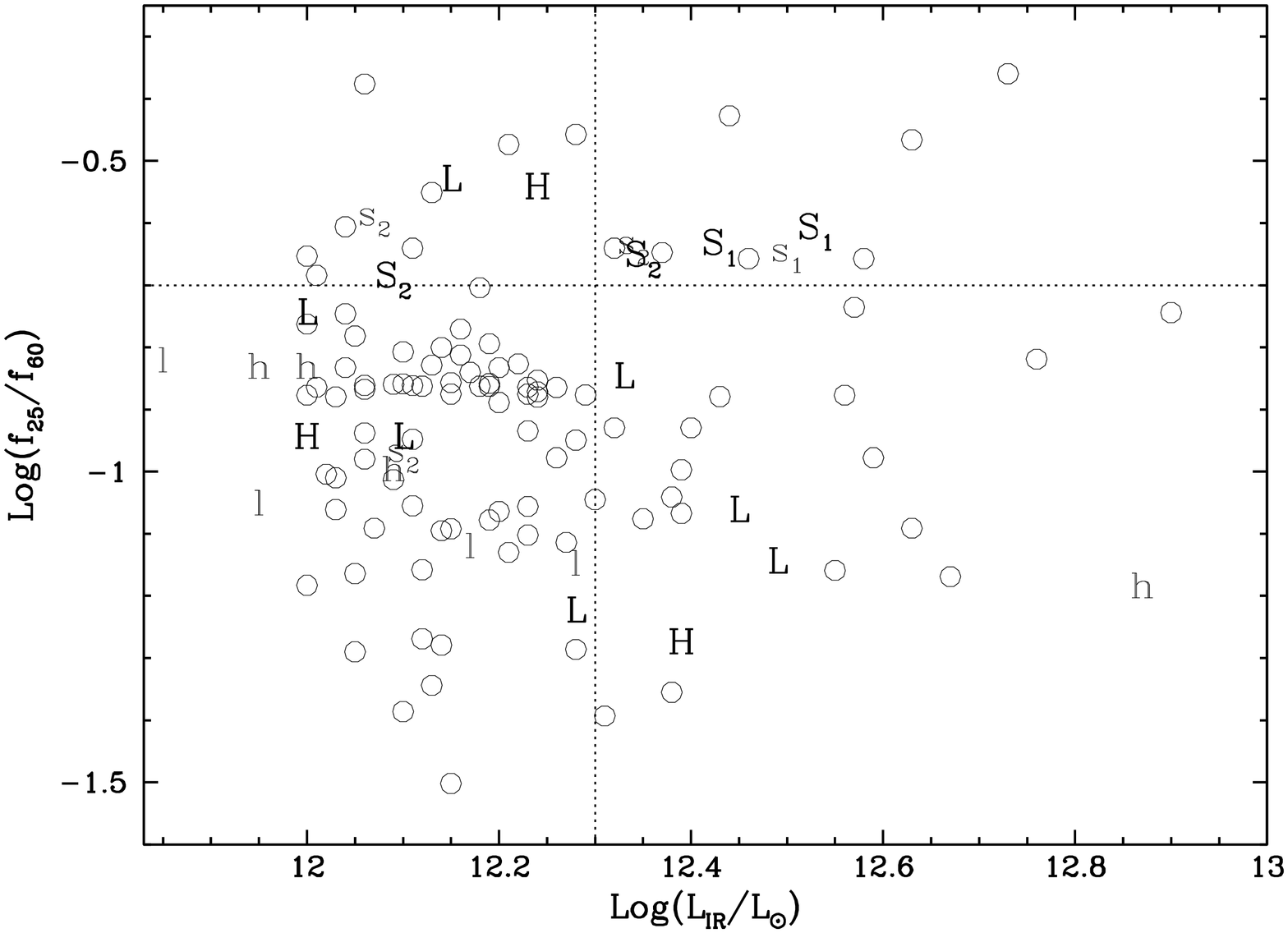}}
\caption{A plot of the logarithm of the {\it IRAS} flux ratio $\rm{f(25~\mu m)/f(60~\mu m)}$ against the logarithm of L$\rm{_{IR}}$ for the 1-Jy sample.  In the top panel, asterisks represent objects observed with {\it Chandra} by \citet{ptak}, stars represent objects observed with {\it XMM-Newton} by \citet{frances}, and filled circles represent objects in our sample, observed with {\it Chandra} and reported in this paper.  Note that a few objects have been observed by two or more groups.  The bottom panel shows the distribution of optical spectral types of ULIRGs observed by {\it Chandra} and {\it XMM-Newton}.  The upper case letters represent the spectral types of our sample while the lower case letters represent the spectral types of the Ptak et al. and Franceschini et al. samples (S$_1$, S$_2$ = type 1 and 2 Seyferts, H = H {\small II} galaxies, and L = LINERs).  The four quadrants are the four parameter bins described in \S~\ref{sec:sample}.  Open circles in both panels show the remaining ULIRGs in the {\it IRAS} 1-Jy sample.}
\label{fig:fluxvl}
\end{figure}

\clearpage

\begin{figure}
\figurenum{2}
\epsscale{0.4}
\hspace{1mm}\rotatebox{-90}{\plotone{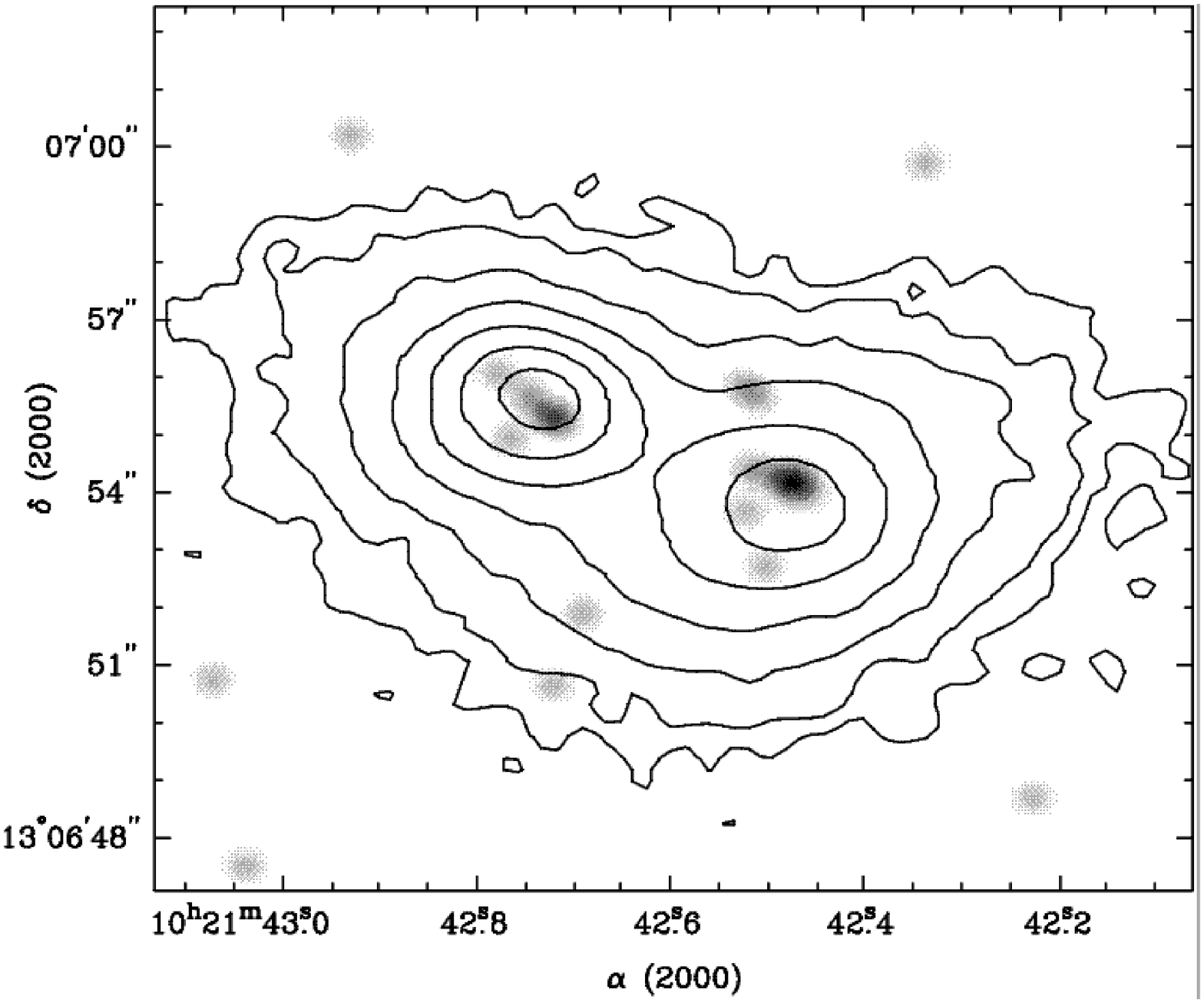}}\hspace{2mm}\rotatebox{-90}{\plotone{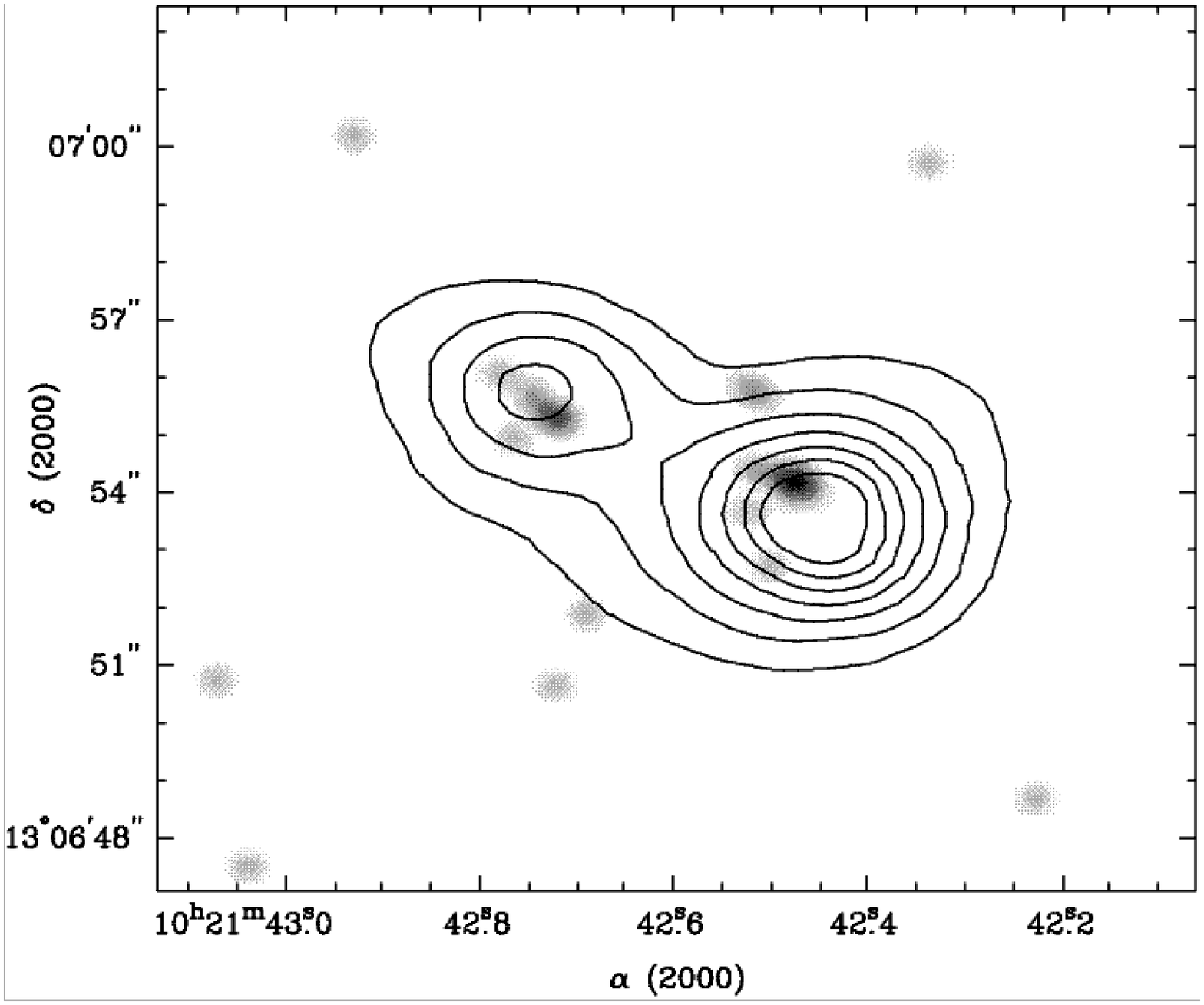}}\vspace{5mm}
\rotatebox{-90}{\plotone{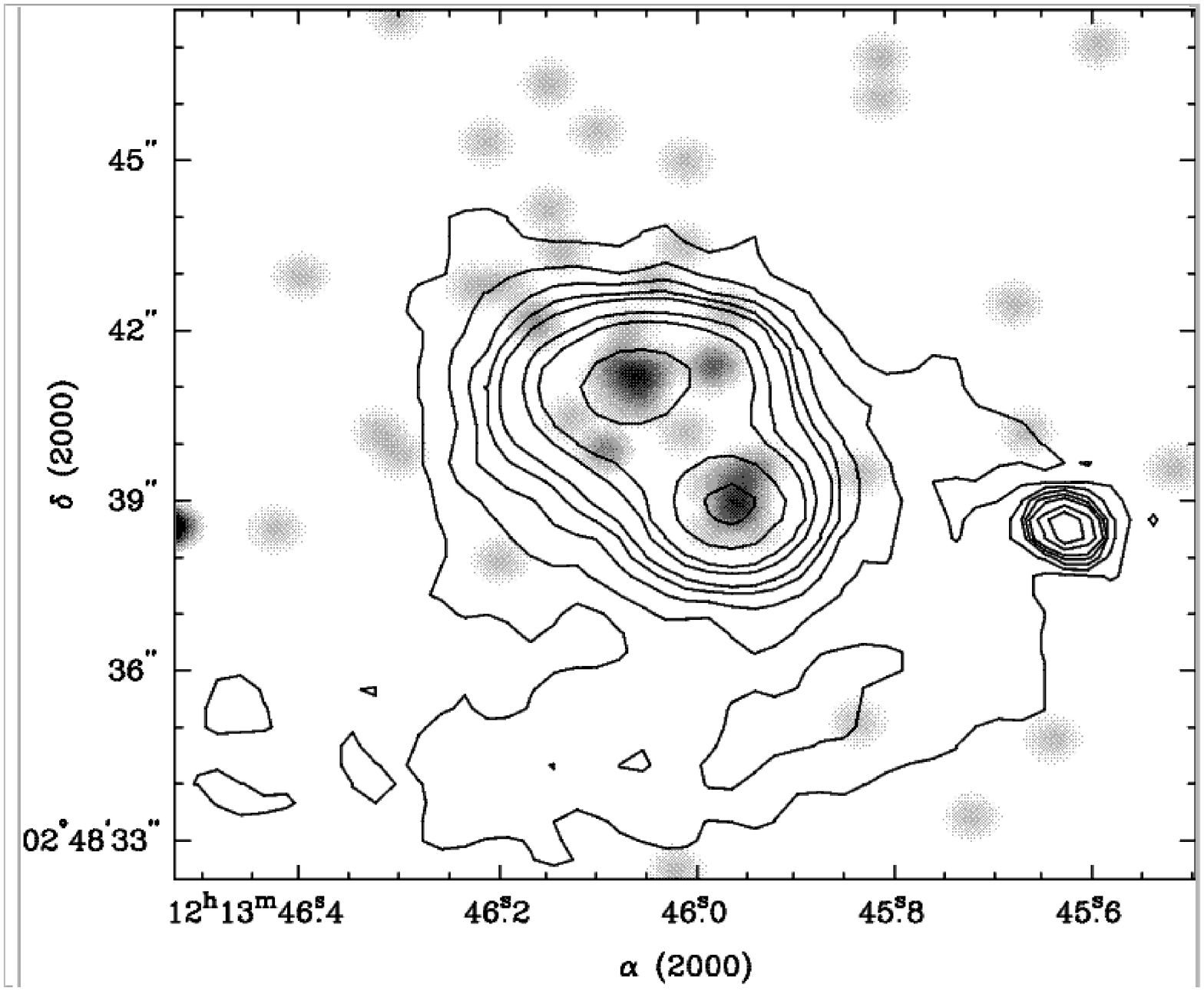}}\hspace{2mm}\rotatebox{-90}{\plotone{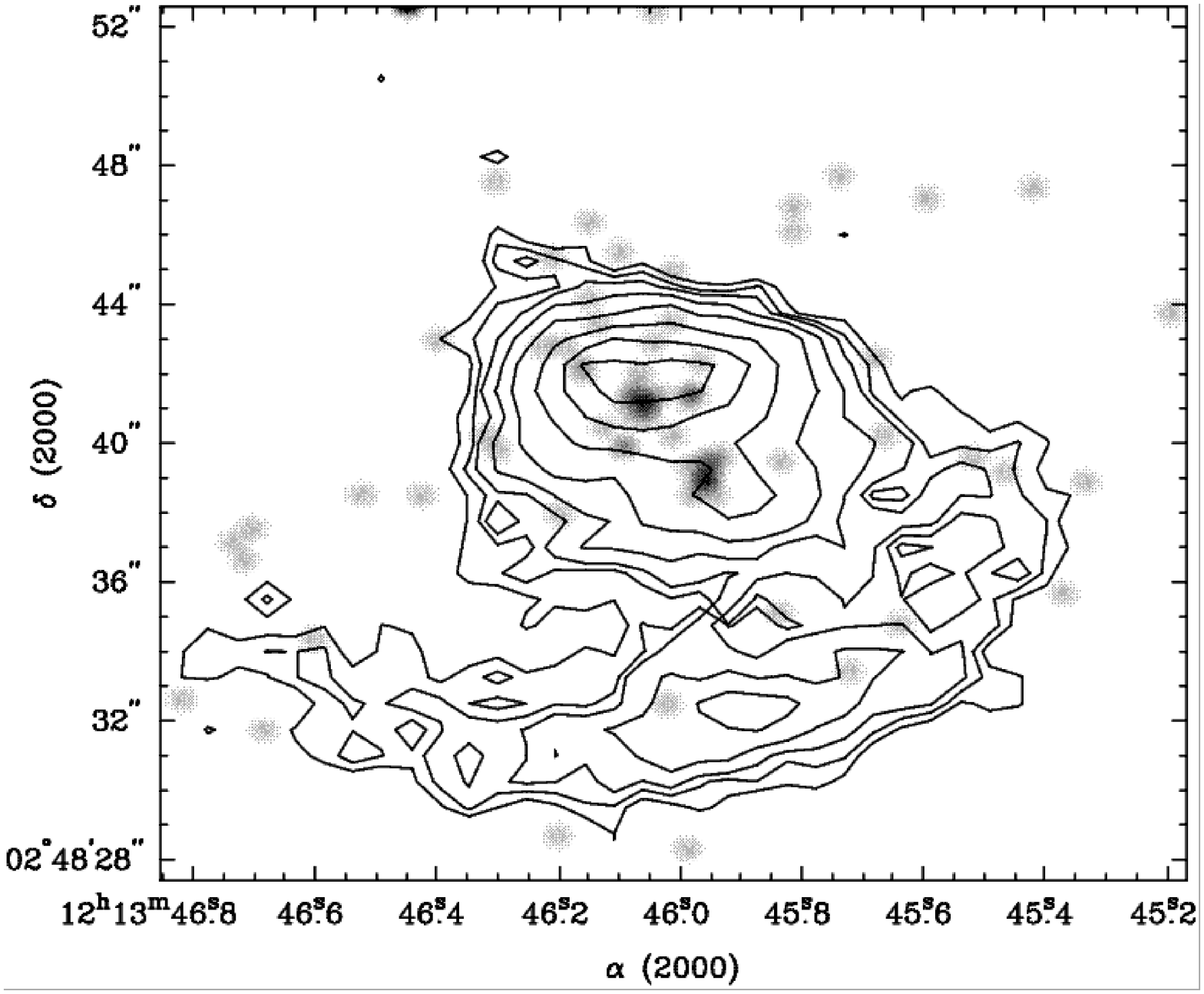}}
\caption{Linear grey scale representation of the X-ray emission from the double nuclei sources F10190+1322 (top) and F12112+0305 (bottom).  The two left panels show X-rays as the grey scale with infrared K$^{\prime}$ band contours.  The two right panels show X-rays as the grey scale with optical R band contours.  The X-ray images have been smoothed to match the resolutions of the IR/optical images.  The apparent ``point sources'' appearing on the edges of the F12112+0305 images are artifacts from the smoothing process; each of these bright ``point sources'' corresponds to only one count and is not a real X-ray source.  The infrared and optical images are from \citet{kim02}.  The linear separations between the two X-ray peaks are 5.6~kpc for F10190+1322 and 3.7~kpc for F12112+0305.}
\label{fig:galpair}
\end{figure}

\clearpage

\begin{figure}
\figurenum{3}
\epsscale{0.4}
\hspace{1.5mm}\rotatebox{-90}{\plotone{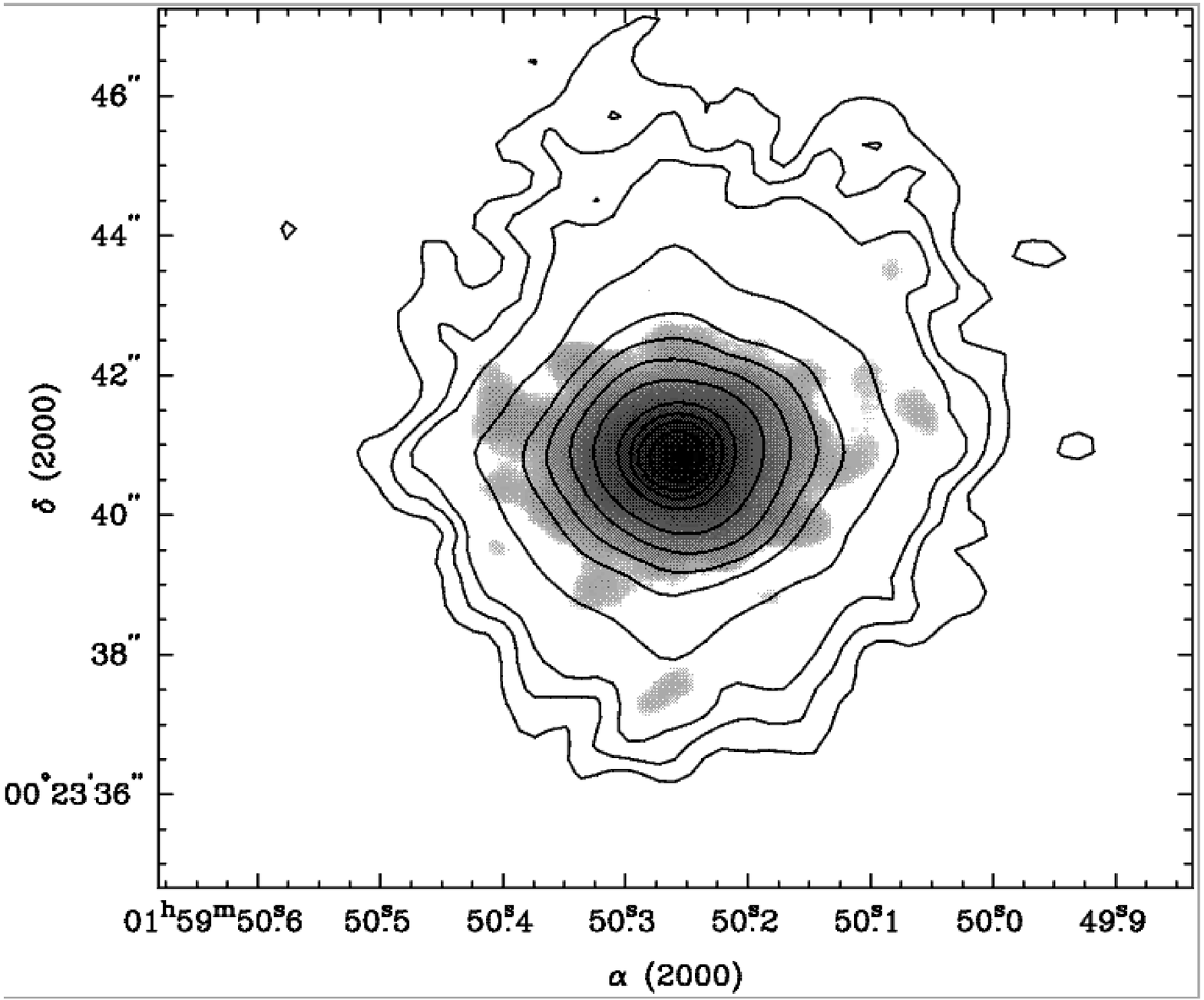}}\hspace{3.5mm}\rotatebox{-90}{\plotone{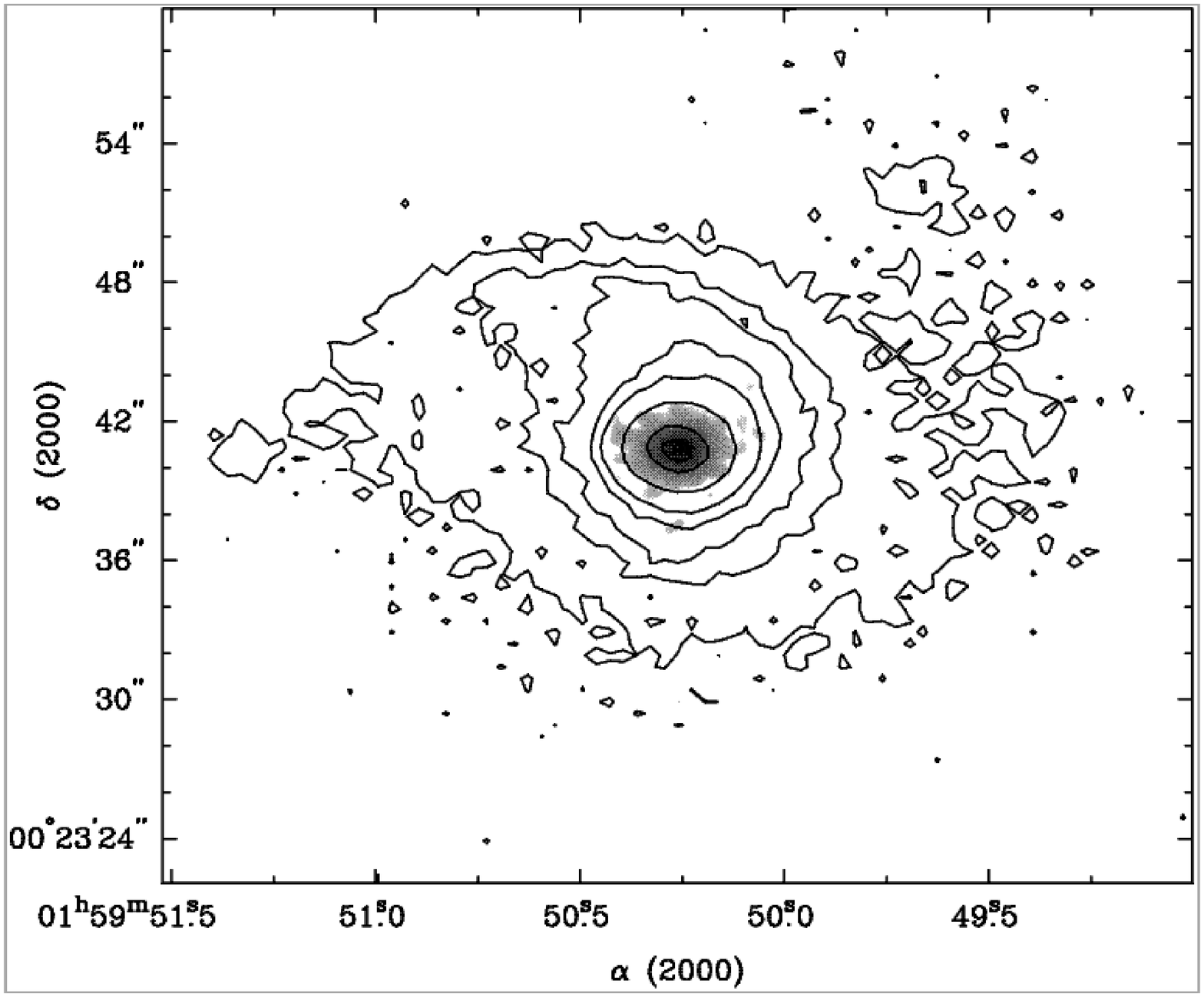}}\vspace{5mm}
\rotatebox{-90}{\plotone{f3c.eps}}\hspace{2mm}\rotatebox{-90}{\plotone{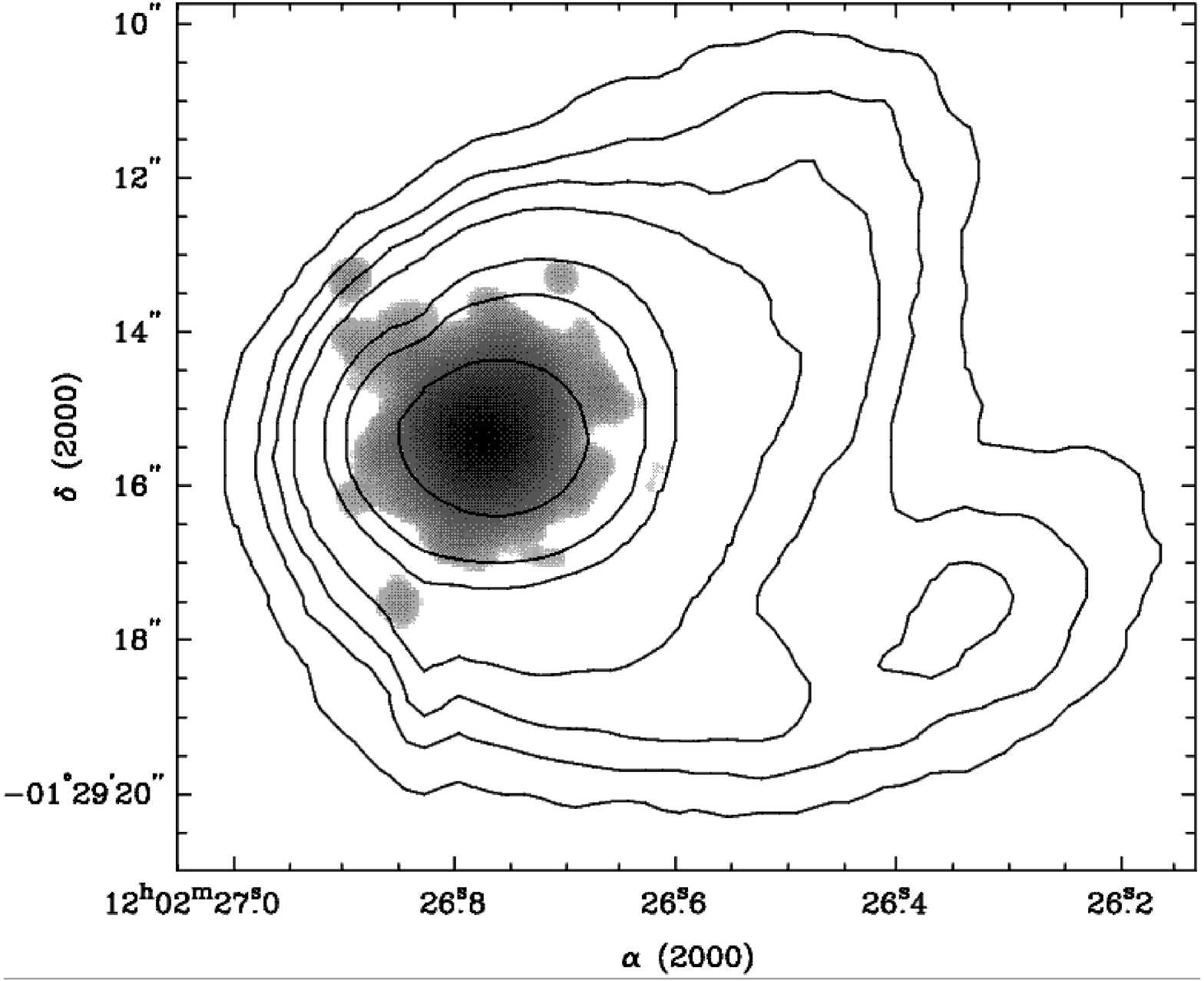}}
\caption{Logarithmic grey scale representations of the X-ray emission from the Seyfert~1 galaxies F01572+0009 (top) and Z11598-0112 (bottom).  The left panels are X-ray grey scale with infrared K$^{\prime}$ band contours.  The right panels are X-ray grey scale with optical R band contours.  Note the difference in spatial scale between the left and right panels.  The X-ray images have been smoothed to match the resolutions of the IR/optical images.  The infrared and optical images are from \citet{kim02}.}
\label{fig:bright}
\end{figure}

\clearpage

\begin{figure}
\figurenum{4}
\center
\epsscale{1.1}
\plotone{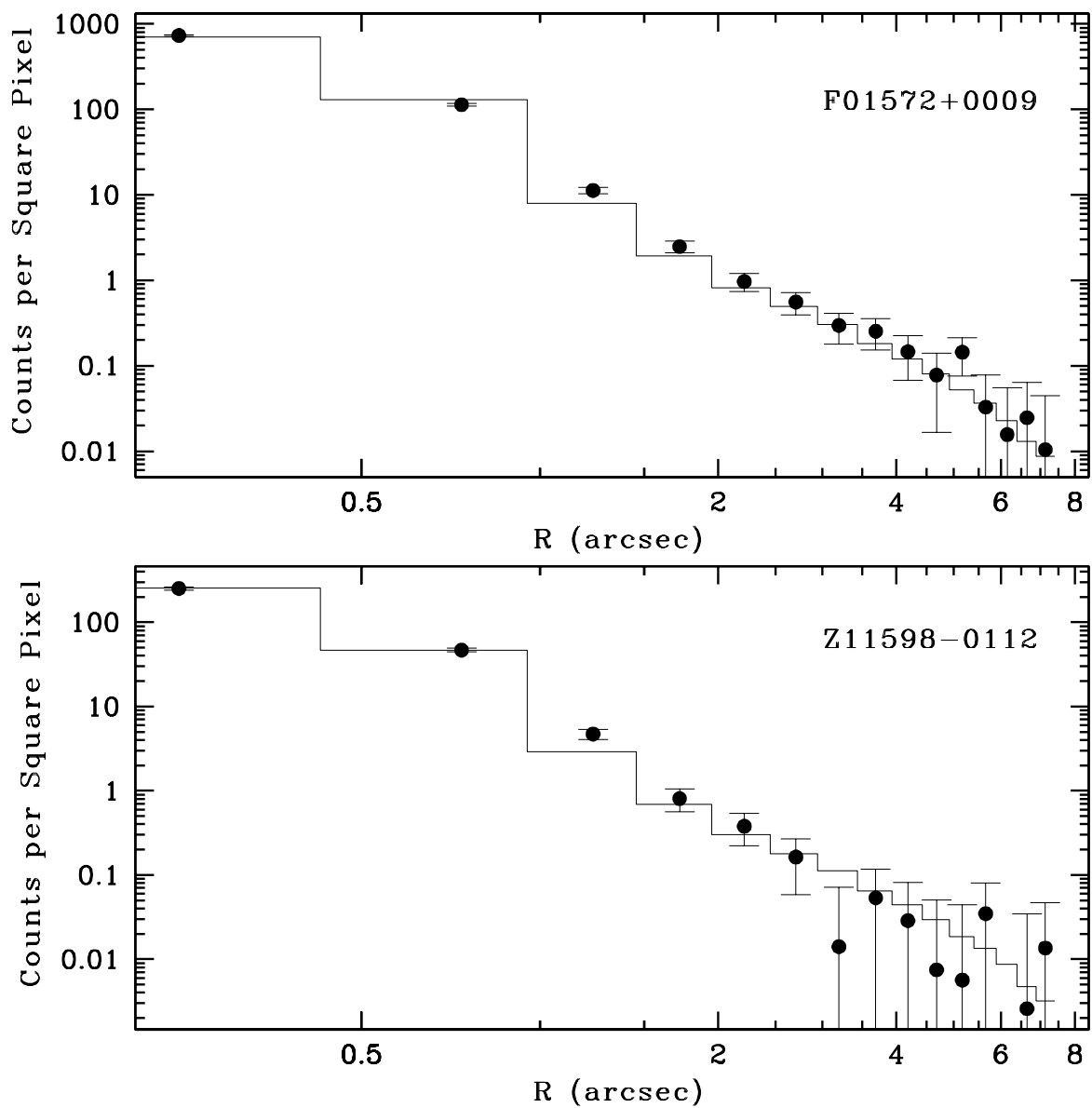}
\caption{Comparison of model PSFs with observed radial profiles for the two bright sources.  The x-axis is distance in arcseconds from the centroid of the emission while the y-axis is the surface brightness in counts per square pixel. The error bars are errors on the net counts per square pixel assuming Poissonian statistics. In each panel, the histogram is the model PSF obtained from the PSF library at 1.0~keV.  The points are observed total counts with energy in the range of 0.5 to 2.0~keV. This diagram shows that the soft X-ray emissions from these two galaxies are unresolved or, at best, marginally resolved.}
\label{fig:psf}
\end{figure}
\begin{figure}
\figurenum{5}
\epsscale{0.8}
\rotatebox{-90}{\plotone{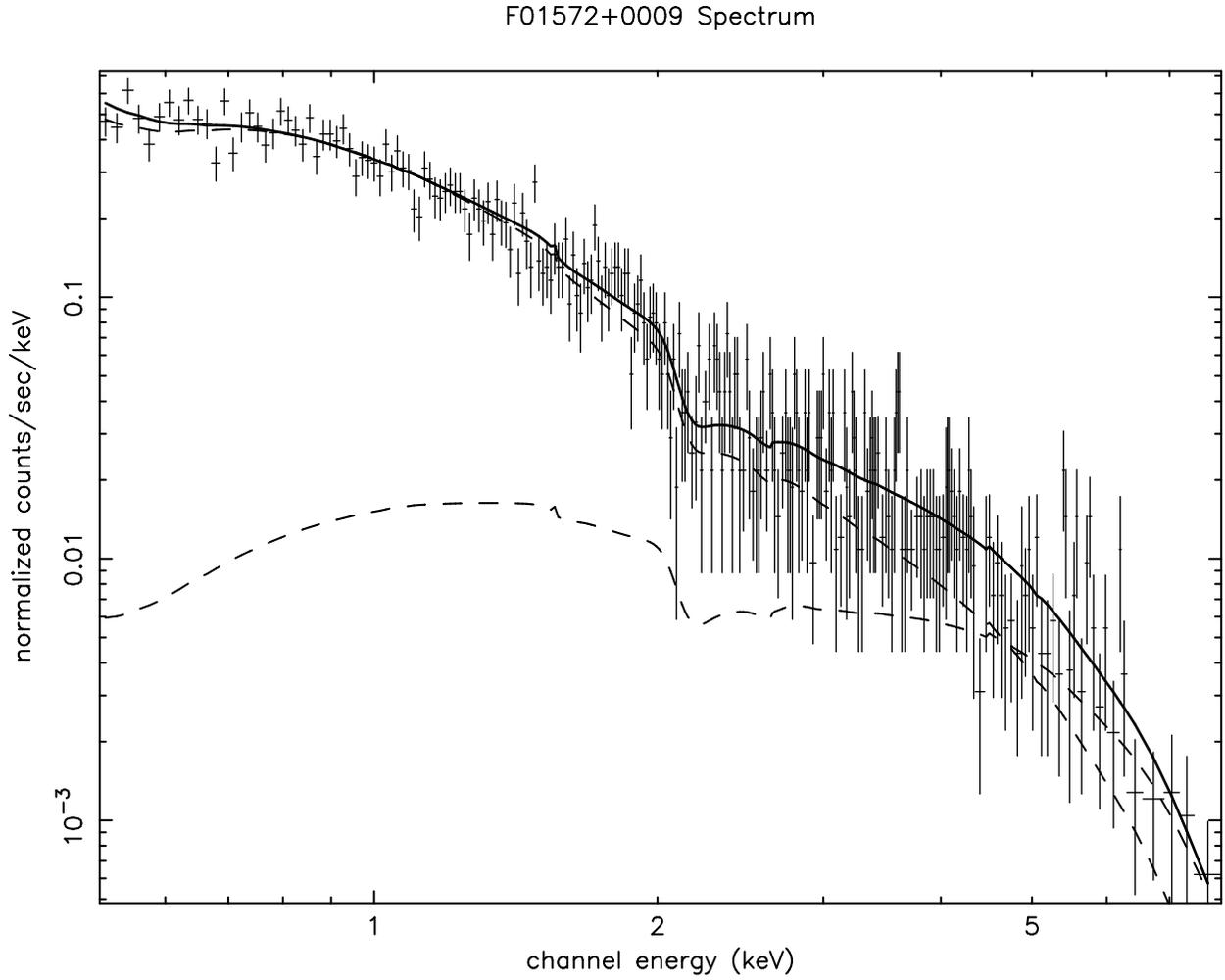}}
\caption{Spectrum of F01572+0009 with at least 3 counts per bin.  The thick solid line is the double power law model, while the dashed lines are the two power law components of the model.  There are hints of emission lines at around 6.0 keV, but the lines are not significant.}
\label{fig:f01572}
\end{figure}

\clearpage

\begin{figure}
\figurenum{6}
\epsscale{0.8}
\rotatebox{-90}{\plotone{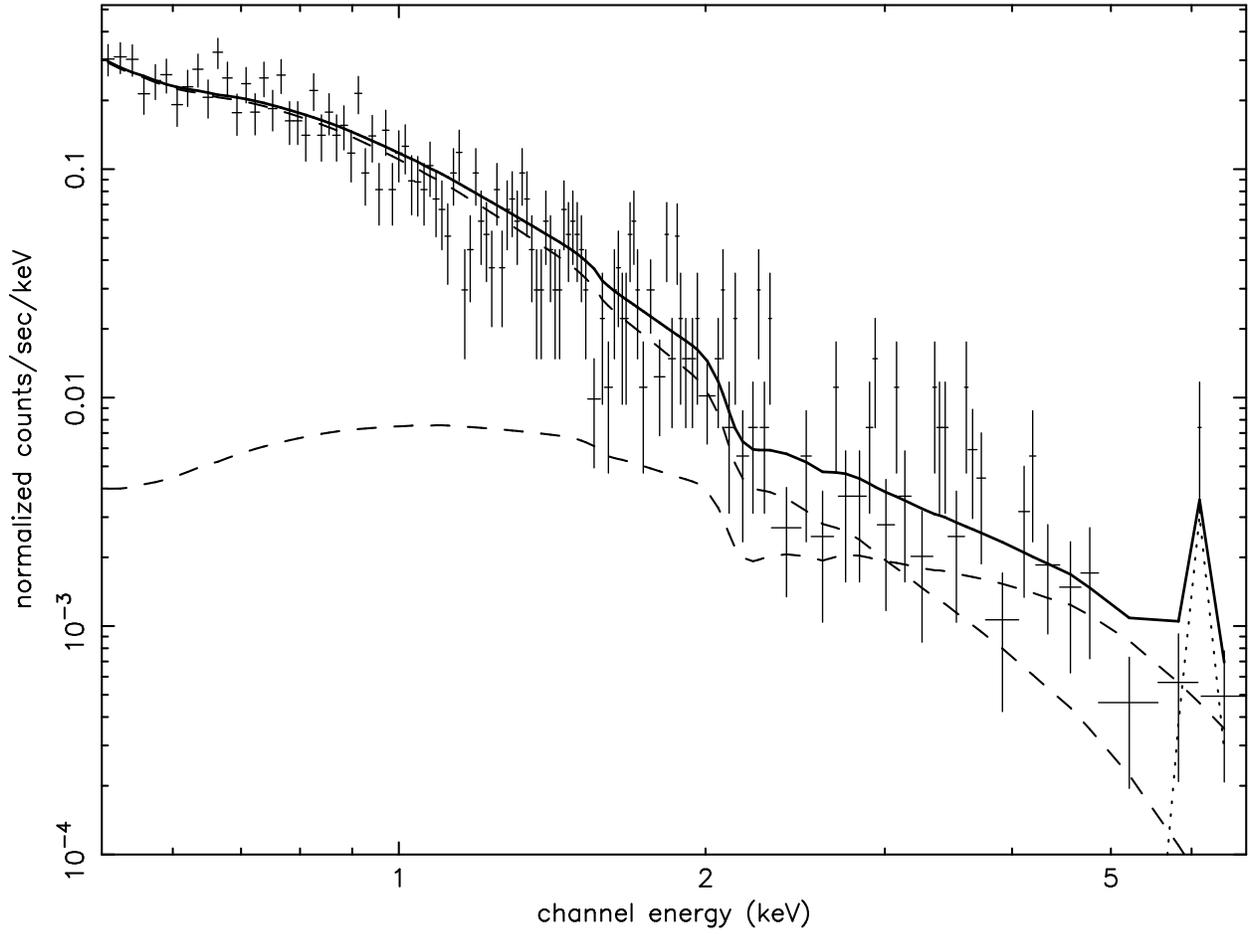}}
\caption{Spectrum of Z11598-0112, with at least 3 counts per bin.  The thick solid line is the model, a combination of two power laws modeling the continuum (the dashed lines) and a narrow Gaussian modeling the emission feature (the dotted line).}
\label{fig:z11598}
\end{figure}

\clearpage

\begin{figure}
\figurenum{7}
\epsscale{0.3}
\begin{center}
\rotatebox{90}{\plotone{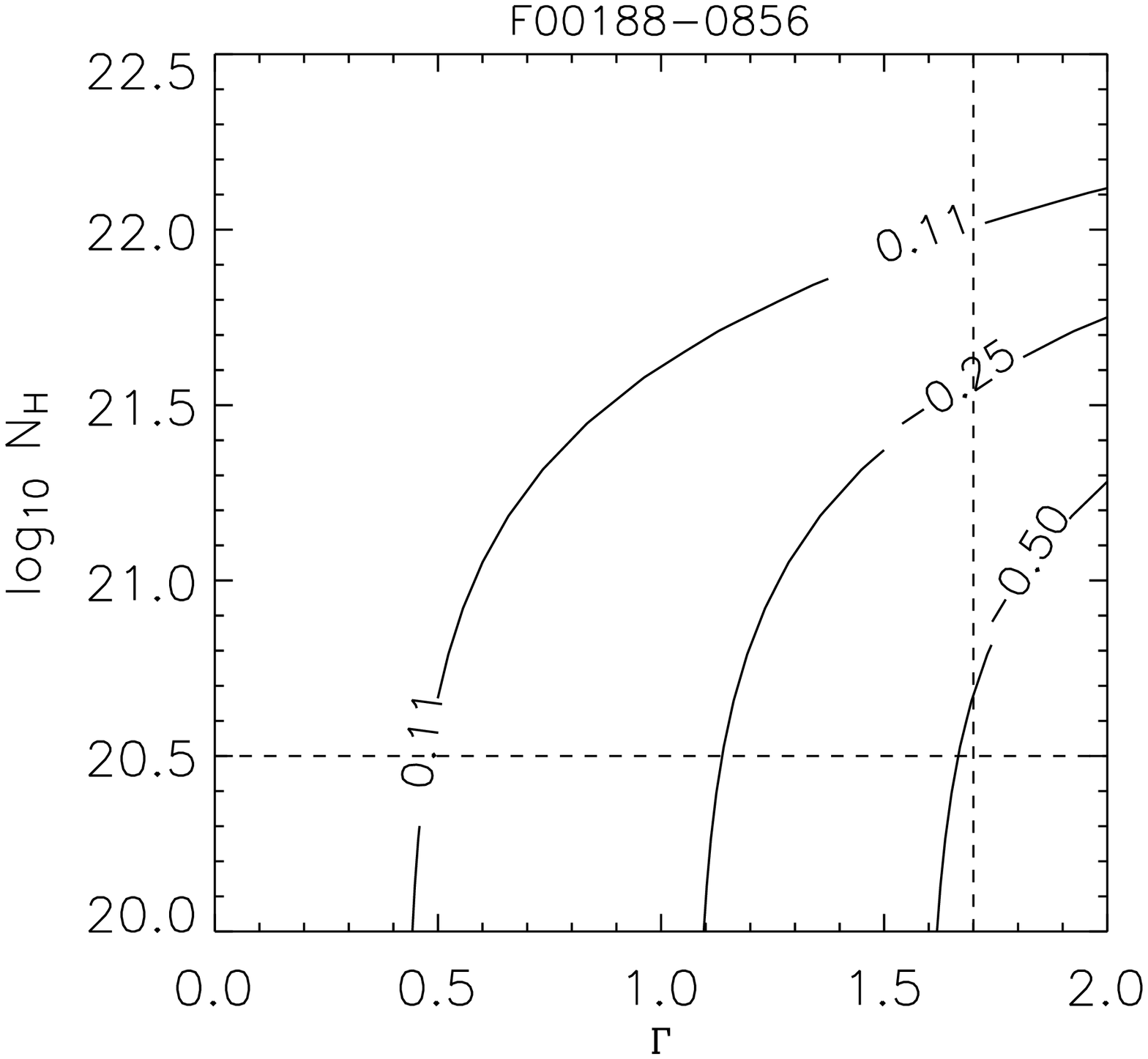}}\rotatebox{90}{\plotone{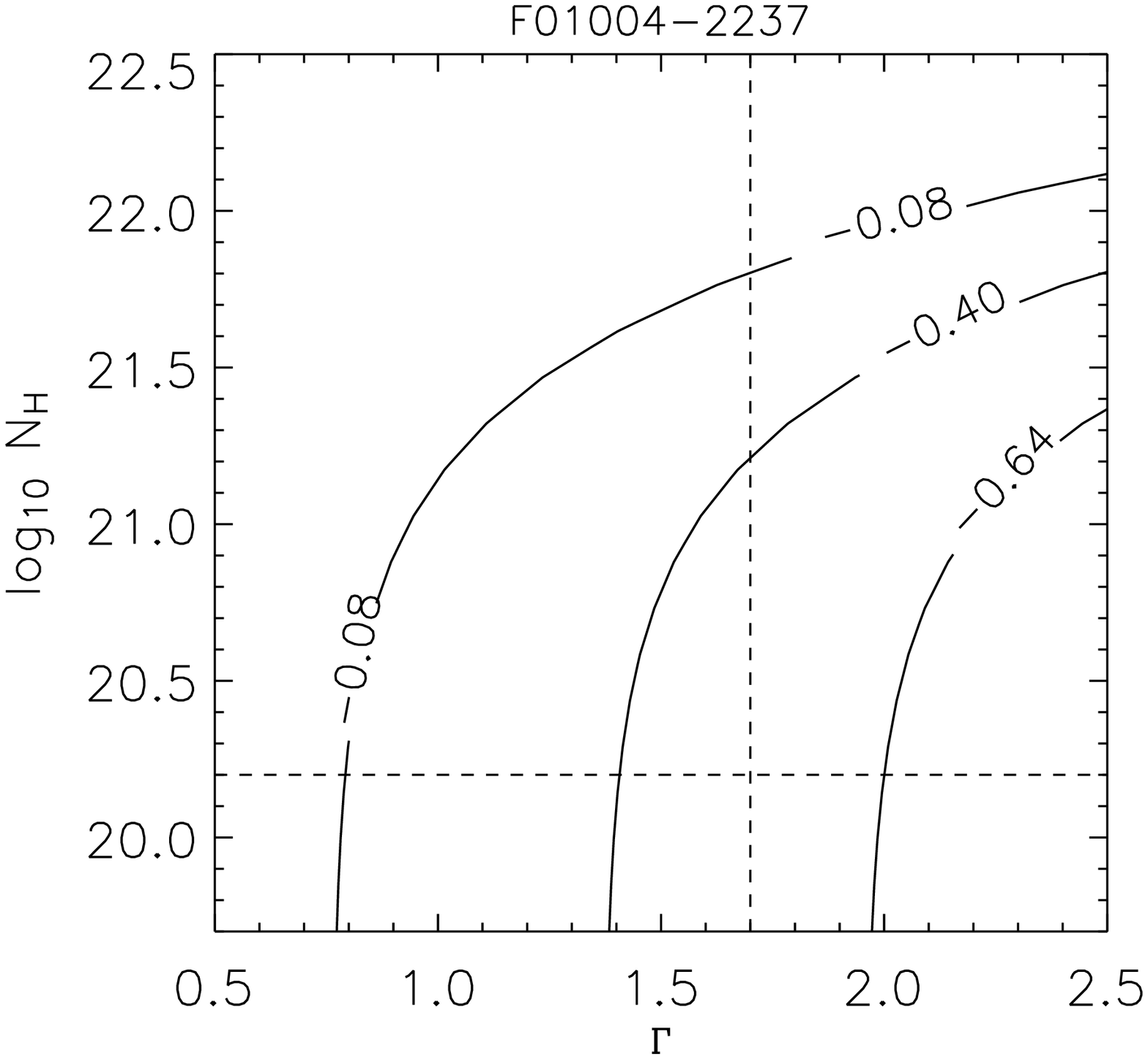}}
\rotatebox{90}{\plotone{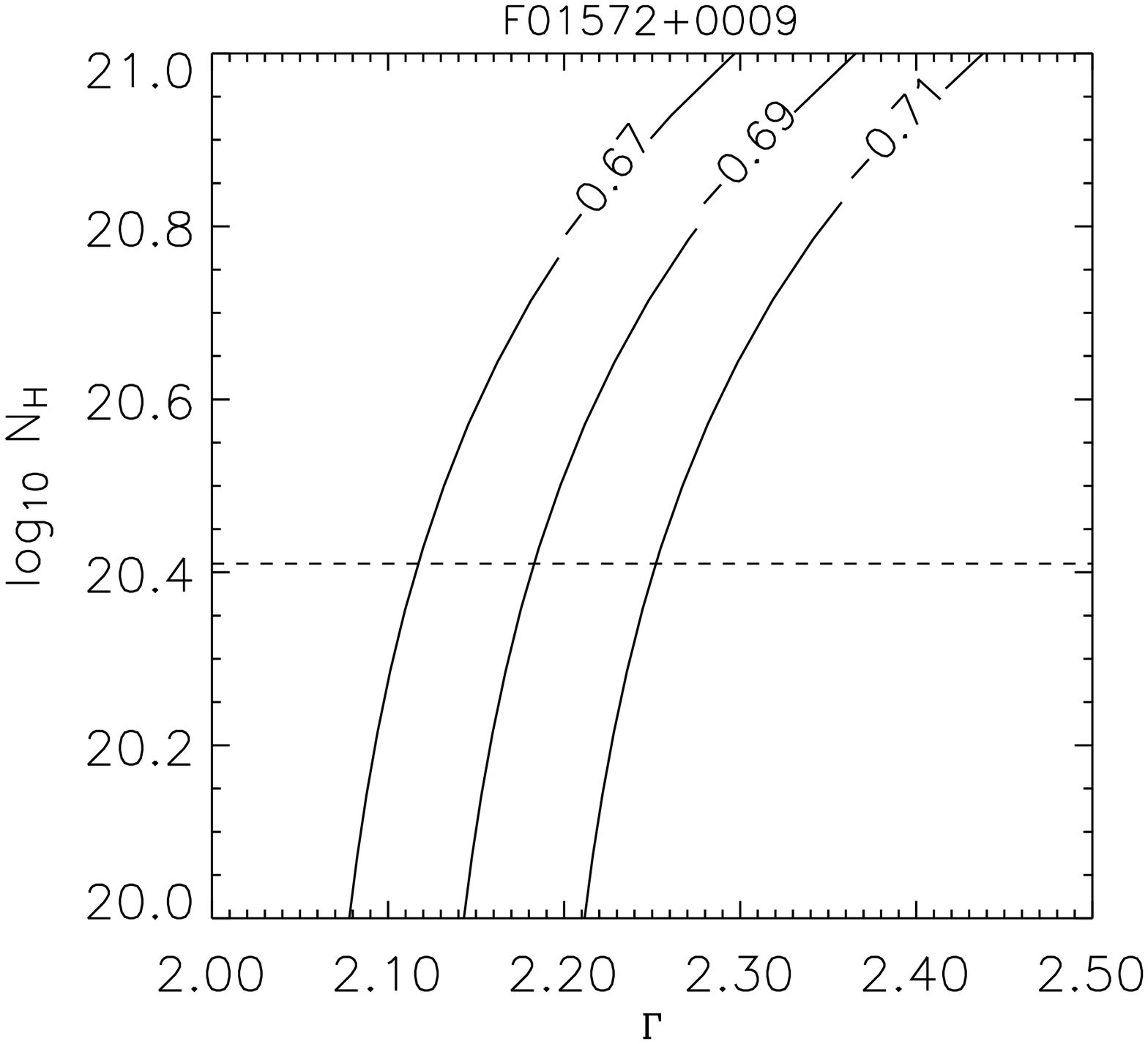}}\rotatebox{90}{\plotone{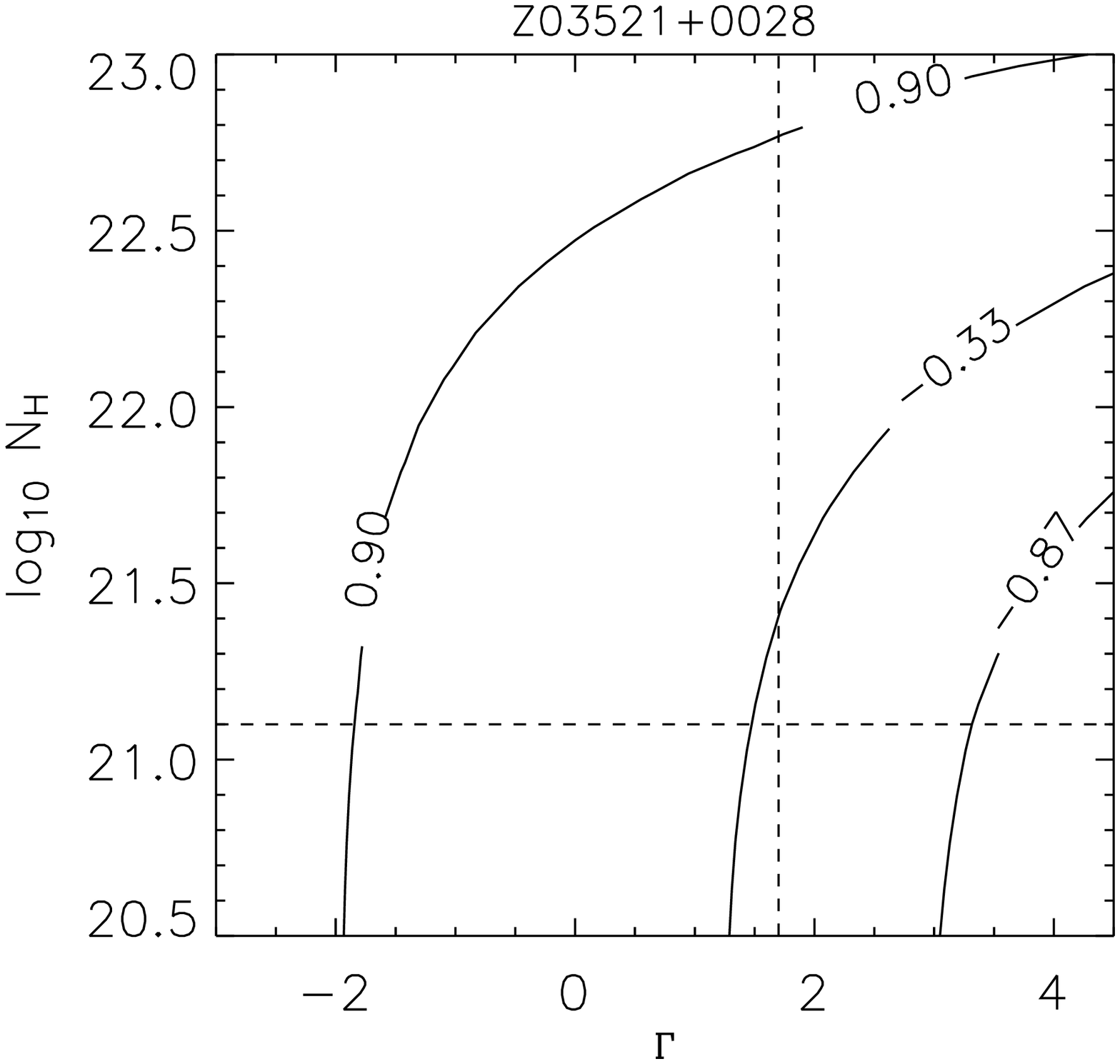}}
\rotatebox{90}{\plotone{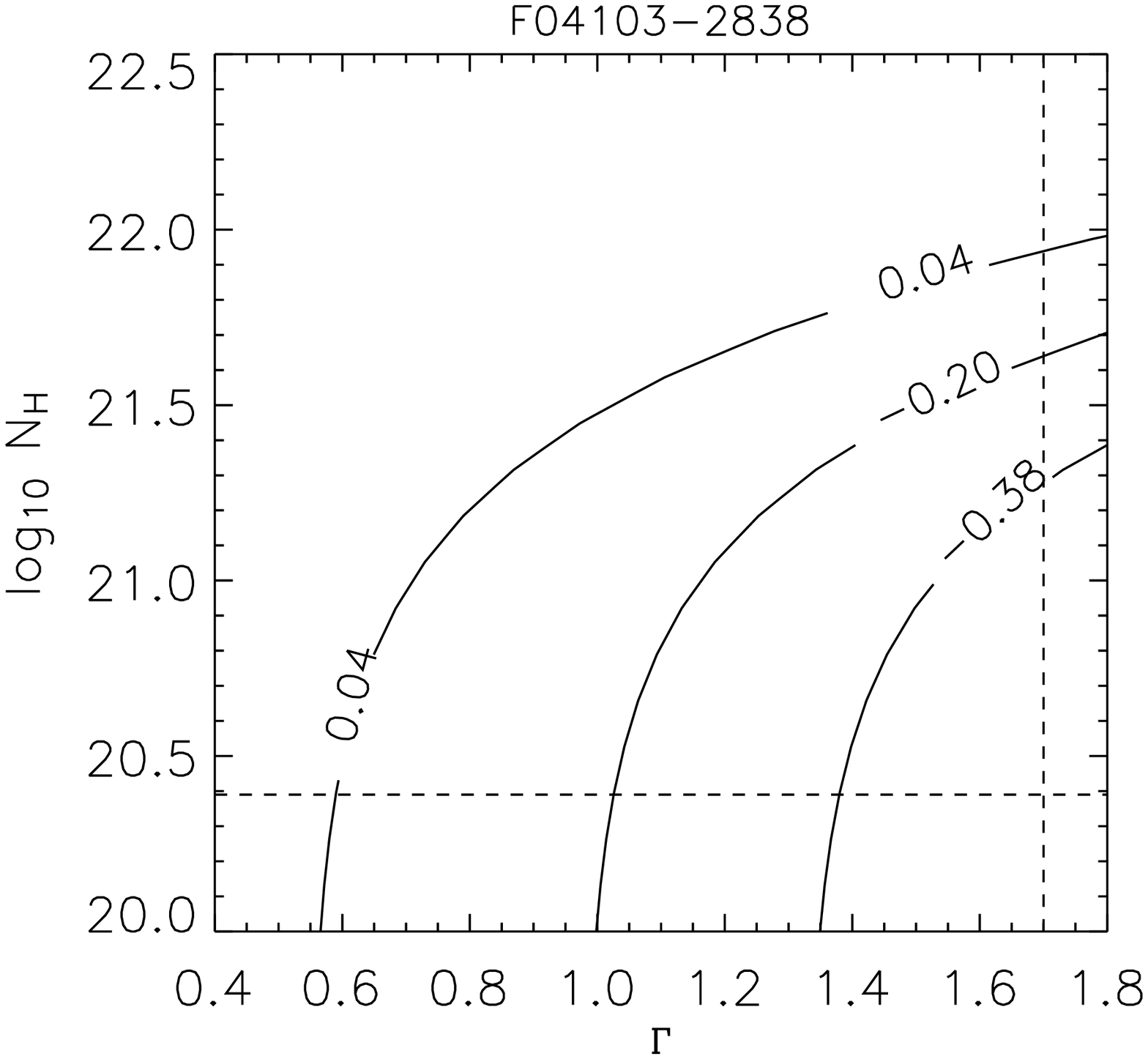}}\rotatebox{90}{\plotone{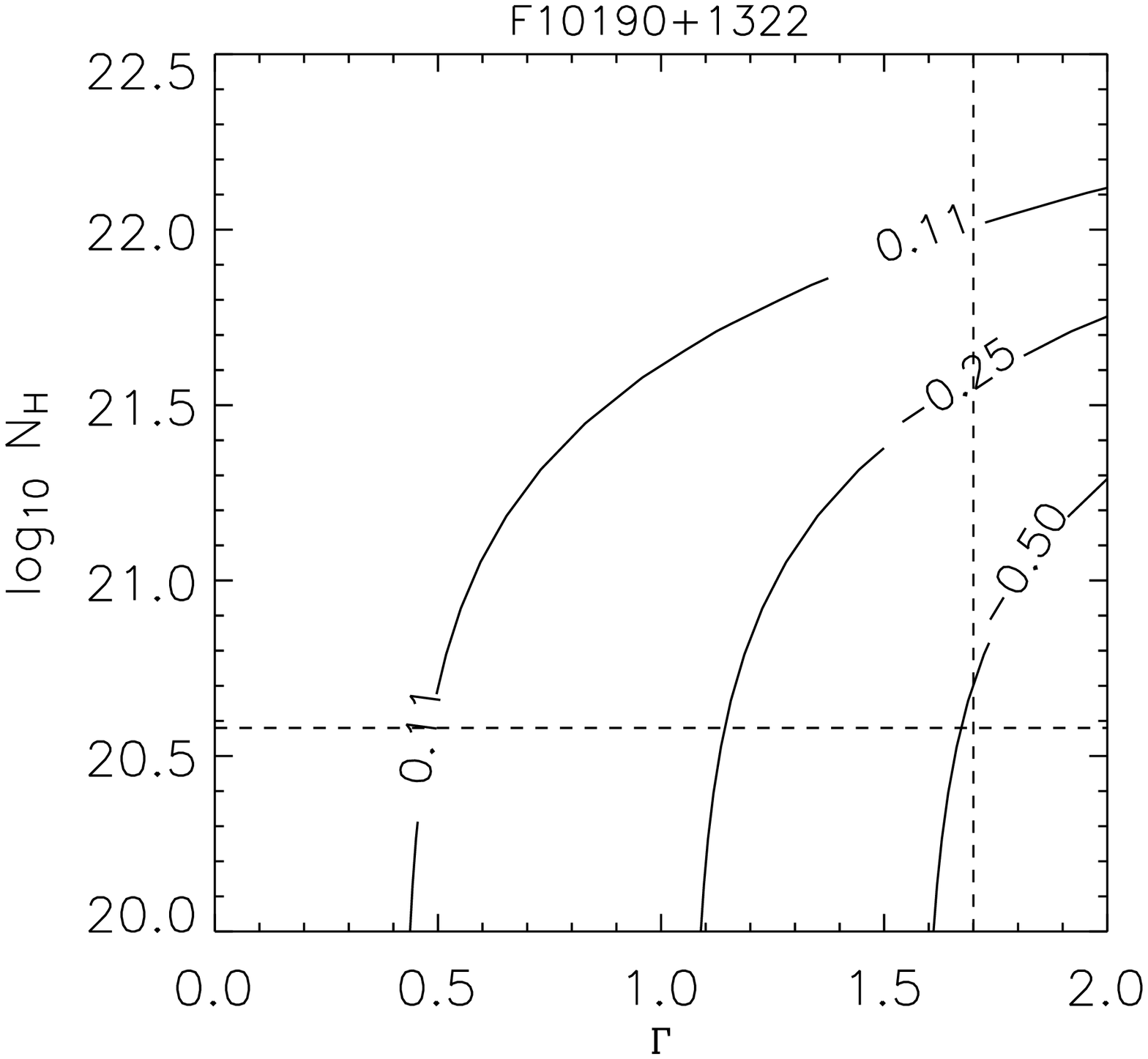}}
\end{center}
\caption{Each plot shows contours of constant hardness ratio (defined in Equation~\ref{eq:hreq}) in the N$\rm{_H}$ versus $\Gamma$ plane.  The middle curve represents, and is labelled with, the observed hardness ratio, while the other two curves represent the hardness ratios 1-$\sigma$ away from the observed value (see Table~\ref{tab:hrs}).  The horizontal dashed line represents the Galactic hydrogen column density, while the vertical dashed line represents $\Gamma$ = 1.7.}
\label{fig:gamma}
\end{figure}

\clearpage

\begin{figure}
\figurenum{7 {\it cont}}
\epsscale{0.3}
\begin{center}
\rotatebox{90}{\plotone{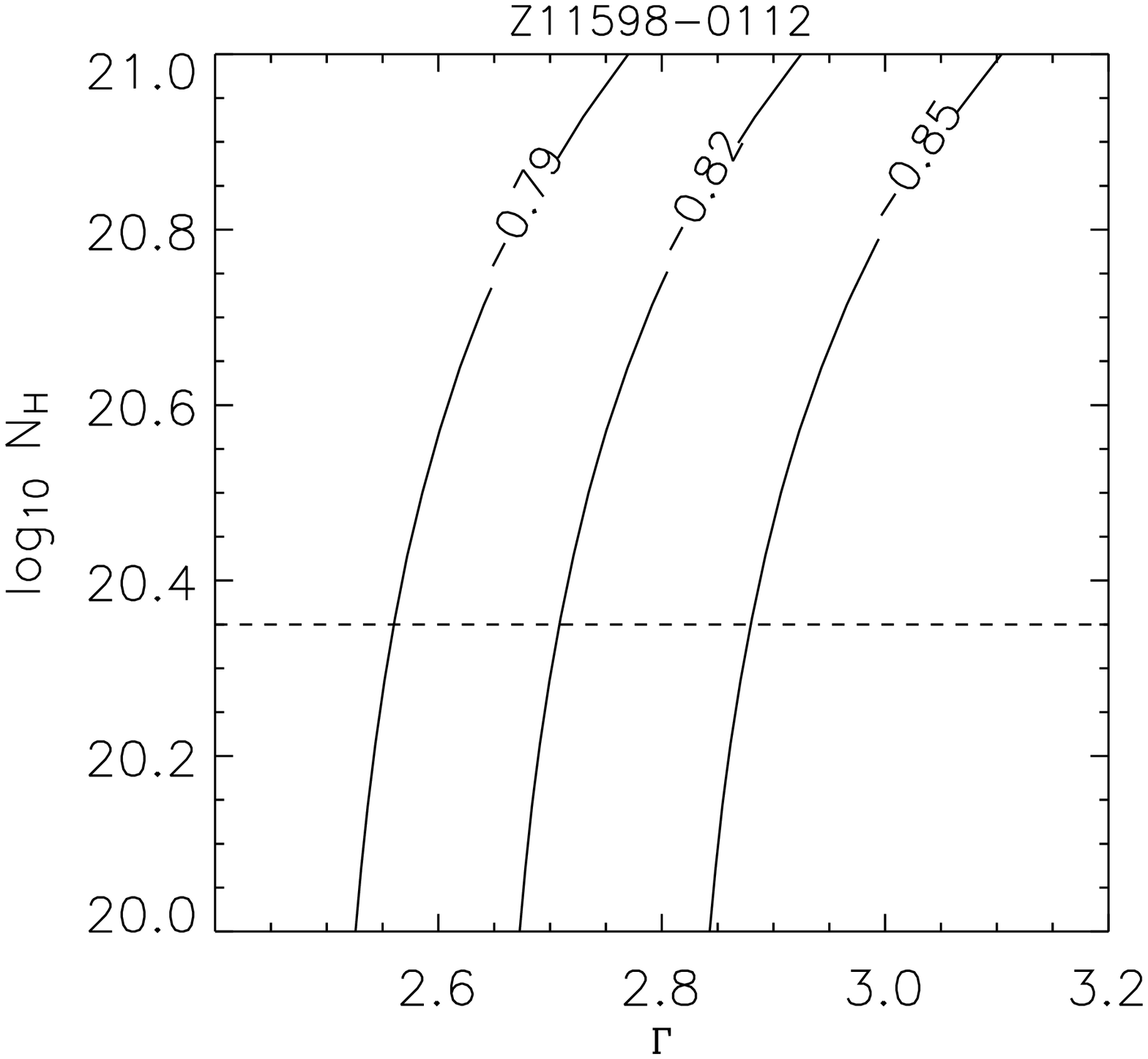}}\rotatebox{90}{\plotone{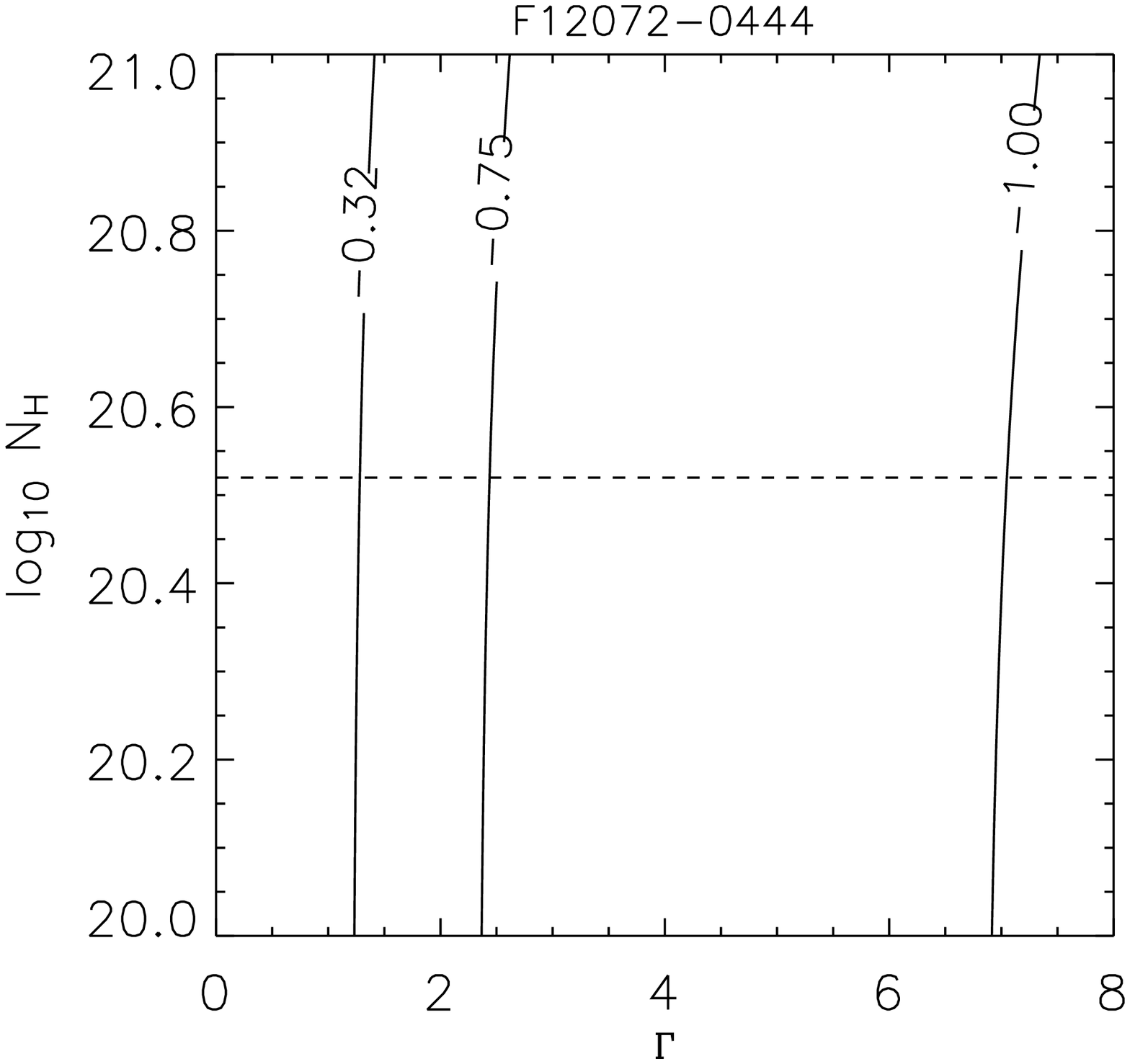}}
\rotatebox{90}{\plotone{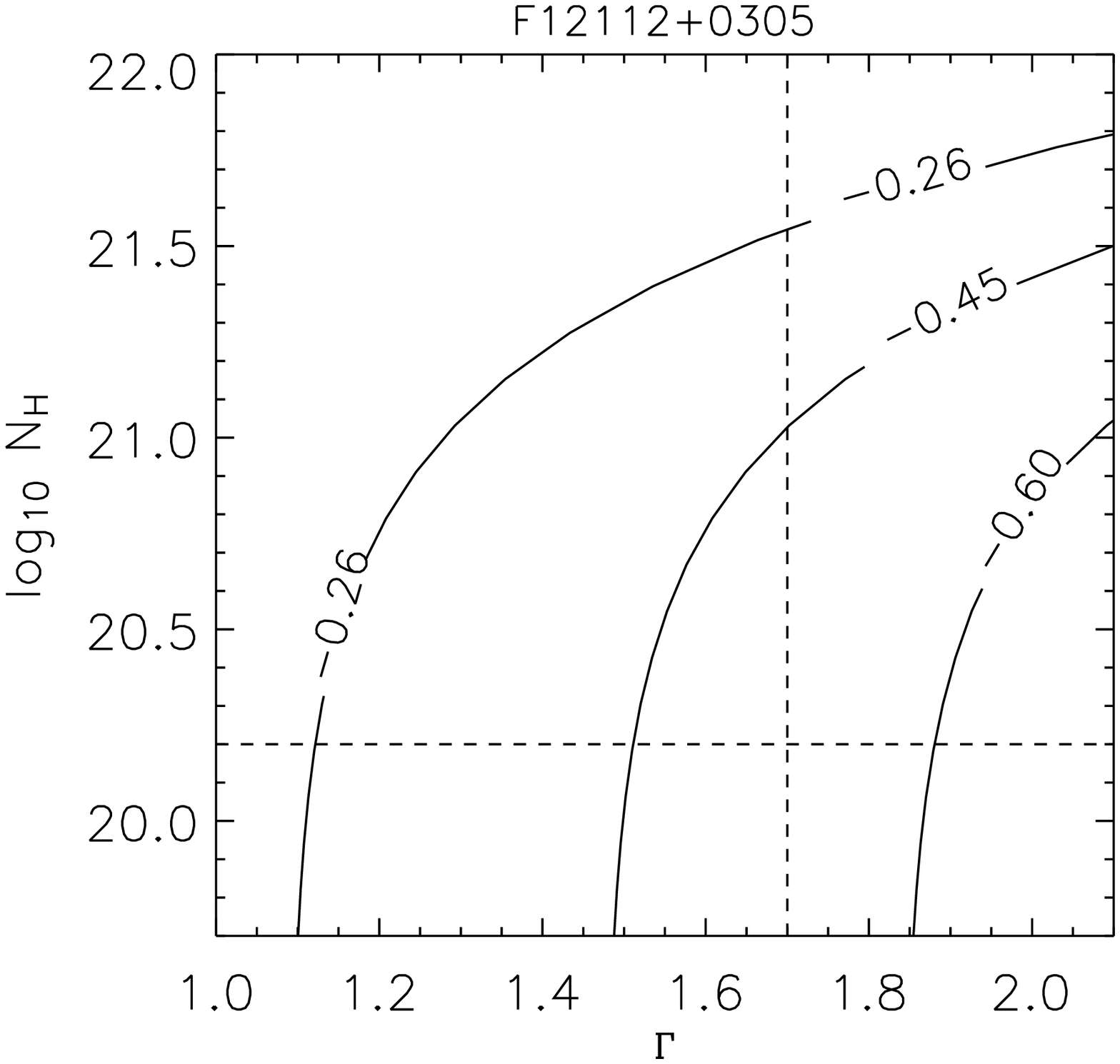}}\rotatebox{90}{\plotone{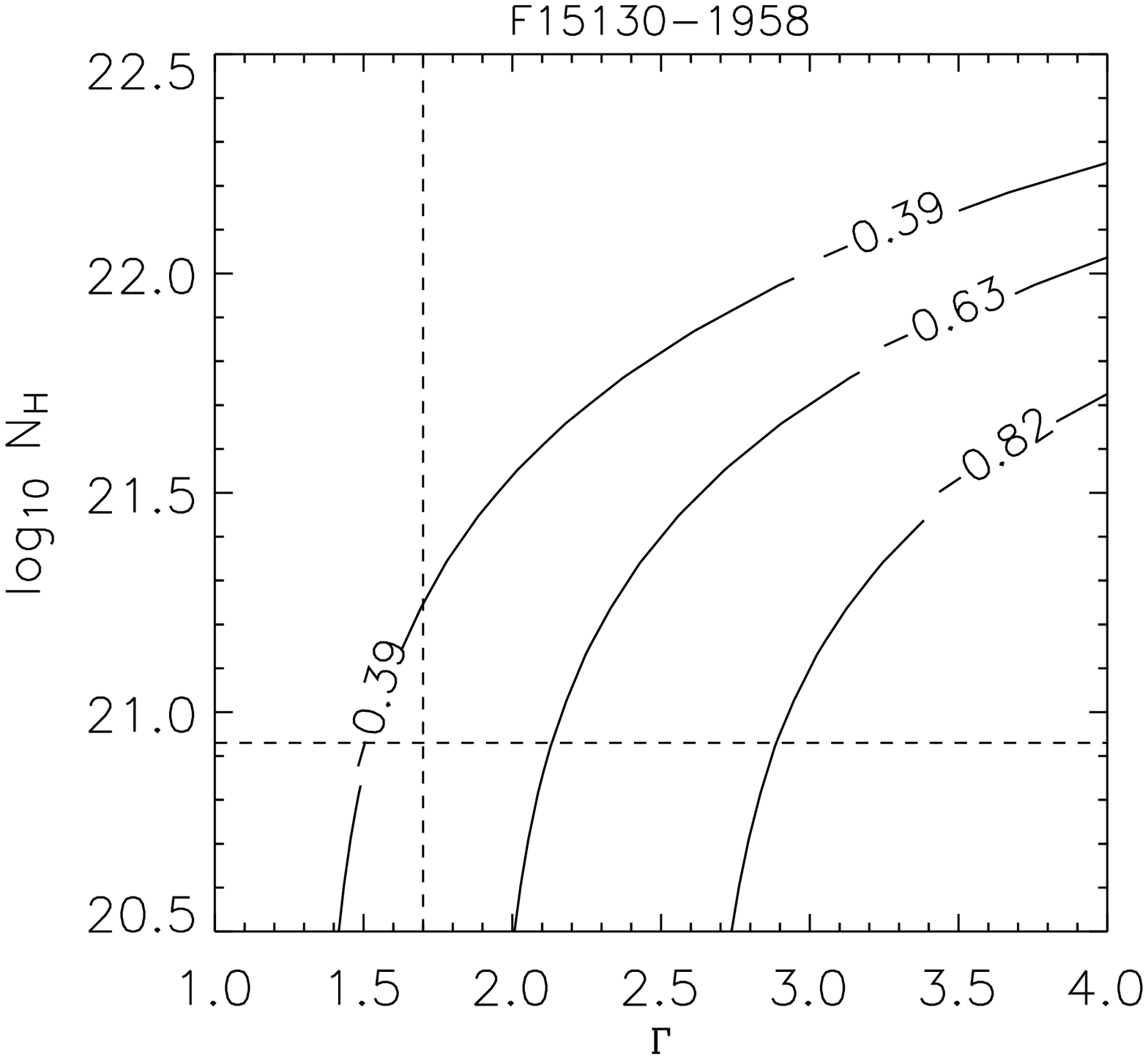}}
\rotatebox{90}{\plotone{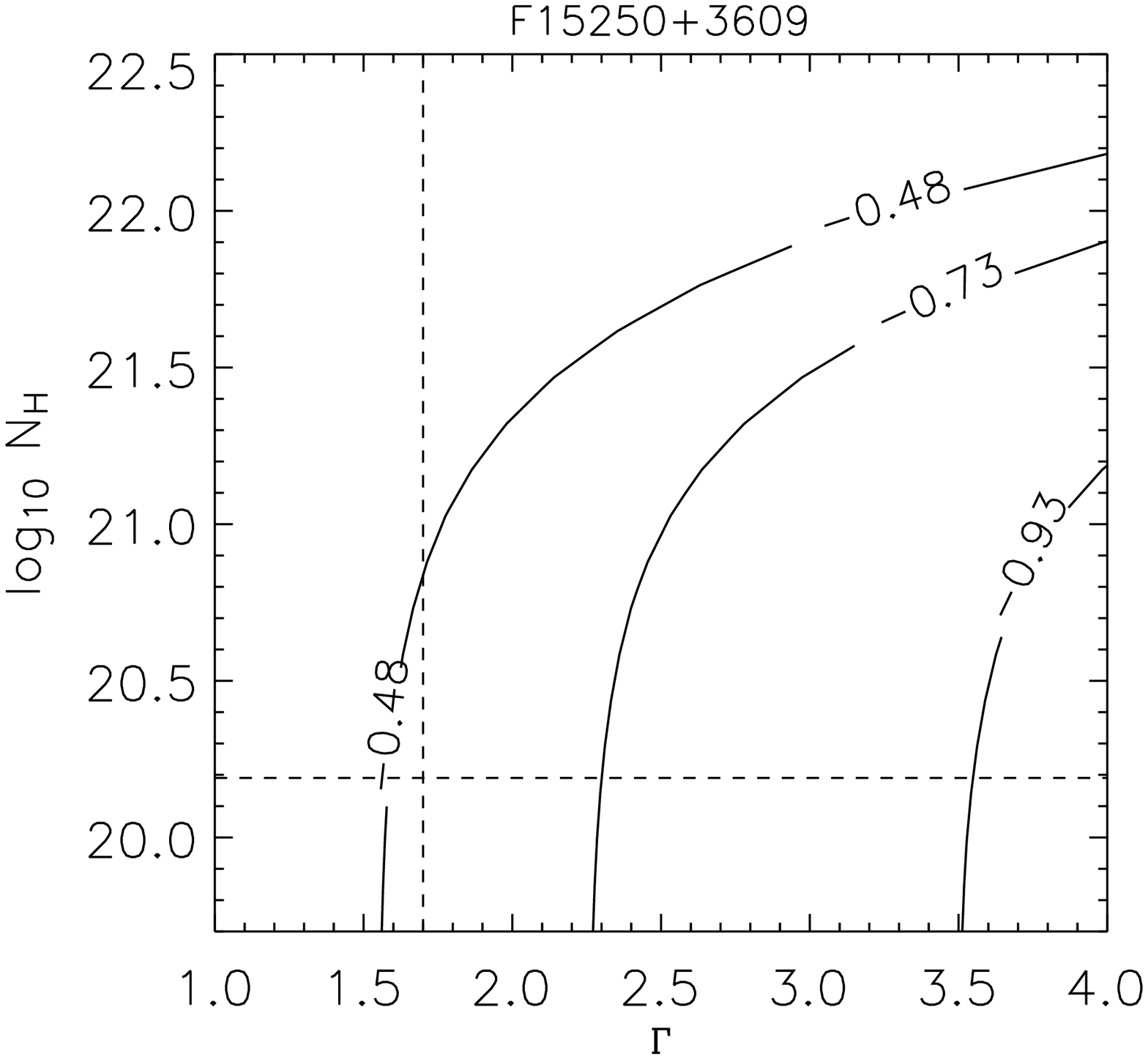}}\rotatebox{90}{\plotone{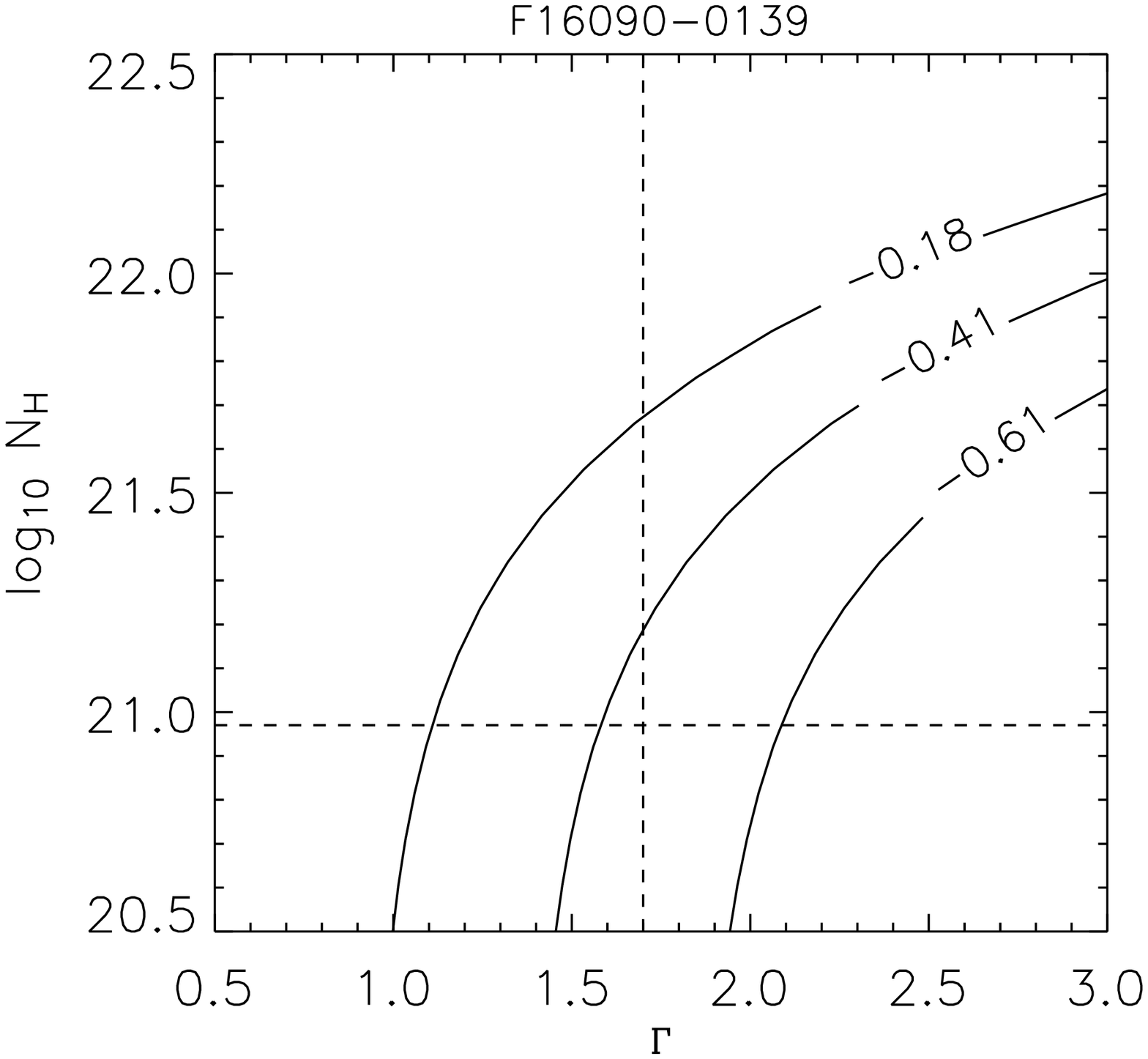}}
\rotatebox{90}{\plotone{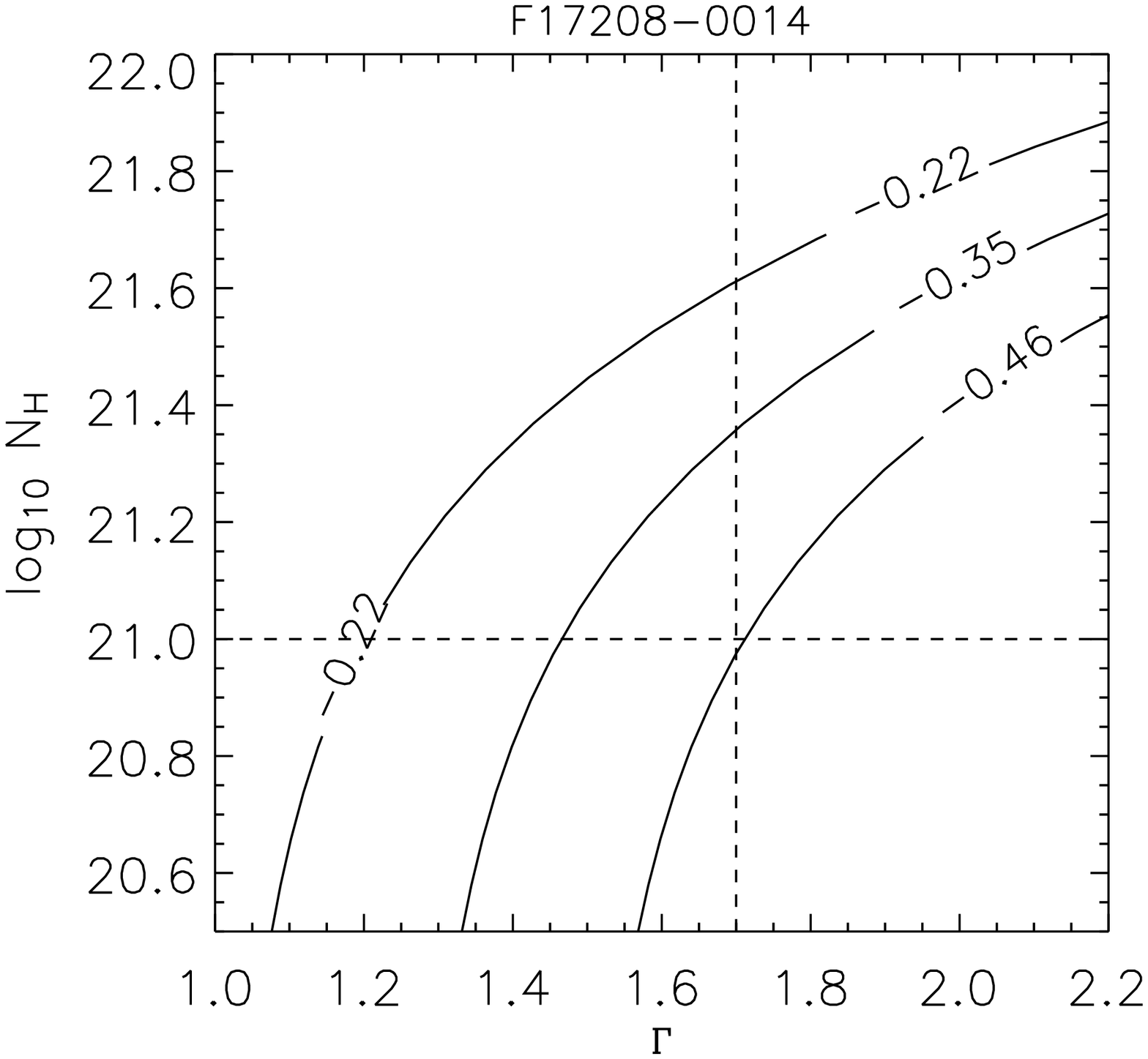}}\rotatebox{90}{\plotone{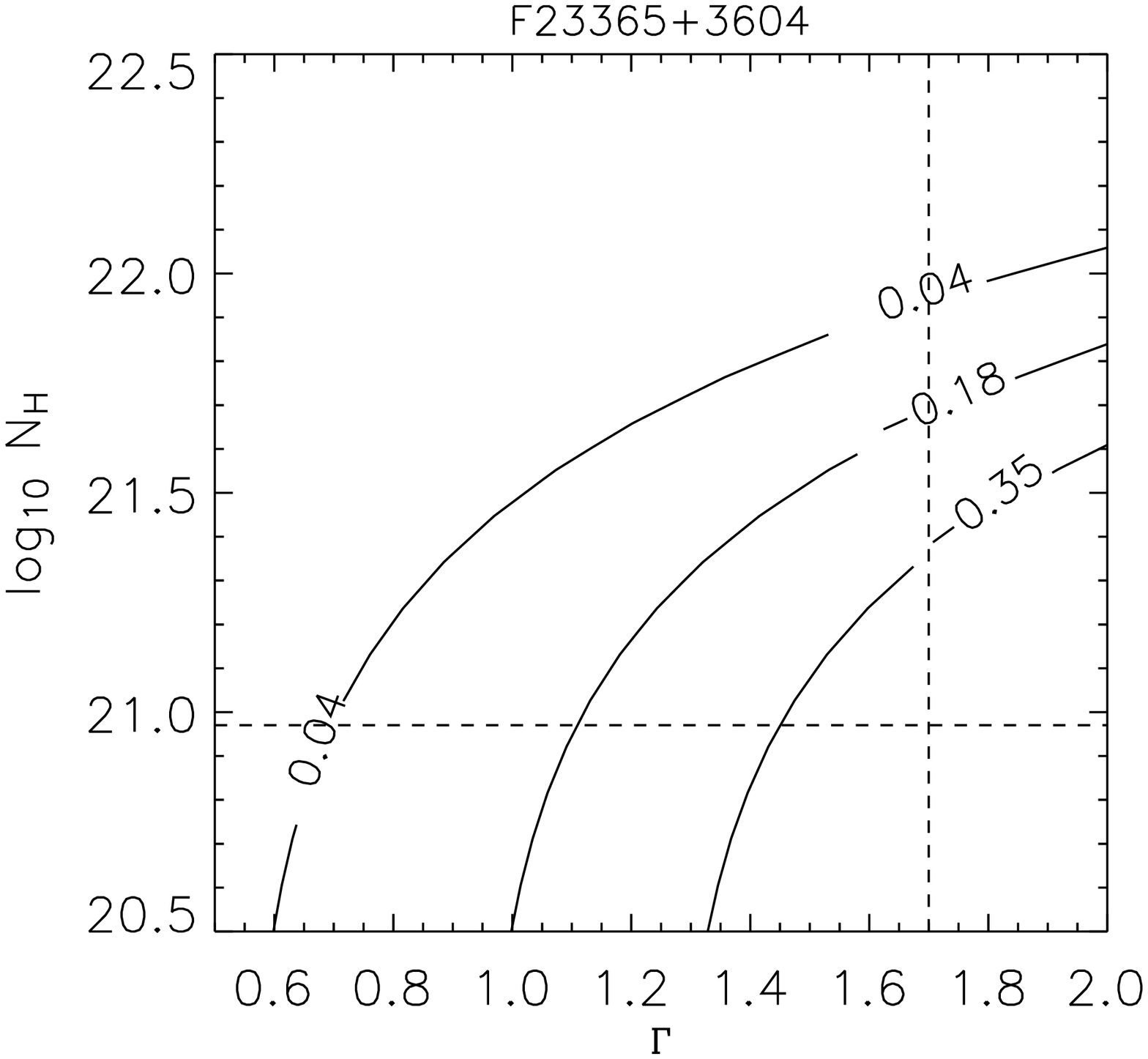}}
\end{center}
\caption{}
\end{figure}

\clearpage

\begin{figure}
\figurenum{8}
\epsscale{0.3}
\begin{center}
\rotatebox{90}{\plotone{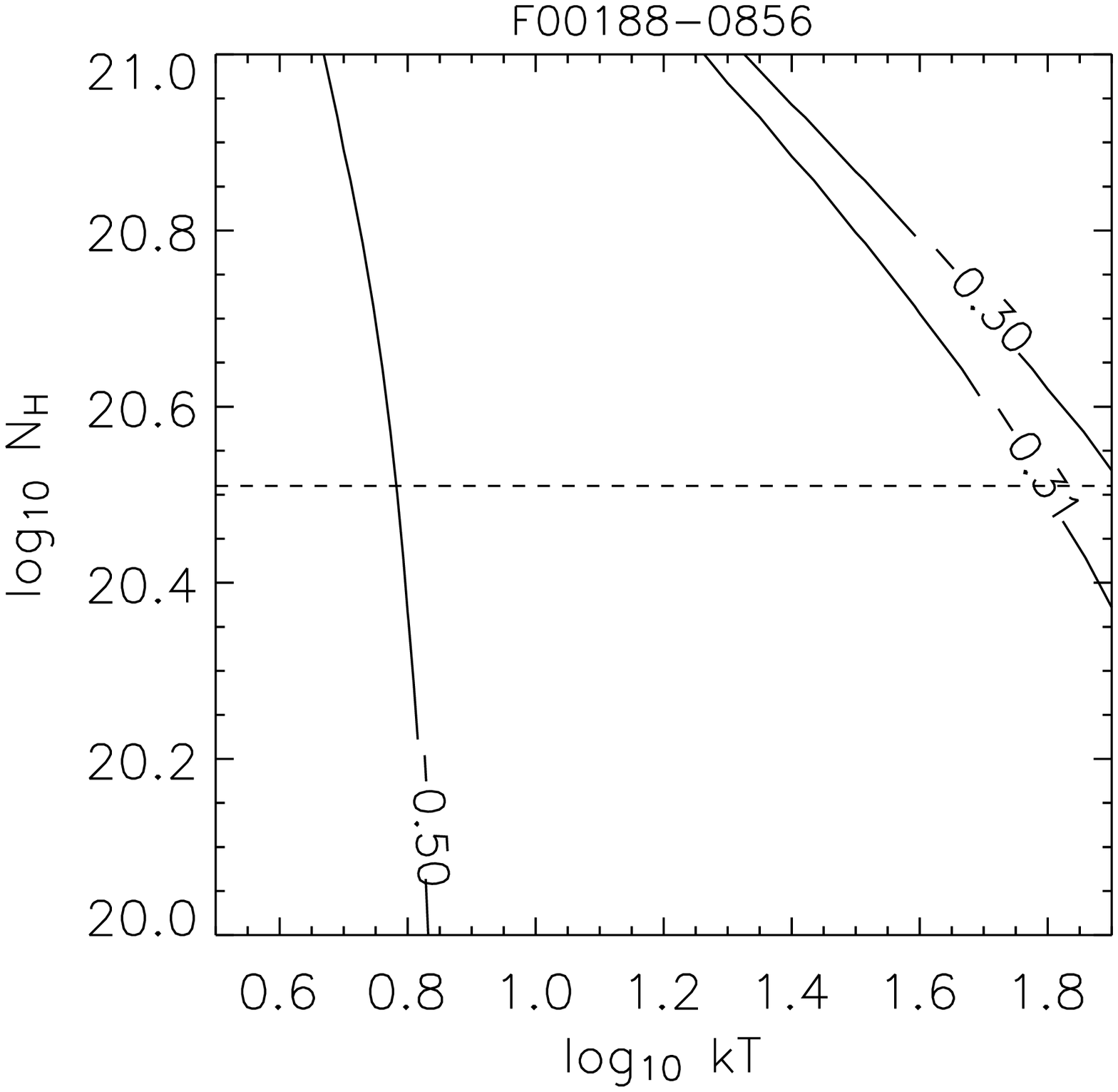}}\rotatebox{90}{\plotone{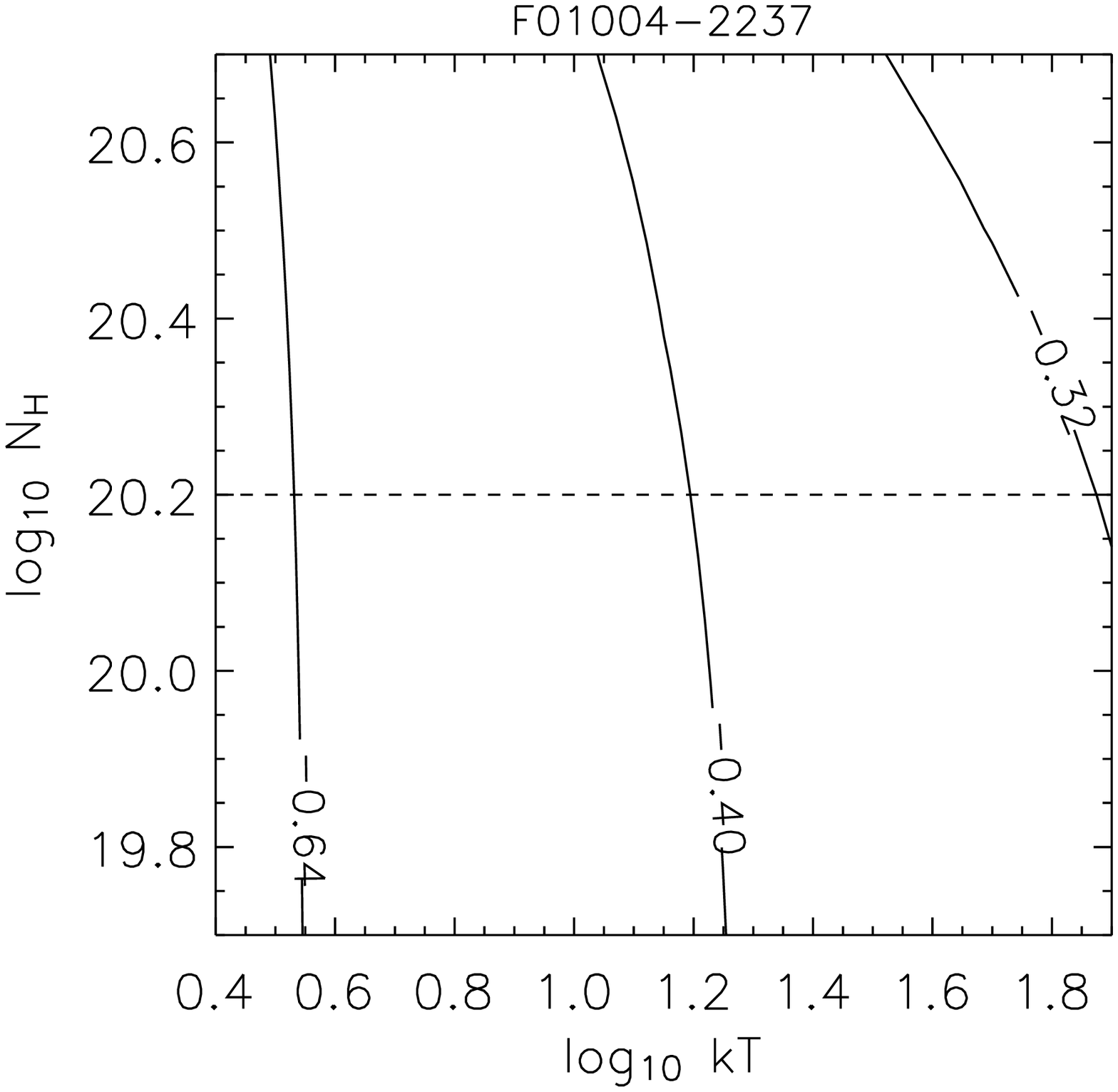}}
\rotatebox{90}{\plotone{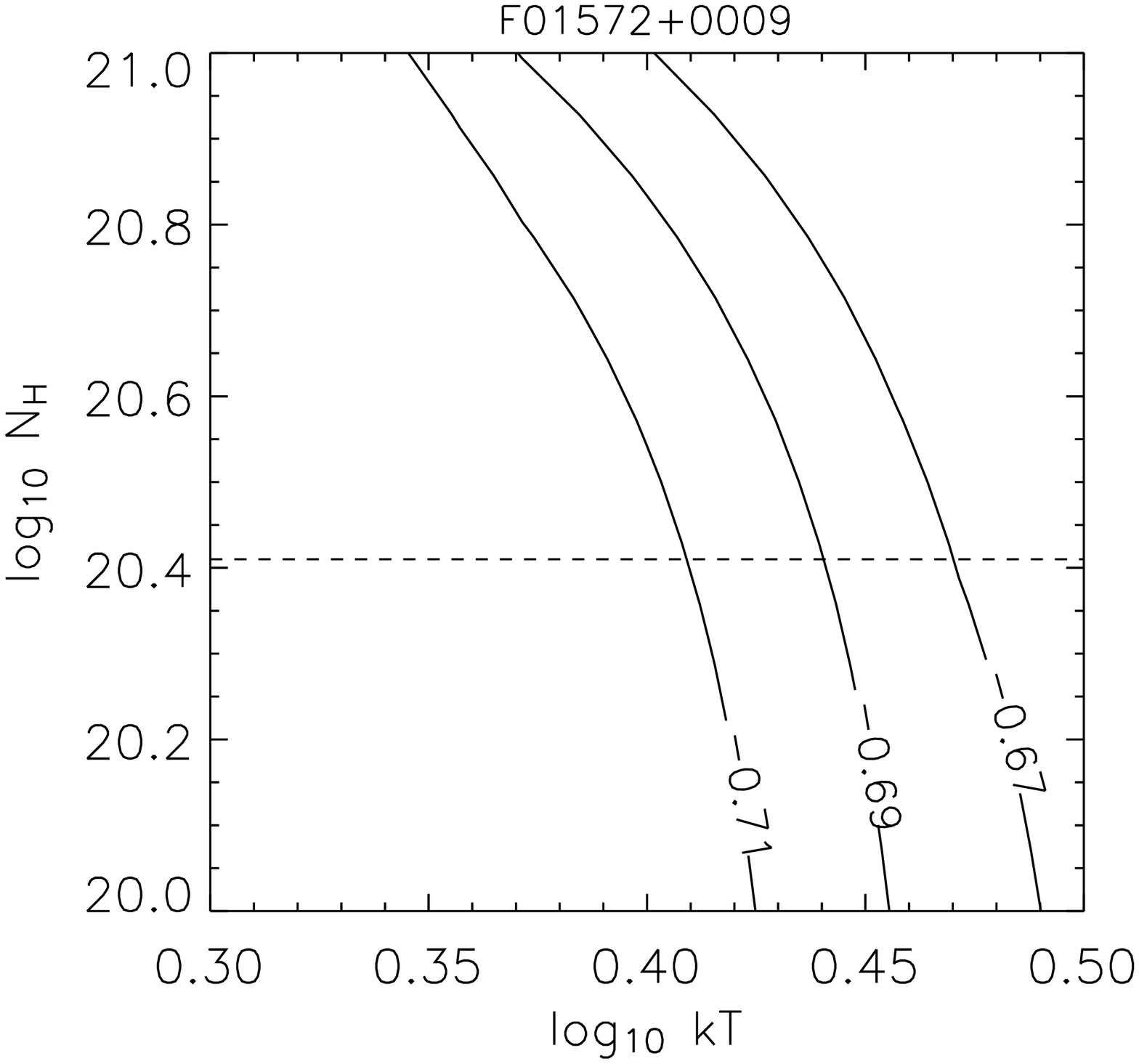}}\rotatebox{90}{\plotone{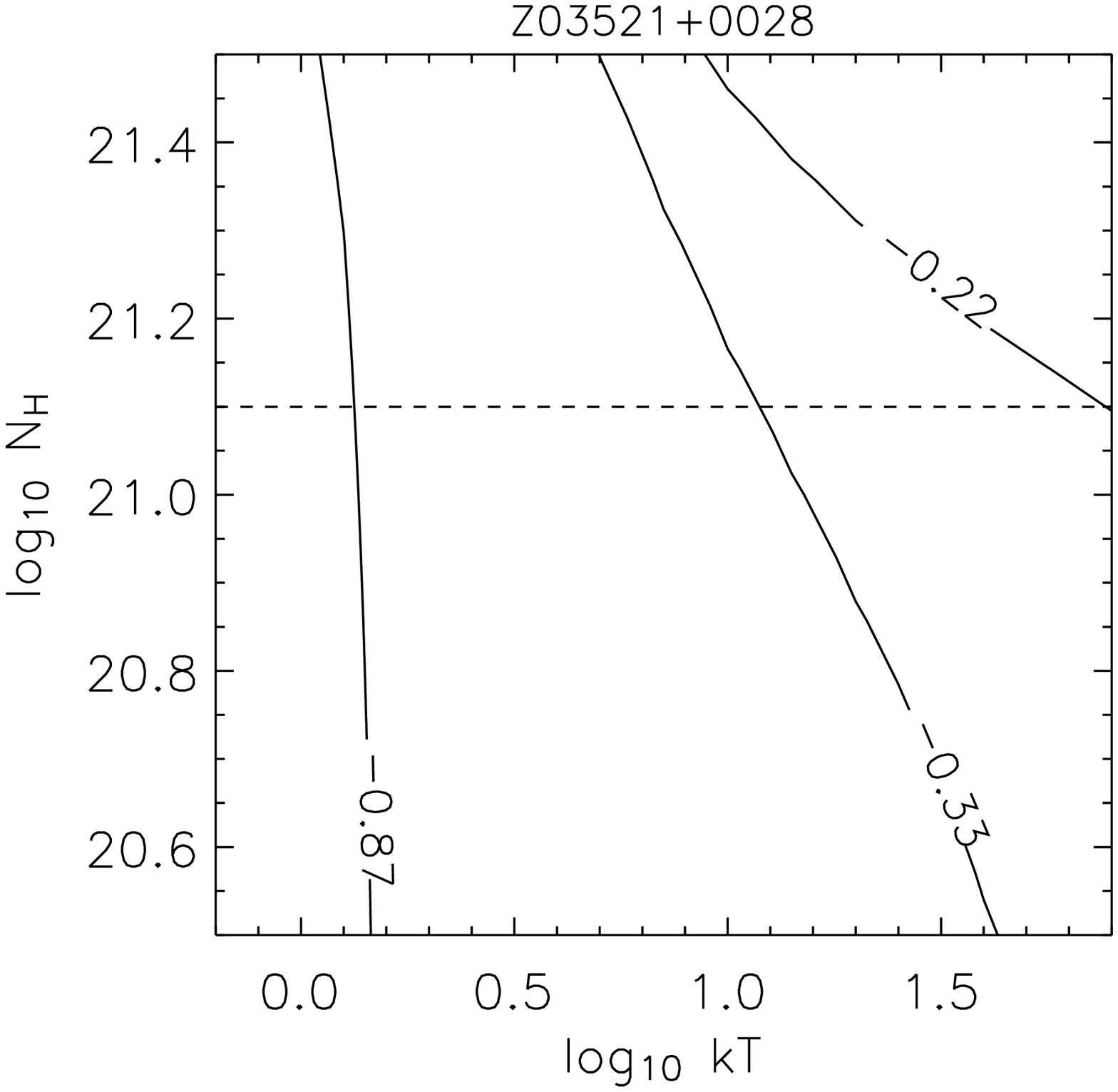}}
\rotatebox{90}{\plotone{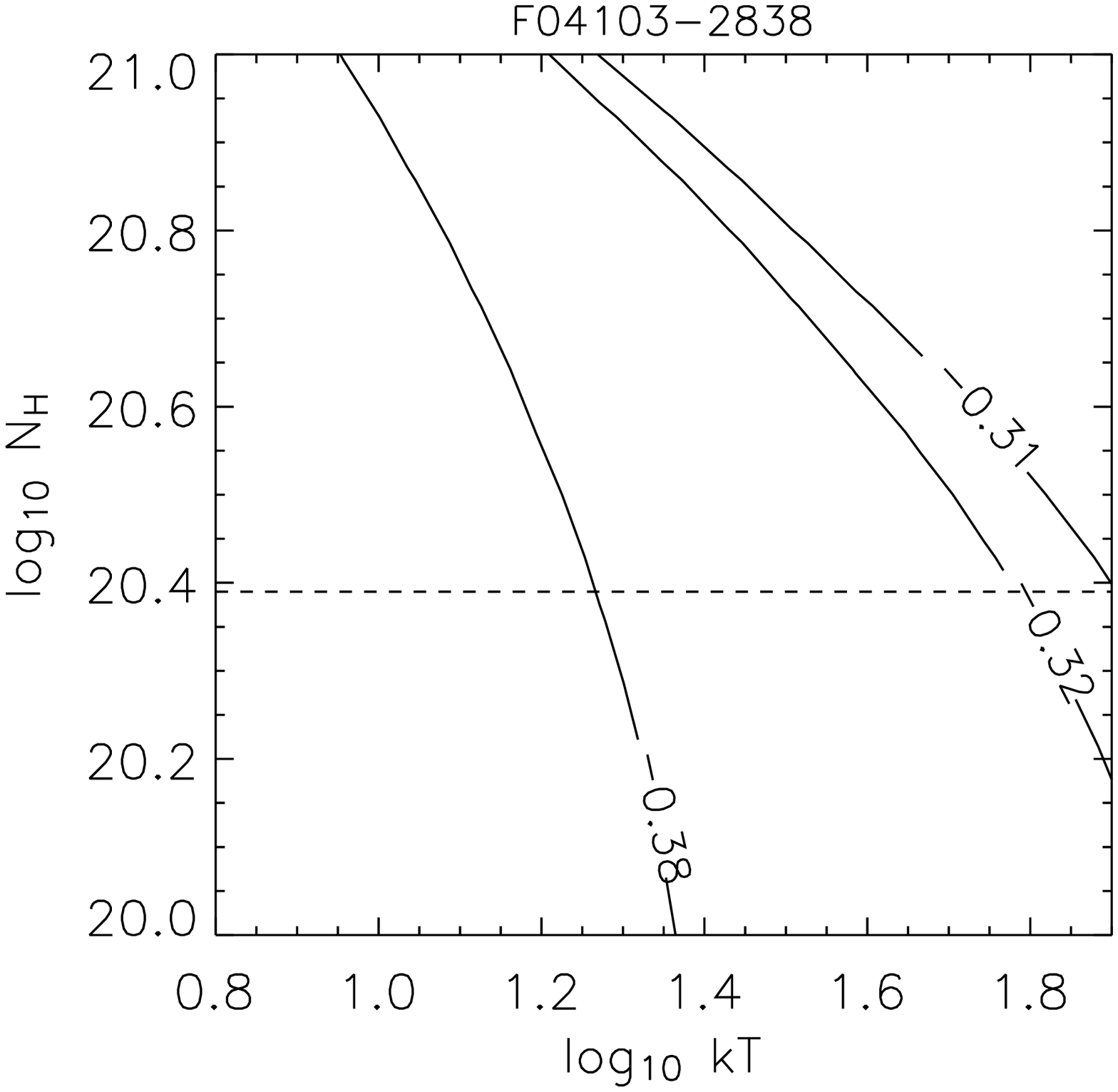}}\rotatebox{90}{\plotone{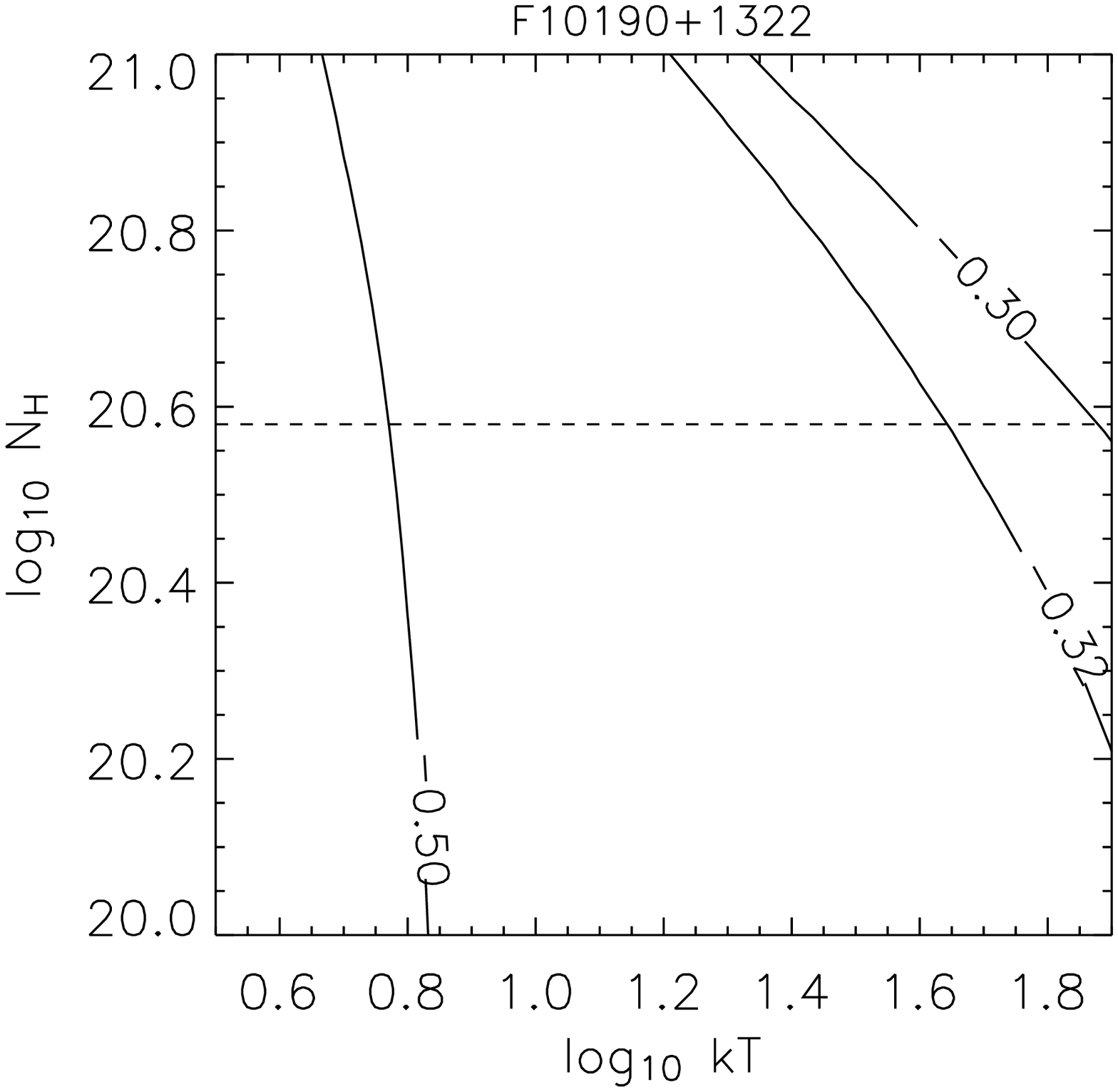}}
\end{center}
\caption{Each plot shows contours of constant hardness ratio (defined in Equation~\ref{eq:hreq}) in the N$\rm{_H}$ versus kT plane.  The middle curve represents, and is labelled with, the observed hardness ratio, while the other two curves represent hardness ratios 1-$\sigma$ away from the observed value (see Table~\ref{tab:hrs}).  The horizontal dashed line represents the Galactic hydrogen column density.}
\label{fig:kt}
\end{figure}

\clearpage

\begin{figure}
\figurenum{8 {\it cont}}
\epsscale{0.3}
\begin{center}
\rotatebox{90}{\plotone{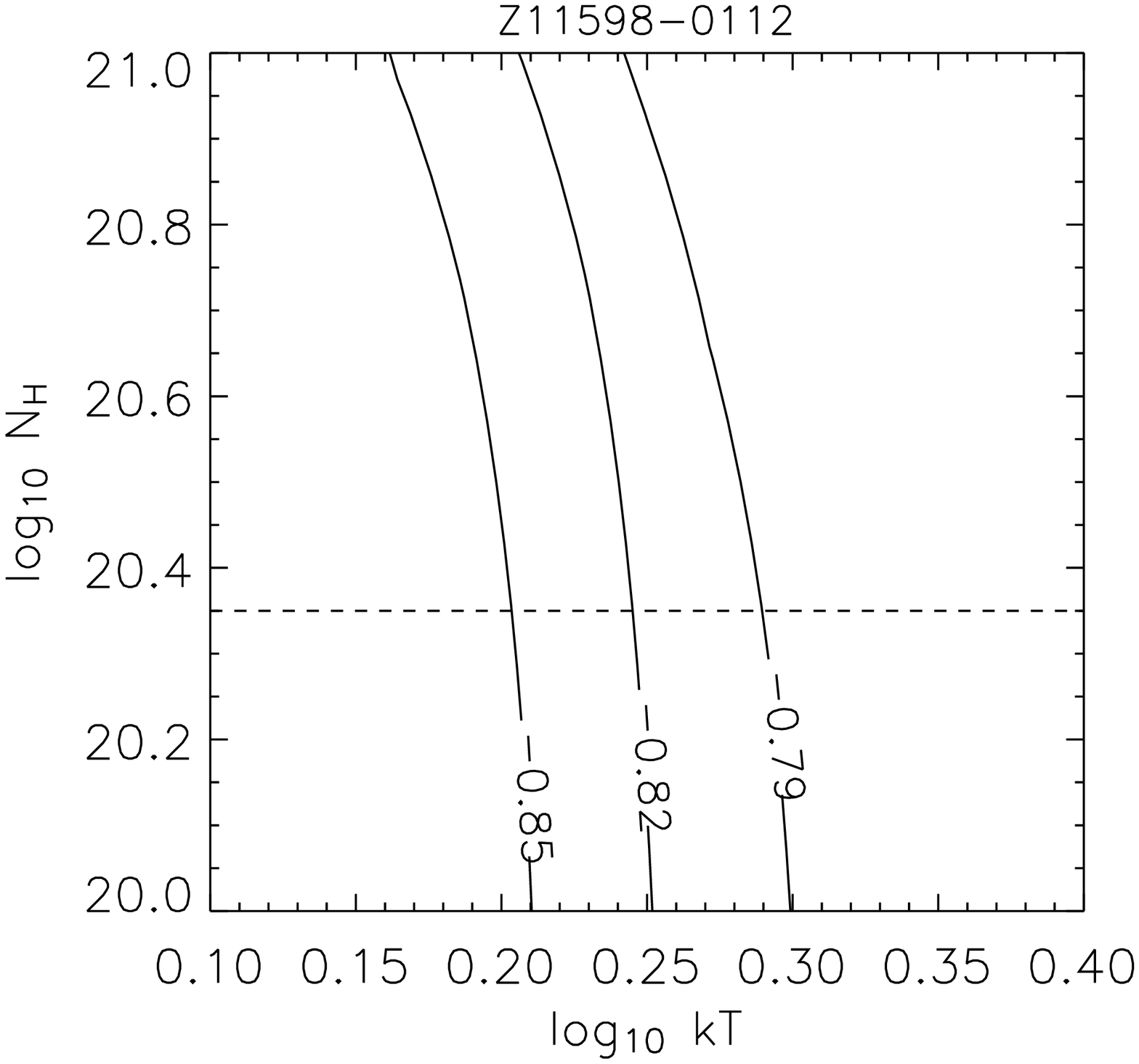}}\rotatebox{90}{\plotone{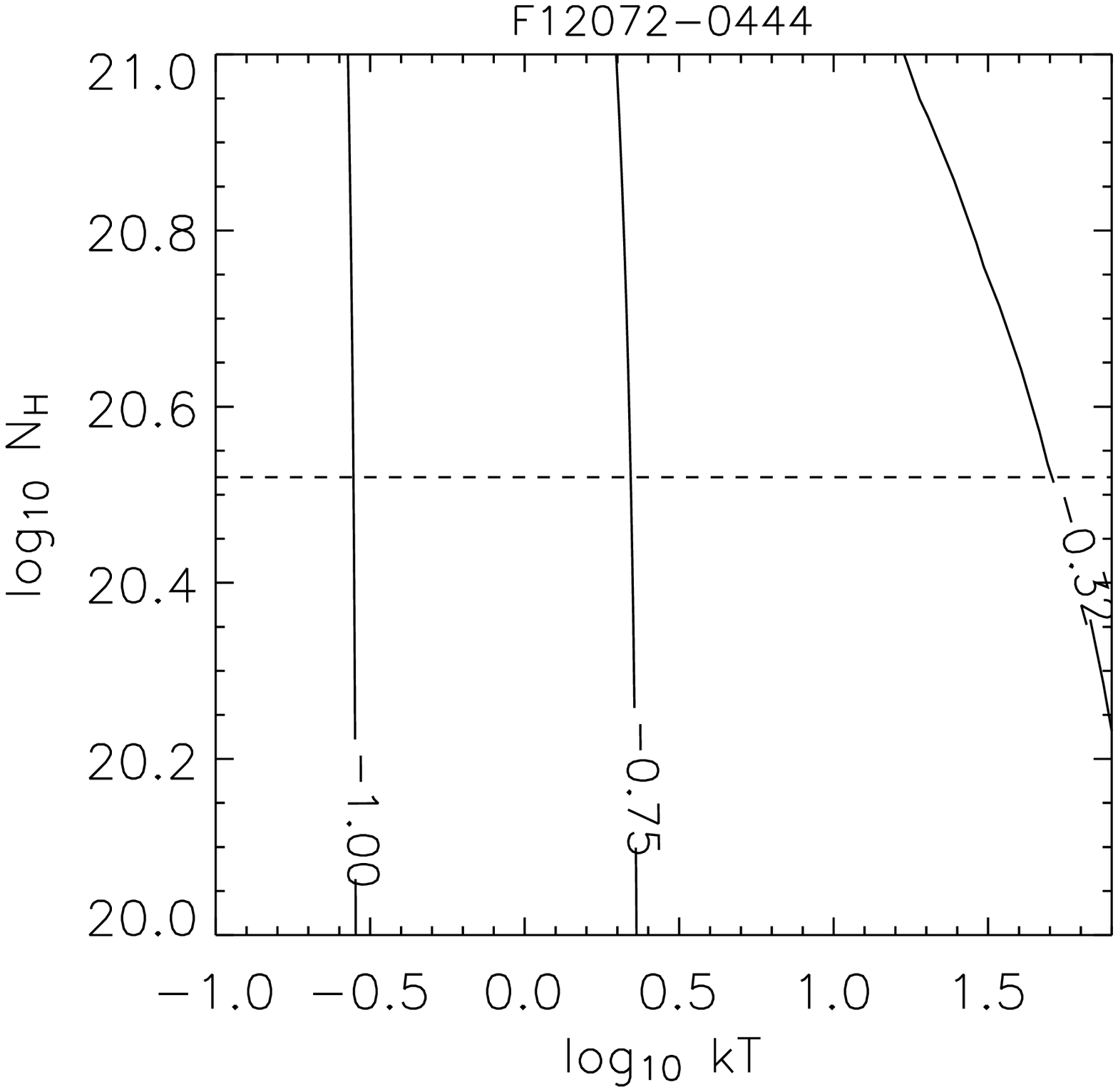}}
\rotatebox{90}{\plotone{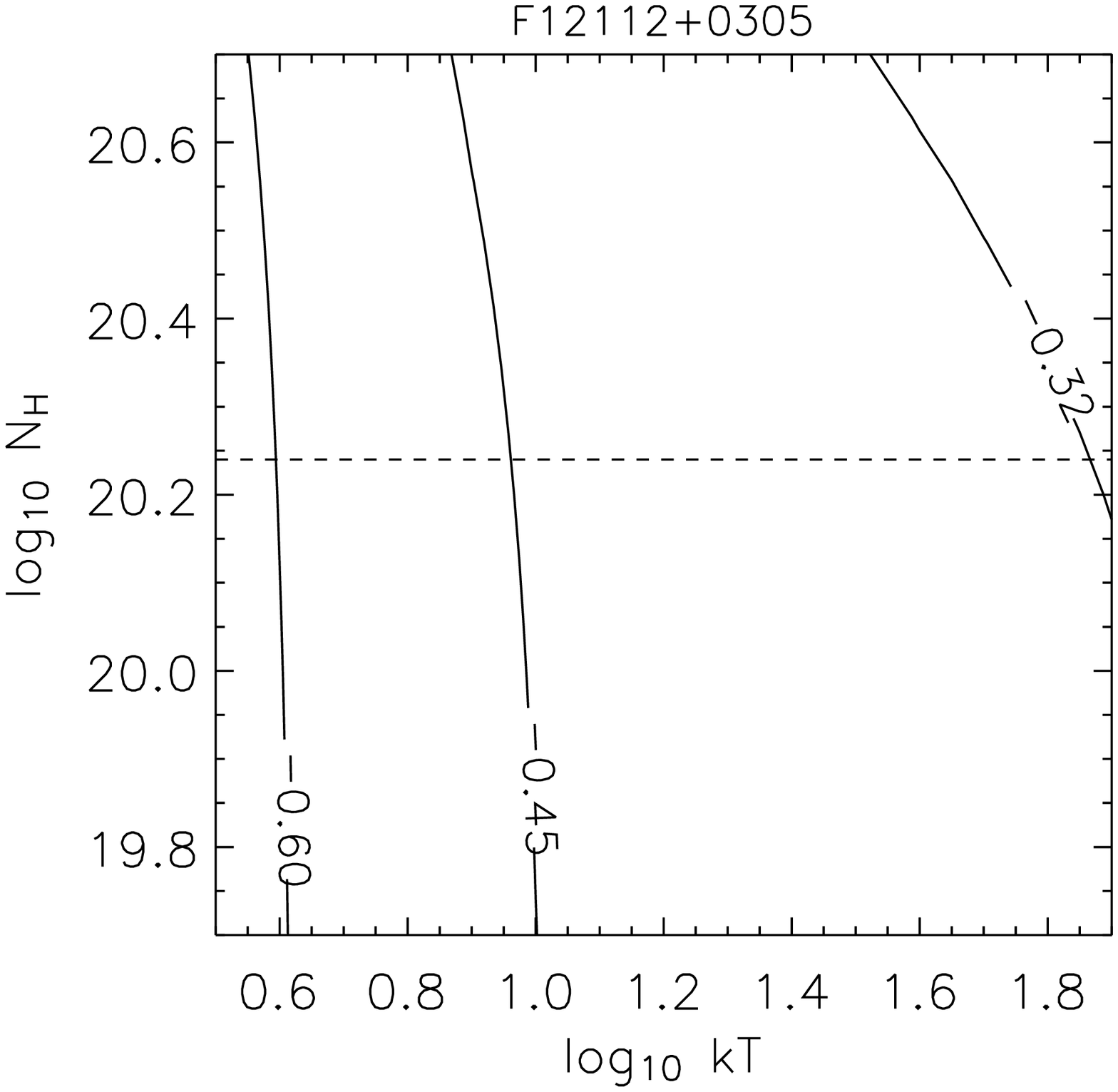}}\rotatebox{90}{\plotone{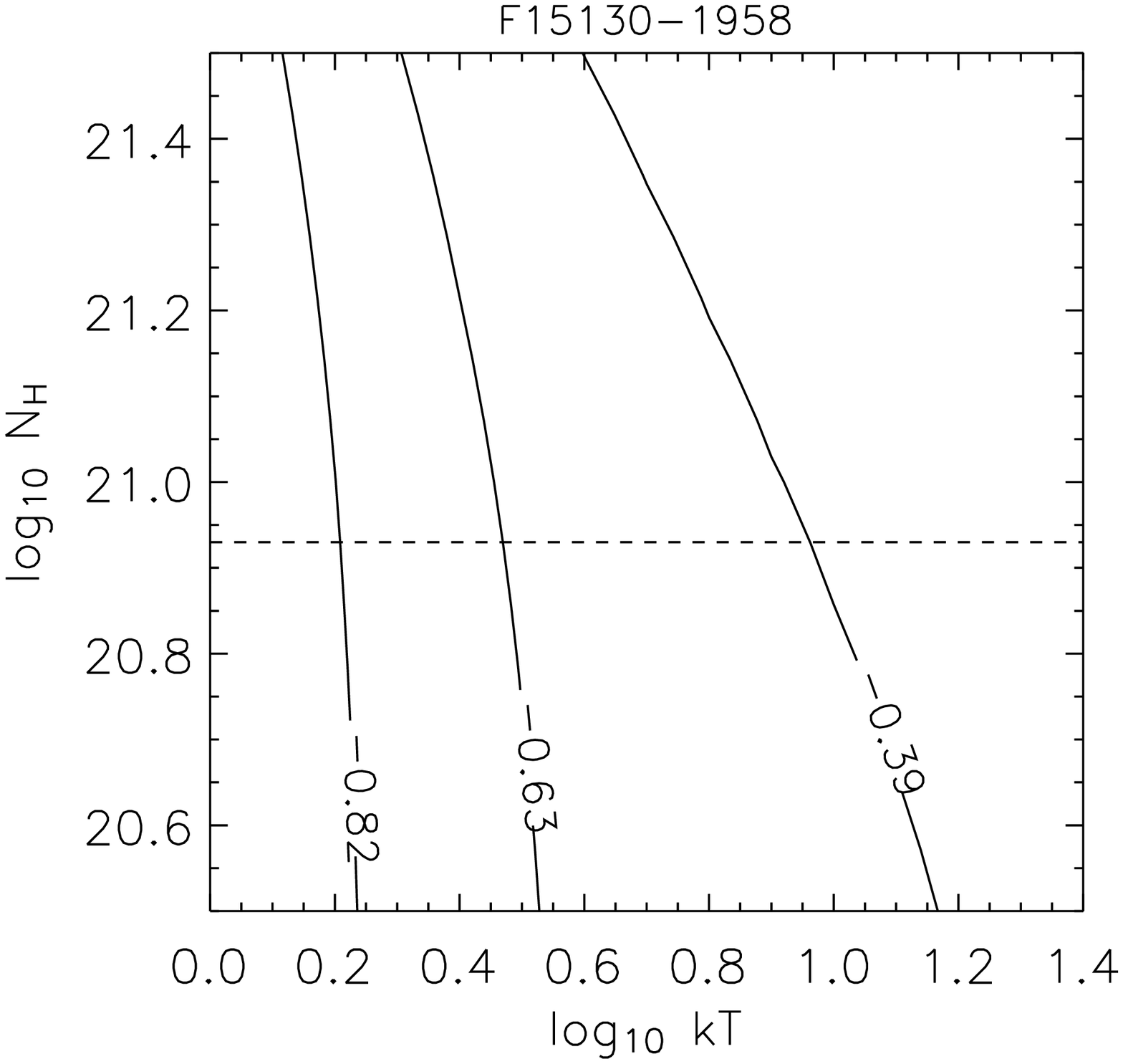}}
\rotatebox{90}{\plotone{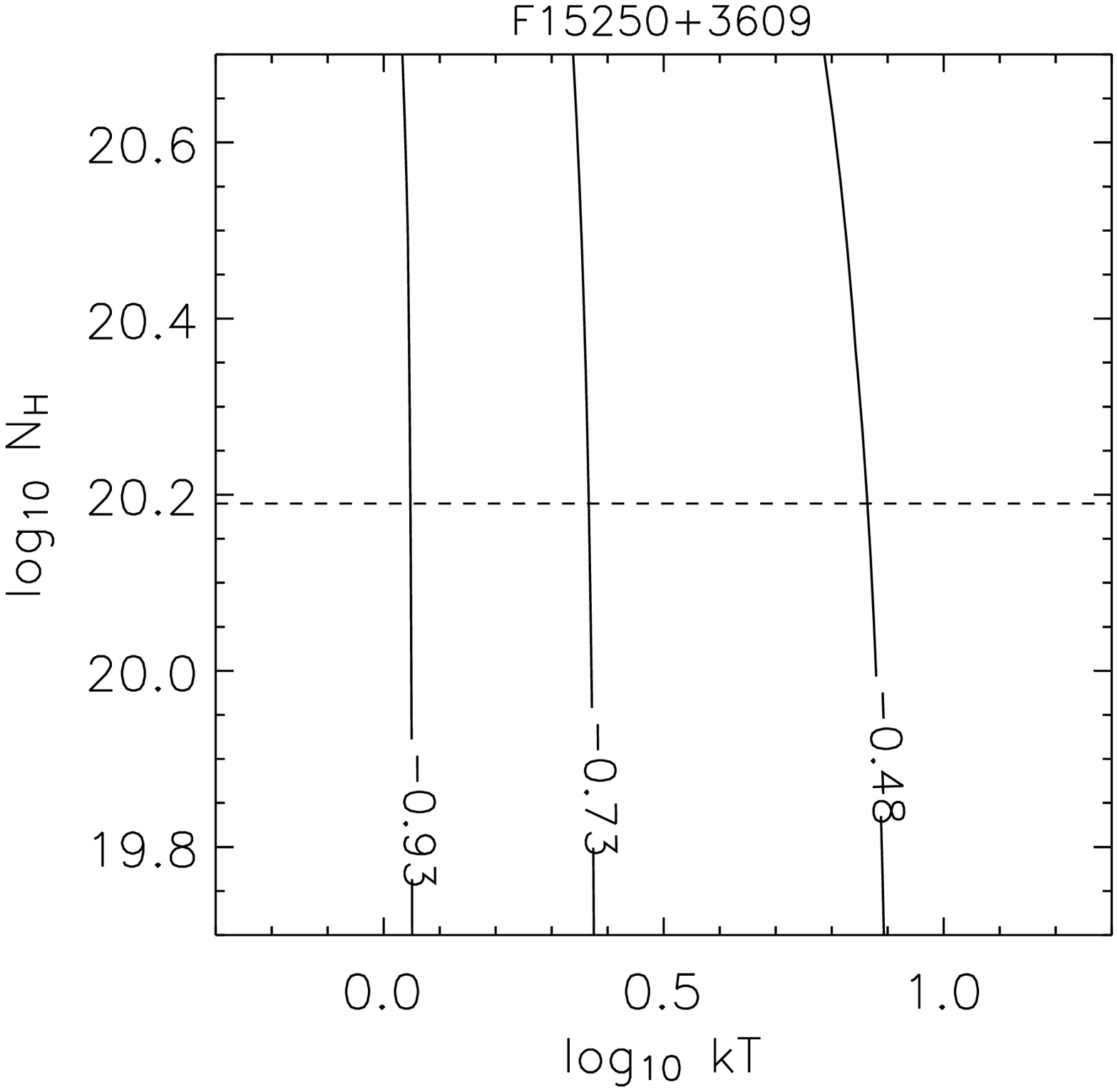}}\rotatebox{90}{\plotone{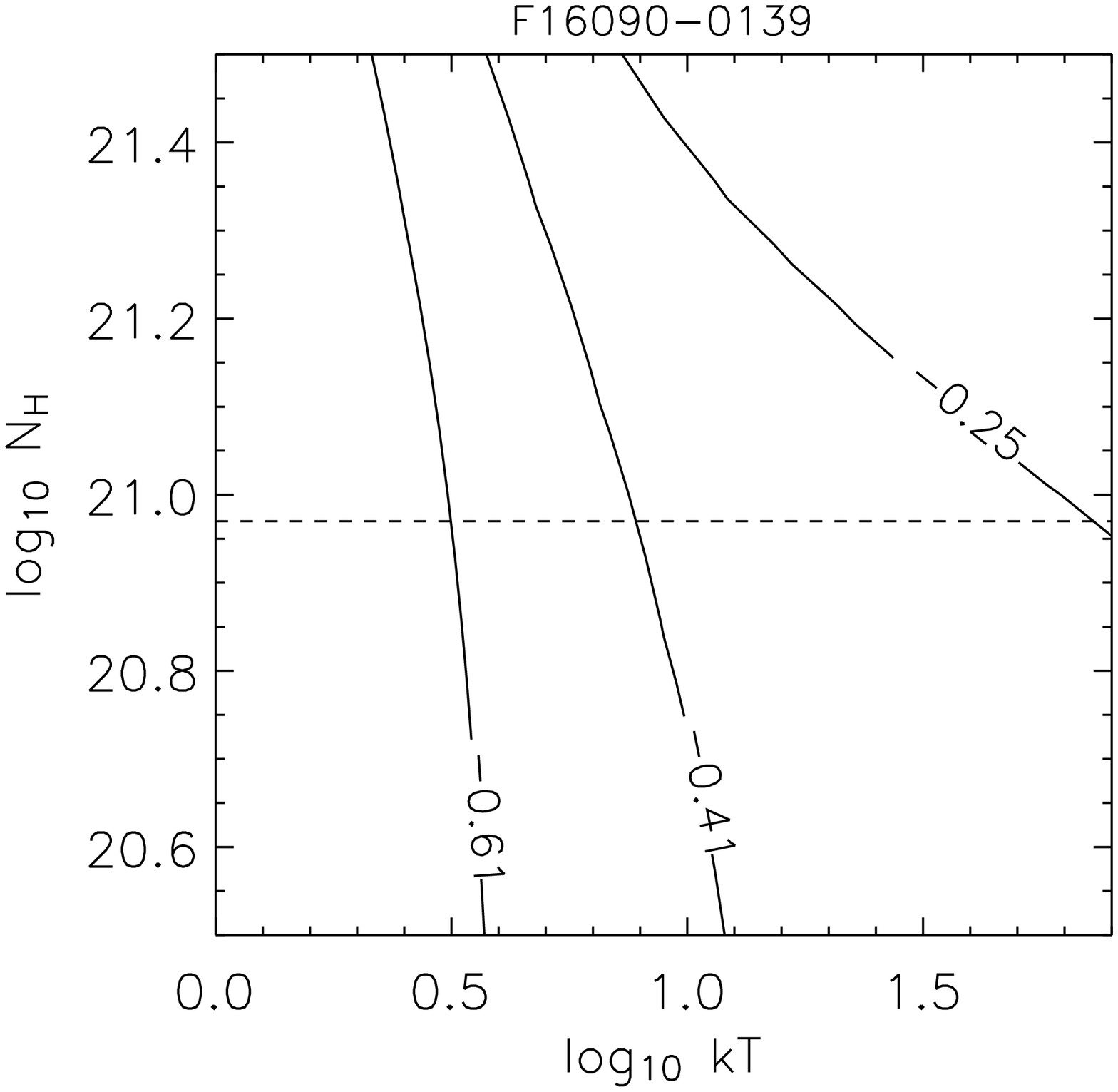}}
\rotatebox{90}{\plotone{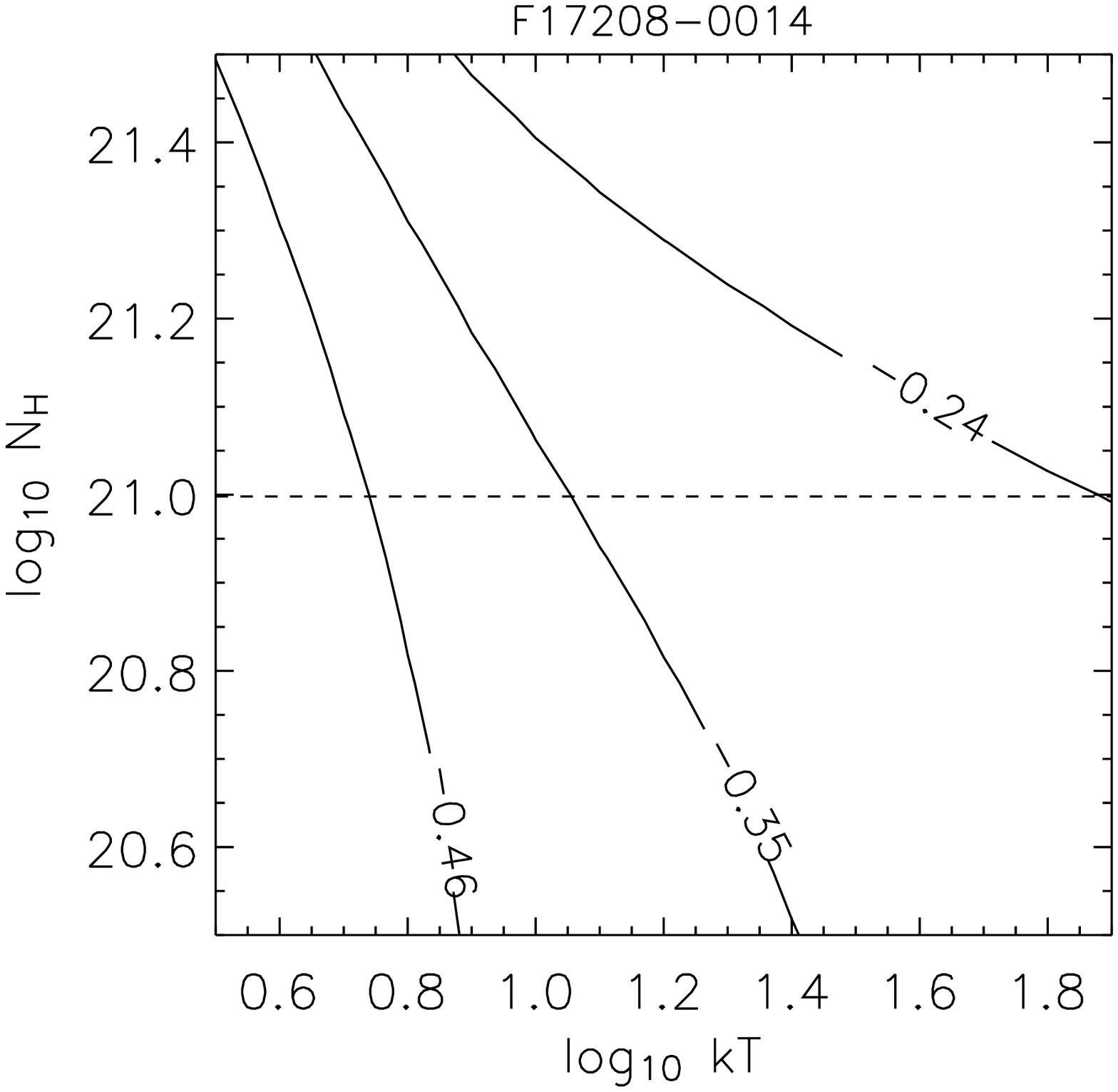}}\rotatebox{90}{\plotone{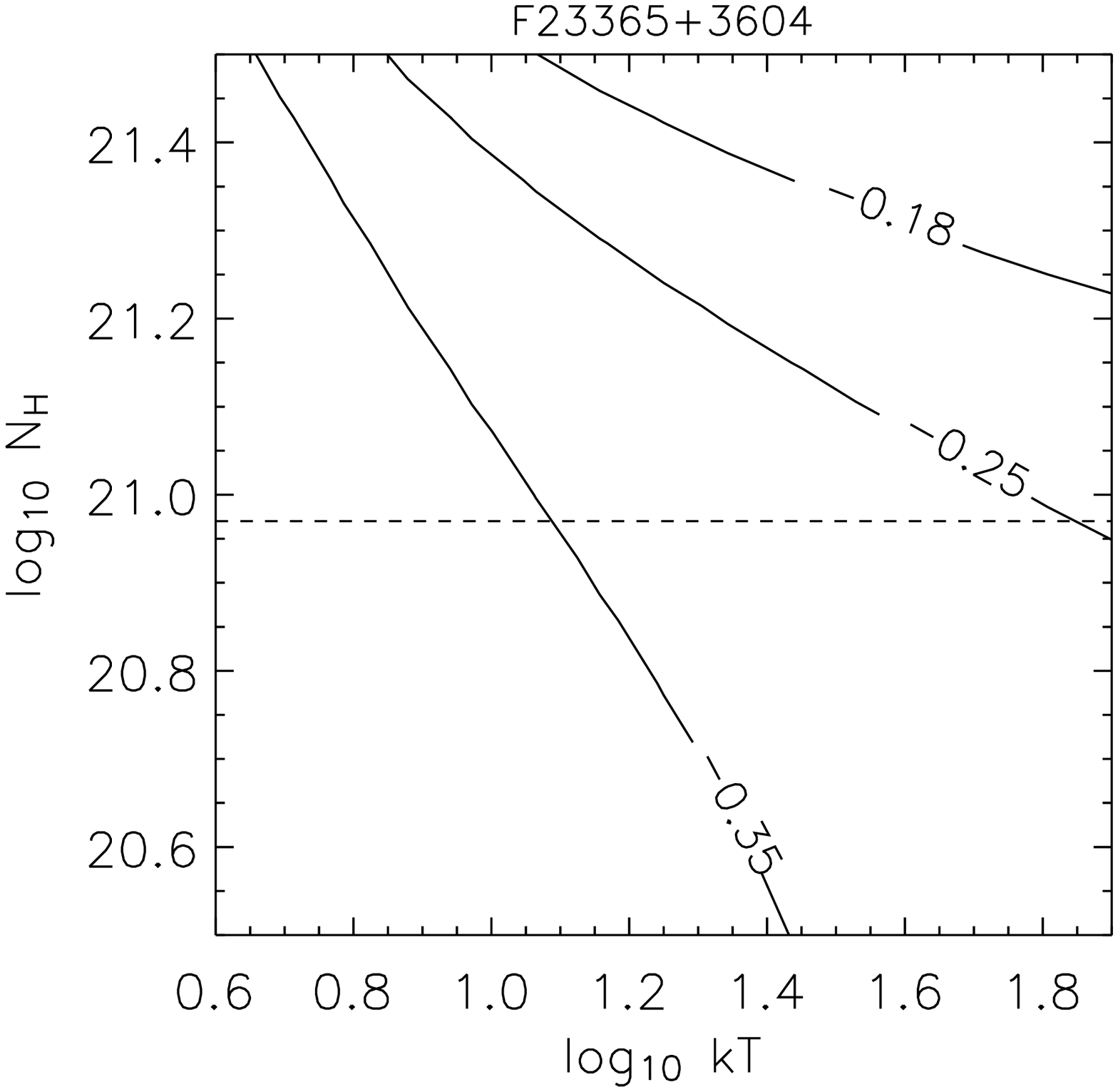}}
\end{center}
\caption{}
\end{figure}

\clearpage

\begin{figure}
\figurenum{9}
\epsscale{0.7}
\rotatebox{-90}{\plotone{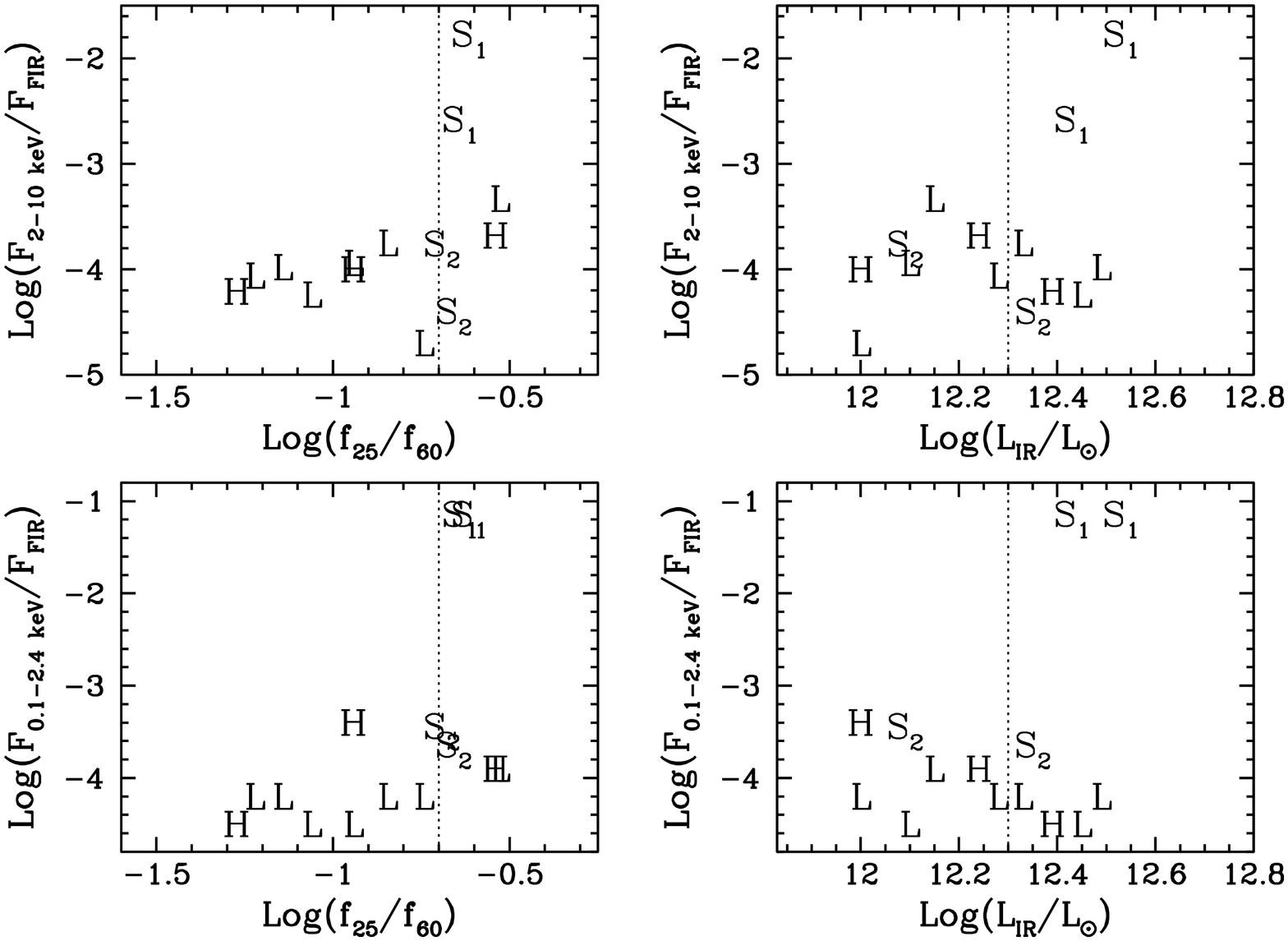}}
\caption{Plots of the log of the ratio of hard X-ray (2.0--10~keV) to far-infrared flux (calculated from Equation~\ref{eq:firflux}) and the log of the ratio of soft X-ray (0.1--2.4~keV) flux to far-infrared flux versus the log of the 25$\mu$m to 60$\mu$m flux ratio and the log of the infrared luminosity between 8 and 1000$\mu$m (the last two being the axes of Figure 1).  The symbols represent the optical spectral classifications.  S$_1$ represents type 1 Seyfert galaxies, S$_2$ type 2 Seyfert galaxies, L LINERs, and H H {\small II} regions.  There is a clear segregation between the two bright Seyfert~1 ULIRGs and the rest of our sample.  The Seyfert~1's have high X-ray luminosity, high infrared luminosity, high X-ray to infrared flux ratios, and ``warm'' colors.  The 2-10~keV flux was calculated from F$\rm{_{HX}}$ in a similar manner to the way in which F$\rm{_{SX1}}$ was calculated from F$\rm{_{SX}}$ (see text, Equation~\ref{eq:xconvert}).  The dotted lines represent the divisions of the infrared colors and luminosity bins that were used in the selection of the sample for observation with {\it Chandra} (\S~\ref{sec:sample} and Figure~\ref{fig:fluxvl}).}
\label{fig:lumin}
\end{figure}

\clearpage

\begin{figure}
\figurenum{10}
\epsscale{0.4}
\rotatebox{-90}{\plotone{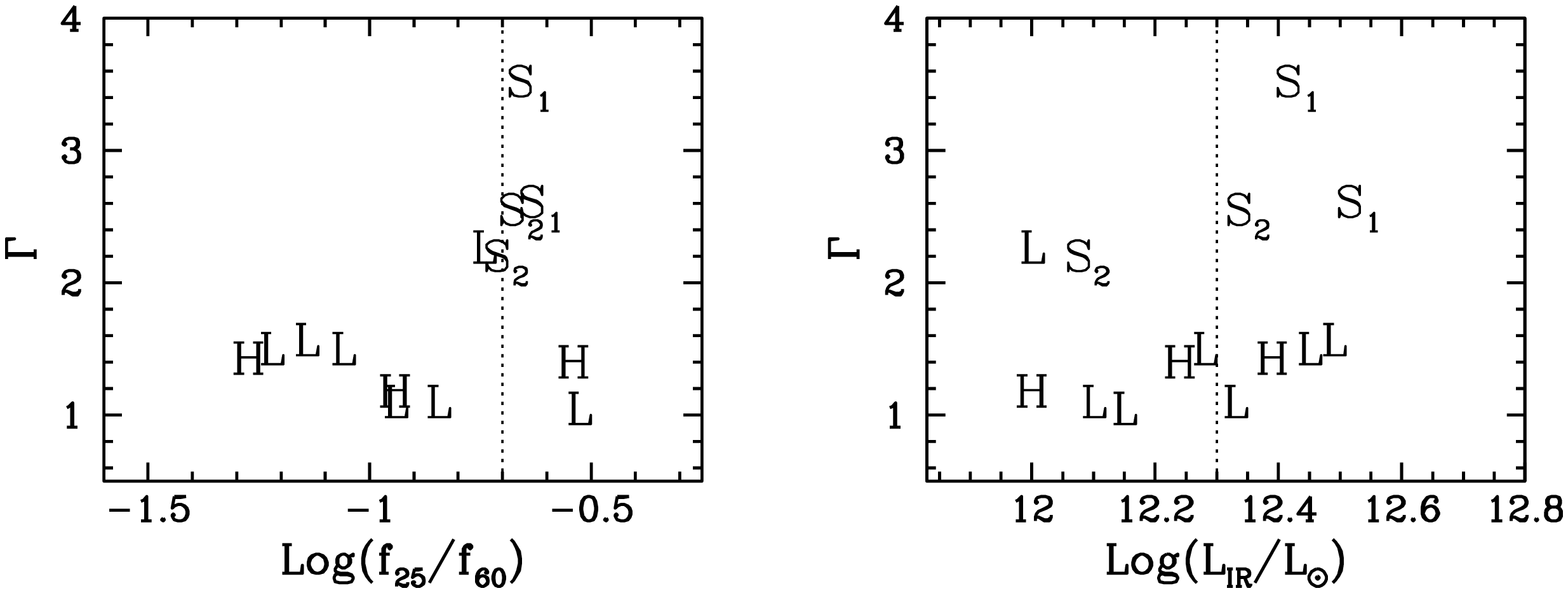}}
\caption{Plot of the photon index ($\Gamma$), assuming the Galactic column, as a function of {\it IRAS} 25--60~$\mu$m colors and infrared luminosities.  With the exception of one LINER, all the non-Seyfert galaxies have photon indices below 2.  There is no clear correlation between the photon index and the infrared flux ratio, or between the photon index and the infrared luminosity.  The plot key is the same as Figure~\ref{fig:lumin} and the dotted lines represent the same divisions of the infrared colors and luminosity bins that were used in the selection of the sample for observation with {\it Chandra} (\S~\ref{sec:sample} and Figure~\ref{fig:fluxvl}).}
\label{fig:gamma_distr}
\end{figure}

\clearpage

\begin{figure}
\figurenum{11}
\epsscale{0.7}
\begin{center}
\rotatebox{-90}{\plotone{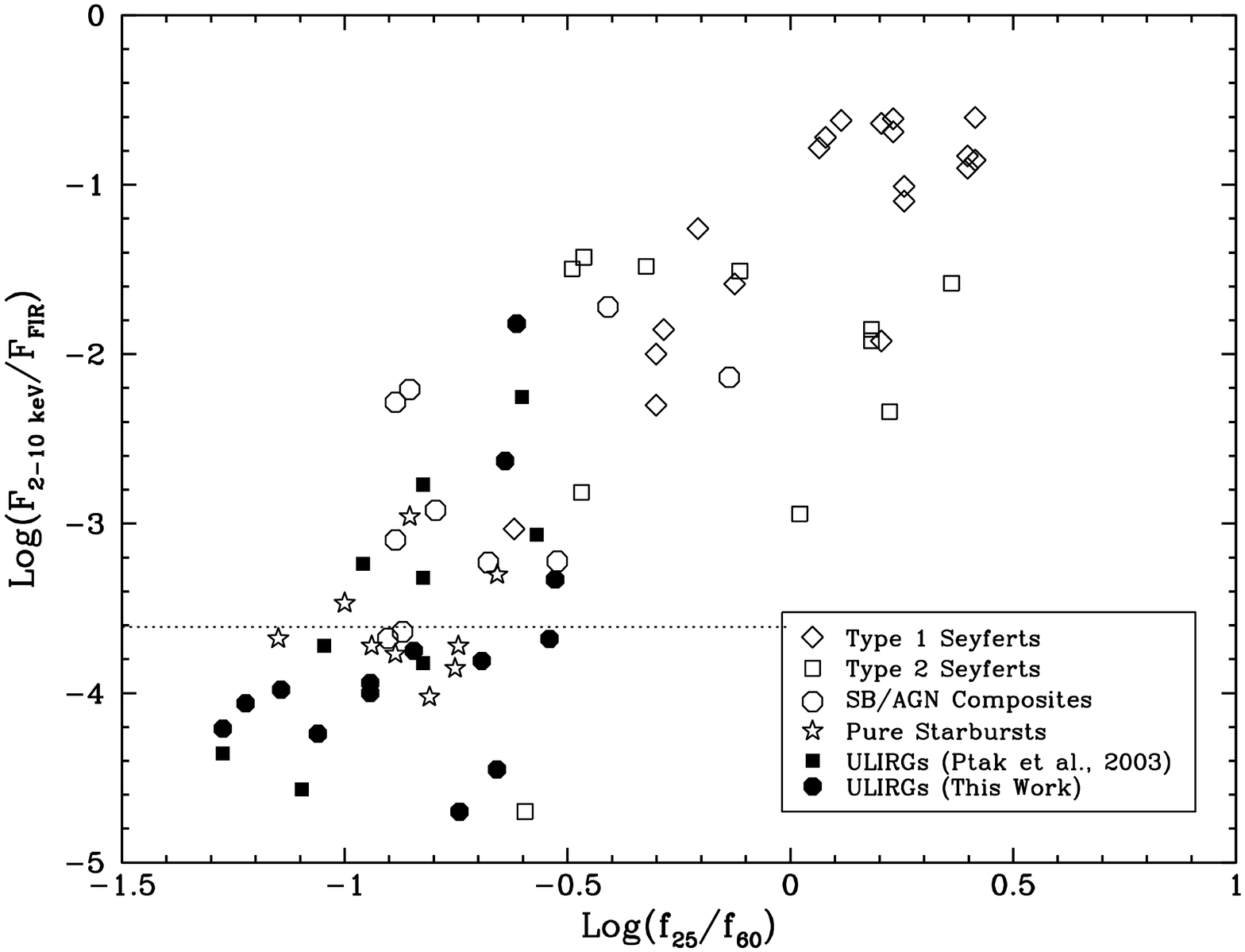}}
\end{center}
\caption{Plot of log($\mathrm {F_{2 - 10~keV}/F_{FIR}}$) vs. log($\rm {f_{25\mu m}/f_{60\mu m}}$).  The type 1 Seyferts in our sample are distributed near the Seyferts, while the others are located among the starbursts and composites.  All members of the 1-Jy sample have Log($\rm {f_{25\mu m}/f_{60\mu m}}$) $\lesssim$ --0.35 (see Figure~\ref{fig:fluxvl}).  Here we have only included the values for the \citet{ptak} ULIRGs derived from their global spectra.  The dotted line represents the average Log($\mathrm {F_{2 - 10~keV}/F _{FIR}}$) values for the pure starbursts.}
\label{fig:ptakplot}
\end{figure}

\clearpage

\begin{figure}
\figurenum{12}
\epsscale{0.7}
\rotatebox{-90}{\plotone{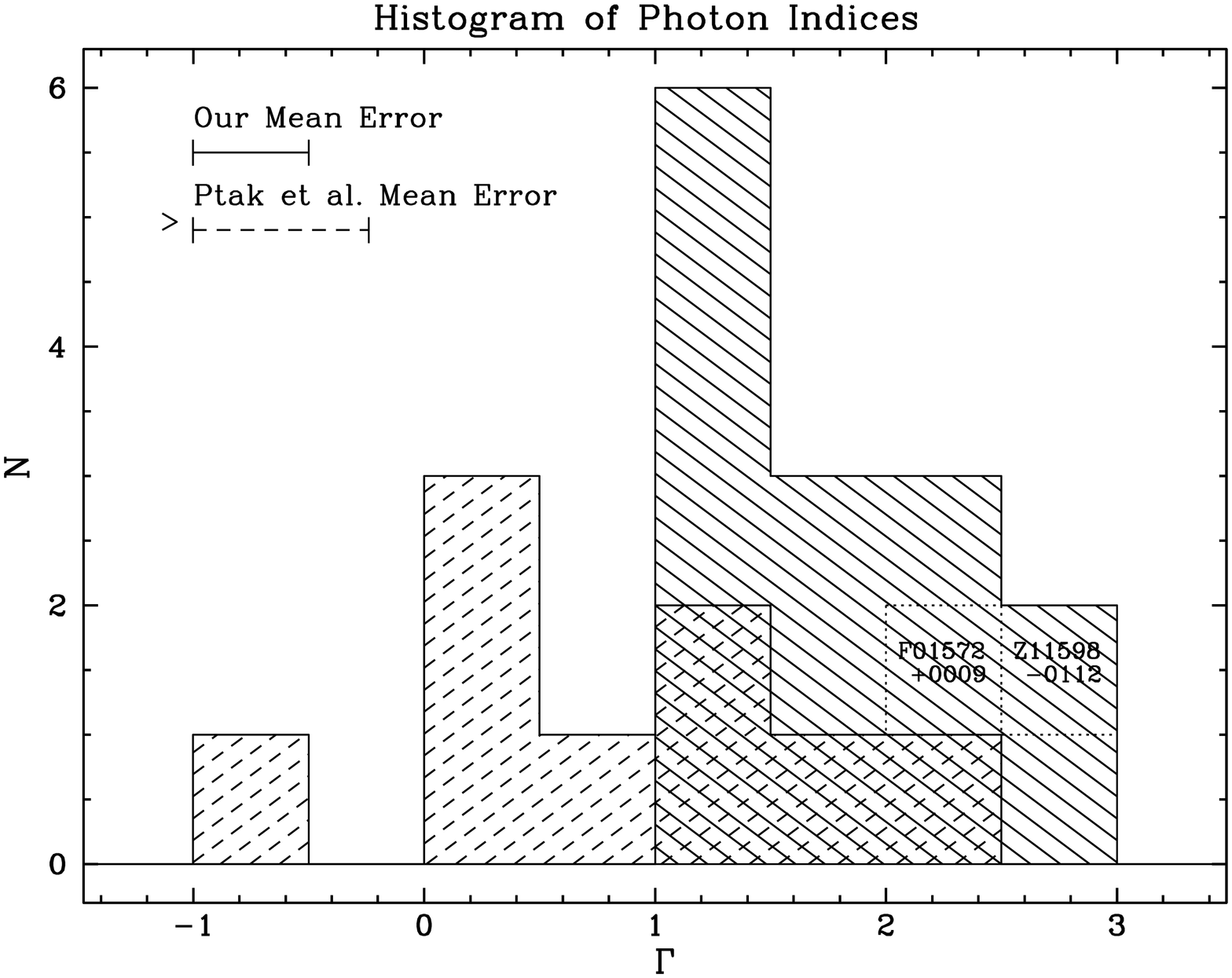}}
\caption{A comparison of the photon indices of a single-power-law model between our sample (solid hashes) and the \citet{ptak} sample (dashed hashes).  The boxes corresponding to the two Seyfert~1 nuclei in our sample (F01572+0009 and Z11598-0112) are indicated; their photon indices were calculated from the ``hardness ratio'' method (Table~\ref{tab:hrs}) as for the rest of the sample.  The histogram indicates that the spectra from our sample are softer than those from \citet{ptak}.  Note that \citet{ptak} quoted error bars only for models with $\chi ^2/dof<1.5$.  Therefore, the mean error for their sample was calculated from models for two of the ten galaxies in their sample.}
\label{fig:hist}
\end{figure}

\clearpage

\begin{figure}
\figurenum{13}
\epsscale{0.7}
\rotatebox{-90}{\plotone{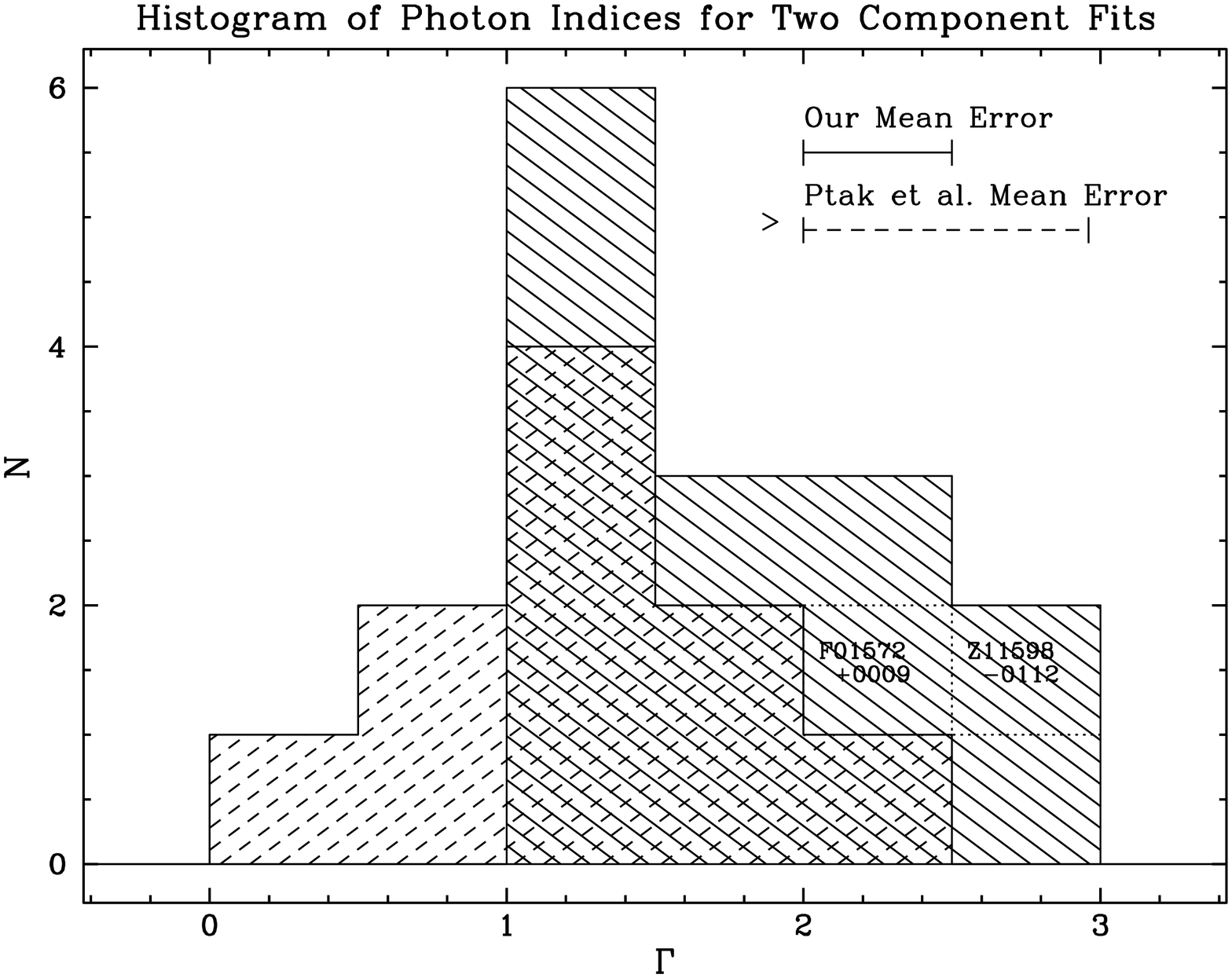}}
\caption{A comparison between the power-law photon indices of our sample (solid hashes) and the \citet{ptak} sample (dashed hashes).  The boxes corresponding to the two Seyfert~1 nuclei in our sample (F01572+0009 and Z11598-0112) are indicated; their single power law photon indices were calculated from the ``hardness ratio'' method (Table~\ref{tab:hrs}) as for the rest of the sample.  The $\Gamma$'s for the \citet{ptak} data are their 2--10~keV results from the plasma plus power law models.  The histogram indicates that the spectra from our sample are softer than those from \citet{ptak}.  \citet{ptak} quoted error bars only for models with $\chi ^2/dof<1.5$.  Therefore, the mean error for their sample was calculated from models for seven of the ten galaxies in their sample.}
\label{fig:hist2}
\end{figure}

\clearpage

\begin{figure}
\figurenum{14}
\epsscale{0.7}
\rotatebox{-90}{\plotone{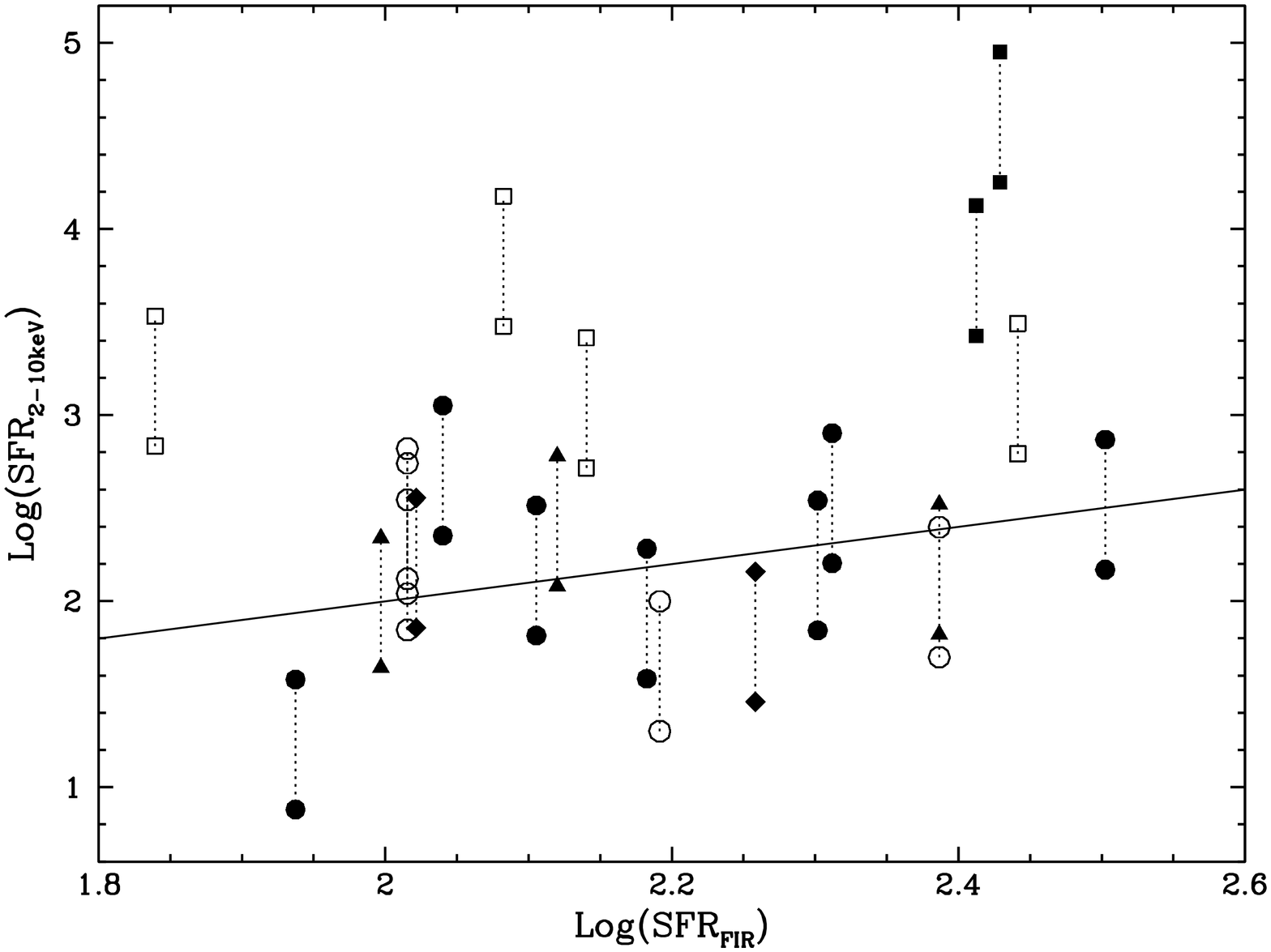}}
\caption{A plot of SFR$_{\mathrm {2 - 10~keV}}$ as a function of SFR$_{\rm FIR}$.  The star formation rates are in solar masses per year.  For each galaxy, two values of SFR$_{\mathrm {2 - 10~keV}}$ (connected by dotted lines) are plotted based on the relations from \citet{ranalli} and \citet{persic2}.  The top value is from the \citet{persic2} relation, while the bottom value is from the \citet{ranalli} relation.  The solid line, represents the line of equality: SFR$_{\mathrm {2 - 10~keV}}$ = SFR$_{\rm FIR}$.  It should be noted here that NGC~6240 (the leftmost pair of open squares) is a luminous infrared galaxy (LIRG) and is used by \citet{ptak} only for comparison purposes because of its high hard X-ray luminosity compared to starburst ULIRGs.  Plot key: filled squares -- optically classified Seyfert~1 ULIRGs from this work, filled diamonds -- Seyfert~2 ULIRGs from this work, filled triangles -- H {\small II} ULIRGs from this work, filled circles -- LINER ULIRGs from this work, open squares -- \citet{ptak} AGN ULIRGs and open circles -- \citet{ptak} starburst ULIRGs.} 
\label{fig:sfr}
\end{figure}

\end{center}
\end{document}